\documentclass[12pt]{book}

\usepackage{epstopdf,amsfonts,amsmath,amssymb}
\usepackage{graphicx}
\DeclareGraphicsExtensions{.eps}

\newcommand\encadremath[1]{\vbox{\hrule\hbox{\vrule\kern8pt
\vbox{\kern8pt \hbox{$\displaystyle #1$}\kern8pt}
\kern8pt\vrule}\hrule}}
\def\enca#1{\vbox{\hrule\hbox{
\vrule\kern8pt\vbox{\kern8pt \hbox{$\displaystyle #1$}
\kern8pt} \kern8pt\vrule}\hrule}}

\newcommand\framefig[1]{
\begin{figure}[bth]
\hrule\hbox{\vrule\kern8pt
\vbox{\kern8pt \vbox{
\begin{center}
{#1}
\end{center}
}\kern8pt}
\kern8pt\vrule}\hrule
\end{figure}
}

\newcommand\figureframex[3]{
\begin{figure}[bth]
\hrule\hbox{\vrule\kern8pt
\vbox{\kern8pt \vbox{
\begin{center}
{\mbox{\epsfxsize=#1.truecm\epsfbox{#2}}}
\end{center}
\caption{#3}
}\kern8pt}
\kern8pt\vrule}\hrule
\end{figure}
}
\newcommand\figureframey[3]{
\begin{figure}[bth]
\hrule\hbox{\vrule\kern8pt
\vbox{\kern8pt \vbox{
\begin{center}
{\mbox{\epsfysize=#1.truecm\epsfbox{#2}}}
\end{center}
\caption{#3}
}\kern8pt}
\kern8pt\vrule}\hrule
\end{figure}
}

\makeatletter
\@addtoreset{equation}{section}
\makeatother
\newtheorem{theorem}{Theorem}[section]

\newtheorem{remark}{Remark}[section]
\newtheorem{proposition}{Proposition}[section]
\newtheorem{lemma}{Lemma}[section]
\newtheorem{corollary}{Corollary}[section]
\newtheorem{definition}{Definition}[section]
\def\br{\begin{remark}\rm\small}
\def\er{\end{remark}}
\def\bt{\begin{theorem}}
\def\et{\end{theorem}}
\def\bd{\begin{definition}}
\def\ed{\end{definition}}
\def\bp{\begin{proposition}}
\def\ep{\end{proposition}}
\def\bl{\begin{lemma}}
\def\el{\end{lemma}}
\def\bc{\begin{corollary}}
\def\ec{\end{corollary}}
\def\beaq{\begin{eqnarray}}
\def\eeaq{\end{eqnarray}}
\newcommand{\proof}{{\noindent \bf proof:}$\quad$ }
\newcommand{\eproof}{ $\square$ }

\newcommand{\be}{\begin{equation}}
\newcommand{\ee}{\end{equation}}
\newcommand{\beq}{\begin{equation}}
\newcommand{\eeq}{\end{equation}}
\newcommand{\bea}{\begin{eqnarray}}
\newcommand{\eea}{\end{eqnarray}}

\newcommand{\Tr}{\operatorname{Tr}}

\newcommand{\ii}{{\rm i}\,}

\newcommand{\Tau}{{\cal T}}

\newcommand{\spcurve}{{\cal S}}
\newcommand{\curve}{{\Sigma}}

\newcommand{\genus}{{\mathfrak g}}

\newcommand{\acycle}{{\cal A}}
\newcommand{\bcycle}{{\cal B}}

\newcommand{\CC}{{\mathbb C}}
\newcommand{\RR}{{\mathbb R}}
\newcommand{\ZZ}{{\mathbb Z}}

\newcommand{\Ker}{\operatorname{Ker}}

\newcommand{\order}{\operatorname{order}}

\newcommand{\Res}{\mathop{\,\rm Res\,}}

\newcommand{\td}{\tilde}
\usepackage[pdftex]{hyperref}

\usepackage{makeidx}
\makeindex
\usepackage{tocbibind}

\hypersetup{colorlinks,urlcolor=magenta,citecolor=red,linkcolor=blue,filecolor=black}

\begin{document}

\sloppy

\pagestyle{empty}
\hfill IPhT-T18/007
\addtolength{\baselineskip}{0.20\baselineskip}
\begin{center}
\vspace{26pt}
{\large \bf {Lectures on compact Riemann surfaces.}}
\newline
\vspace{26pt}


{\sl B.\ Eynard}\footnote{Institut de Physique Th\'{e}orique de Saclay, F-91191 Gif-sur-Yvette Cedex, France.}
\footnote{CRM, Centre de recherches math\'ematiques  de Montr\'eal, Universit\'e de Montr\'eal, QC, Canada.}
 \hspace*{0.05cm}

\end{center}
%
%
\vspace{20pt}

\begin{center}
{\bf Paris--Saclay's IPHT doctoral school Lecture given in winter 2018.
}
\end{center}

$$
\includegraphics[scale=0.6]{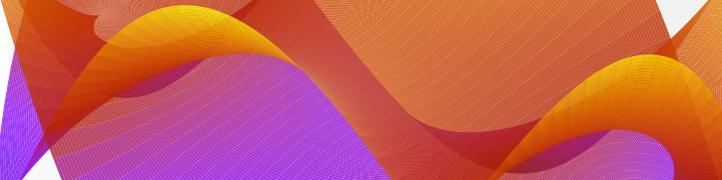}
$$

This is an introduction to the geometry of compact Riemann surfaces.
We largely follow the books \cite{farkas, Fay, MumTata}.
1) Defining Riemann surfaces with atlases of charts, and as locus of solutions of algebraic equations. 
2) Space of meromorphic functions and forms, we classify them with the Newton polygon. 
3) Abel map, the Jacobian and Theta functions. 
4) The Riemann--Roch theorem that computes the dimension of spaces of functions and forms with given orders of poles and zeros.
5) The moduli space of Riemann surfaces, with its combinatorial representation as Strebel graphs, and also with the uniformization theorem that maps Riemann surfaces to hyperbolic surfaces.
6) An application of Riemann surfaces to integrable systems, more precisely finding sections of an eigenvector bundle over a Riemann surface, which is known as the "algebraic reconstruction" method in integrable systems, and we mention how it is related to Baker-Akhiezer functions and Tau functions.


%

\vspace{0.5cm}

\vspace{26pt}
\pagestyle{plain}
\setcounter{page}{1}


\chapter*{Notations}

\begin{itemize}

\item $D(x,r)$ is the open disc of center $x$ and radius $r$ in $\mathbb C$, or the ball of center $x$ and radius $r$ in $\mathbb R^n$.

\item $\mathcal C(x,r)=\partial D(x,r)$ is the circle (resp. the sphere) of center $x$ and radius $r$ in $\mathbb C$ (resp. in $\mathbb R^n$).

\item  $\mathcal C_x$ is a "small" circle around $x$  in $\mathbb C$, or a small circle in a chart around $x$ on a surface, small meaning that it is a circle of radius sufficiently small to avoid all other special points.

\item $T_\tau=\mathbb C/(\mathbb Z+\tau\mathbb Z)$ is the  2-torus of modulus $\tau$, obtained by identifying $z\equiv z+1 \equiv z+\tau$.

\item $\mathbb CP^1 = \overline{\mathbb C} = \mathbb C \cup \{\infty\}$ is the Riemann sphere.

\item $\CC_+$ is the upper complex half--plane $=\{ z \ | \ \Im z>0\}$, it is identified with the Hyperbolic plane, with the metric $ \frac{|dz|^2}{(\Im z)^2}$, of constant curvature $-1$, and whose geodesics are the circles or lines orthogonal to the real axis.

\item $\mathfrak M^1(\curve)$ the $\mathbb C$ vector space of meromorphic forms on $\curve$,

\item $\mathcal O^1(\curve)$ the $\mathbb C$ vector space of holomorphic forms on $\curve$.

\item $H_1(\curve,\mathbb Z)$ (resp. $H_1(\curve,\mathbb C)$) the $\mathbb Z$--module (resp. $\mathbb C$--vector space) generated by homology cycles (equivalence classes of closed Jordan arcs, $\gamma_1\equiv \gamma_2$  if there exists a 2-cell $A$ whose boundary is $\partial A=\gamma_1-\gamma_2$, with 
addition of Jordan arcs by concatenation) on $\curve$.

\item $\pi_1(\curve)$ is the fundamental group of a surface (the set of homotopy classes of closed curves on a Riemann surface with addition by concatenation).

\end{itemize}

\printindex

\tableofcontents

\chapter{Riemann surfaces}
\label{chapsurf}

\section{Manifolds, atlases, charts, surfaces}

\bd[Topological Manifold]
A \textbf{manifold}\index{manifold} $M$ is a second countable\index{second countable} (the topology can be generated by a countable basis of open sets) topological separated space\index{separated space} (distinct points have disjoint neighborhoods, also called \textbf{Haussdorf space}\index{Haussdorf space}), locally \textbf{Euclidian}\index{Euclidian} (each point has a neighborhood homeomorphic to an open subset of $\mathbb R^n$ for some integer $n$).
\ed

\bd[Charts and atlas]
A \textbf{chart}\index{chart} on $M$ is a pair $(V,\phi_V)$, of a neighborhood $V$, together with an homeomorphism $\phi_V:V \to U\subset \mathbb R^n$, called the \textbf{coordinate}\index{coordinate} or the \textbf{local coordinate}\index{local coordinate}.
For each intersecting pair $V\cap V'\neq \emptyset$, the \textbf{transition function}\index{transition function} is the map: $\psi_{U\to U'}:\phi_V(V\cap V') \to \phi_{V'}(V\cap V')$, $x\mapsto \phi_{V'}\circ \phi_V^{-1}(x)$, it is a homeomorphism of Euclidian subspaces, with inverse 
\beq 
\psi_{U\to U'}^{-1}=\psi_{U'\to U}.
\eeq
A countable set of charts that cover the manifold $M$ is called an \textbf{atlas}\index{atlas} of $M$.
Two atlases are said equivalent iff their union is an atlas.
\ed

$$
\includegraphics[scale=0.37]{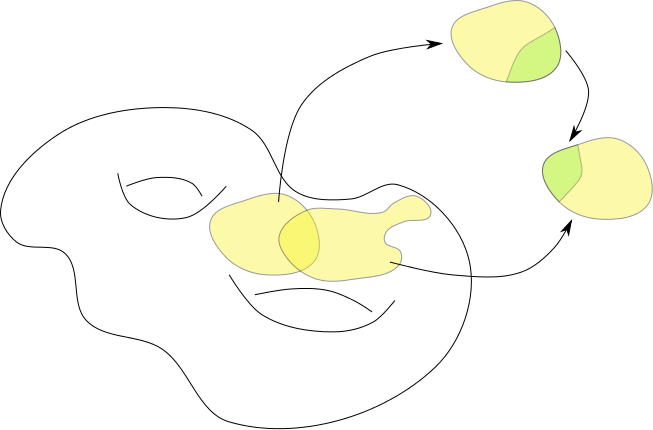}
$$

\bd[Various types of manifold]

$M$ is a topological (resp. smooth, resp. $k$-differentiable, resp. complex)  \textbf{manifold}\index{manifold} if it has an atlas for which all transition maps are continuous (resp. $C^\infty$, resp. $C^k$, resp. holomorphic).

An equivalence class of atlases with transition functions in the given class (smooth, resp. $k$-differentiable, resp. complex) is called a smooth, resp. $k$-differentiable, resp. complex \textbf{structure} on $M$.

\ed

The dimension\index{dimension} $n$ must be constant on each connected part of $M$.
We shall most often restrict ourselves to connected manifolds, thus having fixed dimension.

\begin{itemize}

\item A surface is a manifold of dimension $n=2$.


\item A surface is a \textbf{Riemann surface}\index{Riemann surface} if, identifying $\mathbb R^2=\mathbb C$, each transition map is analytic with analytic inverse.
An equivalence class of analytic atlases on $M$ is called a \textbf{complex structure}\index{complex structure} on $M$.

\item A differentiable manifold is \textbf{orientable}\index{orientable} if, all transition maps $\psi:(x_1,\dots,x_n)\mapsto (\psi_1(x_1,\dots,x_n),\dots,\psi_n(x_1,\dots,x_n))$, have positive Jacobian $\det (\partial \psi_i/\partial x_j )>0$. 
Thanks to Cauchy-Riemann equations, a Riemann surface is always orientable.

\item A manifold is \textbf{compact}\index{compact} if it has an atlas made of a finite number of bounded (by a ball in $\RR^n$) charts.
Every sequence  of points $\{p_n\}_{n\in \mathbb N}$ on $M$, admits at least one adherence value, or also every Cauchy sequence on $M$ is convergent.

\end{itemize}

\subsubsection{Defining a manifold from an atlas}

\bd
An abstract atlas is the data of
\begin{itemize}
\item a countable set $I$,
\item a collection $\{U_i\}_{i\in I}$ of open subsets of $\mathbb R^n$,
\item a collection $\{U_{i,j}\}_{i,j\in I\times I}$ of (possibly empty) open subsets of $\mathbb R^n$ such that $U_{i,j}\subset U_i$, and such that $U_{i,j}$ is homeomorphic  to $U_{j,i}$, i.e. --if not empty-- there exists an homeomorphism $\psi_{i,j}:U_{i,j}\to U_{j,i}$ and an homeomorphism $\psi_{j,i}:U_{j,i}\to U_{i,j}$ such that $\psi_{i,j}\circ \psi_{j,i}=\text{Id}$.
Moreover we require that $U_{i,i}=U_i$ and $\psi_{i,i}=\text{Id}$.
Moreover we require that $U_{j,i}\cap U_{j,k} = \psi_{i,j}(U_{i,j}\cap U_{i,k})$ and that $\psi_{j,k}=\psi_{i,k}\circ \psi_{j,i}$ on $U_{j,i}\cap U_{j,k}$ (if not empty):
\beq\label{eqdefabstractatlas}
\left\{\begin{array}{l}
\psi_{i,j}\circ \psi_{j,i} = \text{Id} \cr
\psi_{j,k}=\psi_{i,k}\circ \psi_{j,i}
\end{array}\right.
\eeq
\end{itemize}
Depending on the type of manifold (topological, smooth, $k$-differentiable, complex), we require all homeomorphisms to be in the corresponding class.
\ed

$$
\includegraphics[scale=0.45]{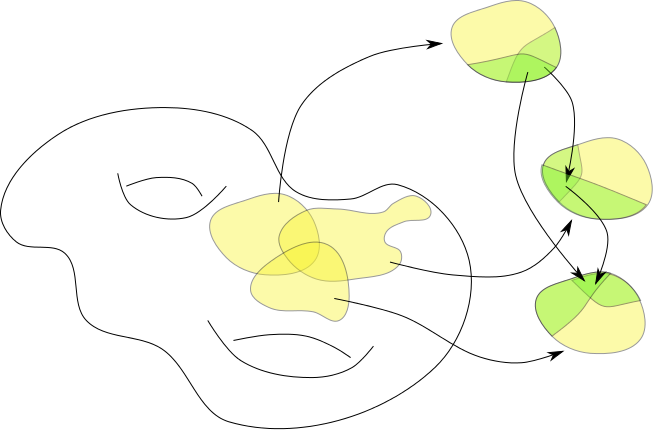}
$$

From an abstract atlas we can define a manifold as a subset of $\mathbb R^n\times I$ quotiented by an equivalence relation:
\bp
\beq
M = \frac{\{(z,i)\in \mathbb R^n\times I \ | \ z\in U_i\} }{(z,i)\equiv (z',j) \quad \text{iff}\quad z\in U_{i,j} \ , \ z'\in U_{j,i} \ , \ \psi_{i,j}(z)=z'}
\eeq
with the topology induced by that of $\mathbb R^n$, is a manifold (resp. smooth, resp. complex, depending on the class of homeomorphisms $\psi_{i,j}$).
\ep

\proof
It is easy to see that this satisfies the definition of a manifold.
Notice that in order for $M$ to be a well defined quotient, we need to prove that $\equiv$ is a well defined equivalence relation, and this is realized thanks to relations \eqref{eqdefabstractatlas}.
Then we need to show that the topology is well defined on $M$, this is easy and we leave it to the reader.
\eproof

All manifolds can be obtained in this way.

\subsection{Classification of surfaces}

We shall admit the following classical theorem:
\bt[Classification of topological compact surfaces]
Topological compact connected surfaces are classified by:
\begin{itemize}
\item the orientability: orientable or non--orientable
\item the Euler characteristics
\end{itemize}
This means that 2 surfaces having the same orientability and Euler characteristic are isomorphic.

\smallskip
\begin{itemize}
\item An orientable surface $\curve$ has an even Euler characteristic of the form
\beq
\chi=2-2\genus
\eeq
where $\genus\geq 0$ is called the genus, and is isomorphic to a surface with $\genus$ holes.
Its fundamental group (non-contractible cycles) is generated by $2\genus$ cycles:
\beq
\pi_1(\curve)\sim \ZZ^{2\genus}.
\eeq
$$
\includegraphics[scale=0.8]{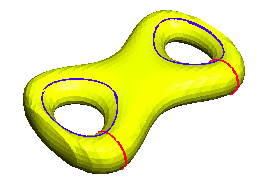}
$$

\item A non-orientable surface $\curve$ has an Euler characteristic 
\beq
\chi=2-k
\eeq 
with $k\geq 1$, it is isomorphic to a sphere from which we have removed $k$ disjoint discs, and glued $k$ M\"obius strips at the $k$ boundaries (this is called $k$  crosscaps).
$$
\includegraphics[scale=0.25]{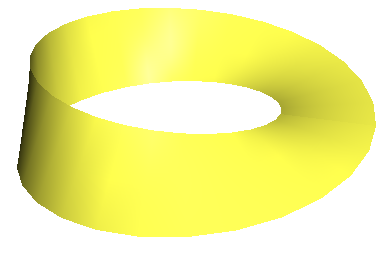}
$$

If $\chi=1$, it is isomorphic to the real projective plane $\RR P^2 $.

If $\chi=0$, it is isomorphic to the Klein bottle.

\end{itemize}

\et

\section{Examples of Riemann surfaces}

\subsection{The Riemann sphere}

$\bullet$
Consider the Euclidian unit sphere in $\mathbb R^3$, the set $\{(X,Y,Z) \ | \ X^2+Y^2+Z^2=1\}$.
Define the 2 charts:
\bea
V_1 = \{(X,Y,Z) \ | \ X^2+Y^2+Z^2=1,\ Z>-\frac35 \}  \ , \quad \phi_1:\ (X,Y,Z)\mapsto  \frac{X+iY}{1+Z} \cr
V_2 = \{(X,Y,Z) \ | \ X^2+Y^2+Z^2=1,\ Z<\frac35 \} \  , \quad 
\phi_2:\ (X,Y,Z)\mapsto  \frac{X-iY}{1-Z} .
\eea
$U_1$ (resp. $U_2$), the image of $\phi_1$ (resp. $\phi_2$), is the open disc $D(0,2)\subset \CC$.
The image by $\phi_1$ (resp. $\phi_2$) of $V_1\cap V_2$ is the annulus $\frac12 < |z| < 2$ in $U_1$ (resp. $U_2$). On this annulus, the transition map
\beq
\phi_2\circ \phi_1^{-1} = \psi: z \mapsto 1/z
\eeq
is  analytic, bijective, and its inverse is analytic.
This defines the \textbf{Riemann sphere}\index{Riemann sphere}, which is a compact (the 2 charts are bounded discs $D(0,2)$), connected (obvious) and simply connected Riemann surface (easy).
The map $\phi_1$ (resp. $\overline{\phi_2}$) is called the {\it stereographic projection}\index{stereographic projection} from the south (resp. north) pole of the sphere to the Euclidian plane $Z=0$ in $\RR^3$, identified with $\CC$.
$$
\includegraphics[scale=0.3]{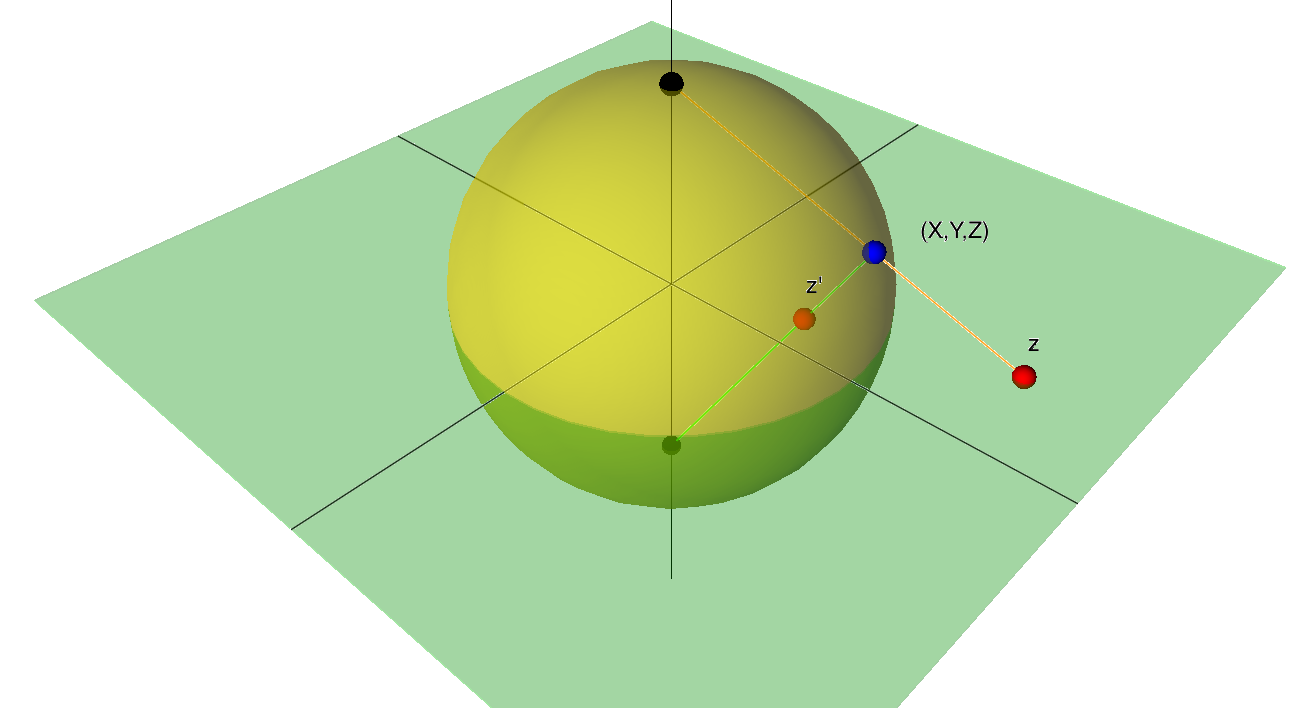}
$$

\smallskip

$\bullet$
Another definition of the Riemann sphere is the \textbf{complex projective plane}\index{complex projective plane}\index{$\mathbb CP^1$} $\mathbb CP^1$:
\beq
\mathbb CP^1 =\frac{ \{(z_1,z_2)\in \mathbb C\times \mathbb C \ | \ (z_1,z_2)\neq (0,0)\}}{(z_1,z_2)\equiv (\lambda z_1,\lambda z_2) \ , \ \forall\ \lambda\in \mathbb C^* }.
\eeq
It has also an atlas of 2 charts, $V_1=\{[(z_1,z_2)] \ | \  z_2\neq 0\}$, $\phi_1: [(z_1,z_2)] \mapsto z_1/z_2$ and $V_2=\{[(z_1,z_2)] \  | \ z_1\neq 0 \}$, $\phi_2:[(z_1,z_2)] \mapsto z_2/z_1$, with transition map $z\mapsto 1/z$ (everything is well defined on the quotient by $\equiv$).

$\mathbb CP^1$ is analytically isomorphic  to the Riemann sphere previously defined.

\smallskip

$\bullet$
Another definition of the Riemann sphere is from an abstract atlas of 2 charts $U_1=D(0,R_1)\subset\mathbb C $ and $U_2=D(0,R_2)\subset\mathbb C$ whose radius satisfy $R_1 R_2>1$.
The 2 discs are glued by the analytic transition map $\psi: z\mapsto 1/z$ from the annulus $\frac{1}{R_2}<|z|<R_1$ in $U_1$ to the annulus $\frac{1}{R_1}<|z|<R_2$ in $U_2$.

In other words, consider the following subset of $\mathbb C\times \{1,2\}$
\beq
\frac{\{ (z,i) \in \CC\times \{1,2\} \ | \ z\in U_i\} }{\equiv}
\eeq
quotiented by the equivalence relation
\beq
(z,i) \equiv (\tilde z,j) \qquad \text{iff} \qquad
i=j \ \text{and} \ z =\tilde z
\quad \text{or} \quad
i+j=3 \ \text{and} \ z \tilde z=1
.
\eeq

This Riemann surface is analytically isomorphic  to the Riemann sphere previously defined.

$\bullet$
Notice that one can choose $R_1$ very large, and $R_2$ very small, and even consider a projective limit $R_1\to \infty$ and $R_2\to 0$, in other words glue the whole $U_1=\mathbb C$ to the single point $U_2=\{0\}$. Notice that the point $z'=0$ in $U_2$ should correspond to the point $z=1/z'=\infty$ in $U_1=\CC$. 
In this projective limit, by adding a single point to $\mathbb C$, we turn it into a compact Riemann surface $\overline{\mathbb C}=\mathbb C\cup \{\infty\}$.
The topology of $\overline{\mathbb C}$ is generated by the open sets of $\mathbb C$, as well as all the sets $V_R=\{\infty\}\cup \{z\in \mathbb C \ | \ |z|>R\}$ for all $R\geq 0$. These open sets form a basis of neighborhoods of $\infty$. With this topology, $\overline\CC$ is compact.

This justifies that the Riemann sphere is called a compactification of $\mathbb C$:
\beq
\mathbb CP^1 = \overline{\mathbb C} = \mathbb C \cup \{\infty\}.
\eeq

\subsection{The torus}

Consider $\tau\in \mathbb C$ with $\Im \tau >0$\index{torus}.
Let
\beq
T_\tau=\mathbb C/(\mathbb Z+\tau \mathbb Z)
\eeq
in other words, we identify $z\equiv z+1 \equiv z+\tau$.
$$
\includegraphics[scale=0.3]{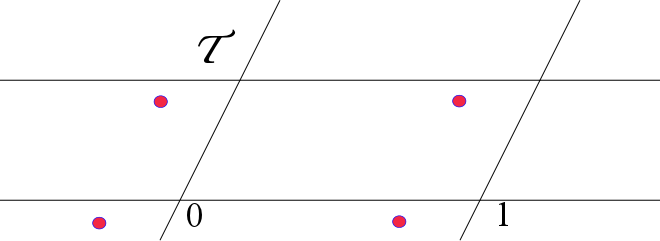}
\qquad \qquad 
\includegraphics[scale=0.3]{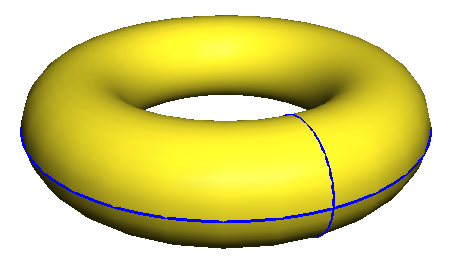}
$$

Each point has a neighborhood homeomorphic to a disc $\subset \mathbb C$.
Transition maps are of the form $z\mapsto z+a+\tau b$ with $a\in \mathbb Z$ and $b\in\mathbb Z$, they are translations, they are analytic, invertible with analytic inverse.

The torus is a Riemann surface, compact, connected, but not simply connected.

\section{Compact Riemann surface from an algebraic equation}

The idea is to show that the locus of zeros (in $\CC\times \CC$) of a polynomial equation $P(x,y)=0$ is a Riemann surface.
This is morally true for all generic polynomials, but there are some subttleties.
Let us start by an example where it works directly, and then see why this assertion has to be slightly adapted.

\subsection{Example}

We start with the polynomial
\beq
P(x,y)=y^2-x^2+4.
\eeq
Consider
\beq
\tilde\curve = \{(x,y) \  | \ y^2-x^2+4=0\} \subset \mathbb C\times \mathbb C
\eeq
which is a smooth submanifold of $\CC\times \CC$.
$$
\includegraphics[scale=0.7]{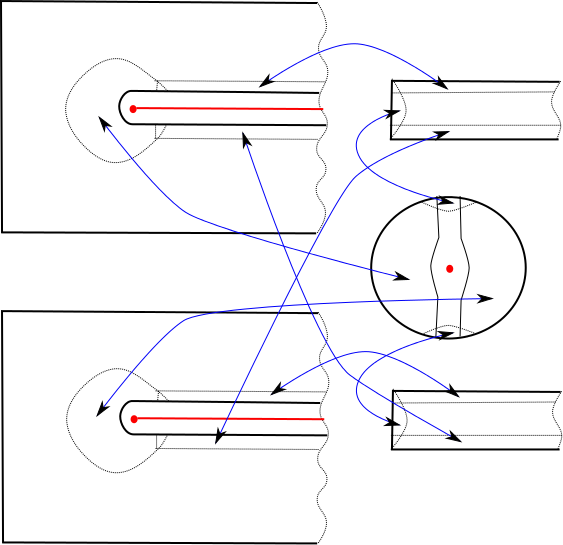}
$$
We can cover it with an atlas of 6 charts as follows
(we choose the square root such that $\sqrt{\mathbb R_+}=\mathbb R_+$, and with the cut on $\mathbb R_-$):
\bea
V_+=\{(x,+\sqrt{x^2-4}) \ | \ x\in U_+=\mathbb C\setminus [-2,2]\}
& \quad , \quad &
\phi_+ : (x,y)\mapsto x \cr
V_-=\{(x,-\sqrt{x^2-4}) \ | \ x\in U_-=\mathbb C\setminus [-2,2]\}
& \quad , \quad &
\phi_- : (x,y)\mapsto x \cr
V_{1}=\{(2+z^2,z \sqrt{4+z^2}) \ | \ z\in U_1=D(0,1)\}
& \quad , \quad &
\phi_1 : (x,y)\mapsto \sqrt{x-2} \cr
V_{-1}=\{(-2+z^2,\ii z \sqrt{4-z^2}) \ | \ z\in U_{-1}=D(0,1)\}
& \quad , \quad &
\phi_{-1} : (x,y)\mapsto \sqrt{x+2} \cr
V_{+-}=\{(x,i\sqrt{4-x^2}) \ | \ x\in U_{+-}= [\frac{-3}{2},\frac32]\times [\frac{-1}{2},\frac12]\}
& \quad , \quad &
\phi_- : (x,y)\mapsto x \cr
V_{-+}=\{(x,-i\sqrt{4-x^2}) \ | \ x\in U_{-+}=[\frac{-3}{2},\frac32]\times [\frac{-1}{2},\frac12]\}
& \quad , \quad &
\phi_- : (x,y)\mapsto x \cr
\eea
We have
$V_+ \cap V_- = \emptyset$ and $V_1 \cap V_{-1} =\emptyset$.
The transition maps on $V_{+-}\cap V_{\pm}$ (resp. $V_{-+}\cap V_{\pm}$) are $x\mapsto x$.
The transition maps on $V_{\pm 1}\cap V_\pm  $ are:
\beq
z \mapsto \pm 2 + z^2 
\eeq
with inverse
\beq
x \mapsto \pm \sqrt{x\mp 2}.
\eeq
All points of $\td\curve$ are covered by a chart, this defines a Riemann surface, it is connected, but it is not simply connected (it has the topology of a cylinder).
However, it is not compact, because two of the charts ($V_+$ and $V_-$) are not bounded in $\CC$.

We shall define a compact Riemann surface $\Sigma$ by adding two points, named $+(\infty,\infty)$ and $-(\infty,\infty)$ to $\td\Sigma$, with two charts as their neighborhoods:
\beq
V_{\pm\infty}=\{(x,\pm\sqrt{x^2-4}) \ | \ |x|>4\}\cup \{\pm(\infty,\infty)\}
 \quad , \quad 
\phi_{\pm\infty} : \begin{array}{l}
 (x,y)\mapsto 1/x  \cr \pm(\infty,\infty)\mapsto 0
 \end{array}
 .
\eeq
Their images $U_{\pm\infty} = D(0,\frac14) $ are discs in $\CC$.

$V_{+\infty}$ (resp. $V_{-\infty}$) intersects $V_+$ (resp. $V_{-}$), and for both, the transition map is
\beq
x\mapsto 1/x.
\eeq
The Riemann surface $\Sigma$ is then compact, connected and simply connected.
Therefore topologically it is a sphere.
Indeed there is a holomorphic bijection (with holomorphic inverse) with the Riemann sphere:
\bea
\mathbb CP^1 &\to & \Sigma \cr
z &\mapsto & (z+1/z,z-1/z) .
\eea

In fact, there is the theorem (that we admit here, proved below as theorem \ref{thg0RS}):\\
\textbf{Theorem \ref{thg0RS} (Genus zero = Riemann sphere)}
{\em
Every simply connected (i.e. genus zero) compact Riemann surface is isomorphic to the Riemann sphere.
}

\subsection{General case}

For every polynomial $P(x,y)\in \mathbb C[x,y]$, let $\td\curve$ be the locus of its zeros in $\mathbb C\times \mathbb C$:
\beq
\td\curve = \{(x,y) \ | \ P(x,y)=0 \} \subset \mathbb C\times \mathbb C.
\eeq

The idea is that we need to map every neighborhood in $\td\curve$ to a neighborhood in $\CC$, and for most of the points of $\td\curve$, we can use $x$ as a coordinate, provided that $x$ is locally invertible.
This works almost everywhere on $\td\curve$ except at the point where $x^{-1}$ is not locally analytic. 
Near those special points we can't use $x$ as a coordinate, and we shall describe how to proceed.

\smallskip

$\bullet$ First let us consider the (finite) set of singular points 
\beq
\td\curve_{\text{sing}} = \{(x,y) \ | \ P(x,y)=0 \ \text{and} \ P'_y(x,y)=0 \},
\eeq
and the set of their $x$ coordinates, to which we add the point $\infty$:
\beq
x_{\text{sing}}=
x(\td\Sigma_{\text{sing}})\cup\{\infty\} \subset \CC P^1.
\eeq
Remark that $x_{\text{sing}}-\{\infty\}$ is the set of solutions of a polynomial equation
\beq
x\in x_{\text{sing}}-\{\infty\} \quad \Leftrightarrow \quad
0=\Delta(x) = \operatorname{Discriminant}(P(x,.)) = \operatorname{Resultant}(P(x,.),P'_y(x,.)),
\eeq
which implies that it is a finite set of isolated points.

$\bullet$  Then choose a connected simply connected set of non--intersecting Jordan arcs in $\CC P^1$, linking the points of $x_{\text{sing}}$, i.e. a simply connected graph $\Gamma\subset \CC P^1$ (a tree) whose vertices are the points of $x_{\text{sing}}$. 
Define $\curve_0=\td\curve\setminus x^{-1}(\Gamma)$ by removing the preimage of $\Gamma$. 
Let $d=\deg_y P$, we define $d$ charts as $d$ identical copies of $\CC  \setminus \Gamma$ as
\beq
U_1=U_2=\dots=U_d = \{ x | (x,y)\in \curve_0\}  = \mathbb C\setminus \Gamma.
\eeq
Each $U_i$ is open, connected and simply connected. 
The $U_i$s play the same role as $U_+$ and $U_-$ in the previous example with $d=2$.
Let $x_0$ be a generic interior point in $\mathbb C\setminus \Gamma$.
The equation $P(x_0,y)=0$ has $d$ distinct solutions, let us label them (arbitrarily) $Y_1(x_0),\dots,Y_d(x_0)$.
For each $i=1,\dots,d$, $Y_i$ can be unambiguously analytically extended to the whole $U_i$ (because it is simply connected), and thus there is an analytic map on $U_i$: $x\mapsto Y_i(x),\ i=1,\dots,d$.
We then define the charts $V_i\subset \td\Sigma$ by
\beq
V_i = \{(x,Y_i(x)) \ | \ x\in U_i\}
\quad , \quad
\phi_i:
\begin{array}{rl}
V_i & \to U_i \cr
(x,Y_i(x)) & \mapsto x
\end{array} .
\eeq
This generalizes the two charts $V_\pm$ in the previous example.
Then we need to define charts that cover the neighborhood of singular points, and the neighborhood of edges of the graph $\Gamma$.

\smallskip

$\bullet$ Consider the most generic sort  of singular point $(a,b)$, such that $P'_y(a,b)=0$, but $P'_x(a,b)\neq 0$ and $P''_{yy}(a,b)\neq 0$.
$(a,b)$ is called a regular  \textbf{ramification point}\index{ramification point} and 
$a$ is called a \textbf{branch point}\index{branch point}.

In that case, there are 2 charts, let us say $V_i,V_j$ with $i\neq j$, that have $(a,b)$ at their boundary. 
Due to our most--generic--assumption, the map $V_i \to \mathbb C, \ (x,y)\mapsto \sqrt{x-a}$ (resp. $V_j \to \mathbb C , \ (x,y)\mapsto -\sqrt{x-a}$) is analytic in a neighborhood of $(a,b)$ in $V_i$ (resp. $V_j$).
We thus define a new chart for each singular point $(a,b)$, as a neighborhood of this point.
It intersects $V_i$ (resp. $V_j$) with transition map $z\mapsto (a+z^2 , Y_i(a+z^2))$ (resp. $z\mapsto (a+z^2 , Y_j(a+z^2))$).
In other words we choose $\sqrt{x-a}$ as a  local coordinate near $(a,b)$.

Also, this defines the "\textbf{deck transformation}"\index{deck transformation} at the singular point: the permutation (here a transposition) $\sigma_{a} = (i,j)$.

$\bullet$ Consider an open edge $e$ of $\Gamma$ (open means excluding the vertices), its boundary consists of the $x$-images $a,a'$ of 2 singular points, each with a permutation $\sigma_{a},\sigma_{a'}$.
In the generic case the 2 transpositions have to coincide, and we associate this transposition $\sigma_e=\sigma_{a}=\sigma_{a'}$ to the edge $e$. 

Now consider a tubular neighborhood $U_{e}$ of $e$ in $\CC\setminus x_{\text{sing}}$, and that contains no other edges.
$U_e\cap U_i$ is disconnected and consists of 2 pieces $U_{e,i,\pm}\subset U_i$.
By pulling back to $\Sigma$ by $x^{-1}$, we get $V_{e,i,\pm}\subset V_i$, and we define
\beq
V_{e,i} = V_{e,i,+}\cup V_{e,\sigma_e(i),-} \cup \{(x,Y_{i}(x)) \ | \ x\in e\},
\quad
\phi_{e,i}: (x,y)\mapsto x.
\eeq
the chart $V_{e,i}$ is an open connected domain of $\td\Sigma$ and $\phi_{e,i}$ is analytic.
The transition maps $x\mapsto x$ are analytic with analytic inverse.

This is the generalization of the charts $V_{+-}$ and $V_{-+}$ in the example above.
It consists of gluing neighborhoods of the 2 sides of an edge, to neighborhoods obtained by the permutation $\sigma_e$.

$\bullet$
For generic polynomials $P$, all singular points are of that type, and we get a Riemann surface, non-compact (this was the case for the example $y^2-x^2+4=0$).

$\bullet$ We can make it compact by adding new points at $\infty$, as we did for the example above, but many subttleties can occur at $\infty$.

This shows that algebraic curves are generically Riemann surfaces, that can be compactified.

In fact we shall see below in section \ref{secprojalgcurve} that the converse is almost true: every compact Riemann surface can be algebraically immersed into $\CC P^2$ (we have to replace $\CC^2$ by $\CC P^2$ to properly compactify at $\infty$).
Generically this immersion is in fact an embedding.

\subsection{Non--generic case: desingularization}

\index{desingularization}
Sometimes the singular points are not generic, this can also be the case near $\infty$.
Like we did in the example, where we added new points to $\curve$ to make it compact in neighborhoods of $\infty$, we can desingularize all singular points by adding new points, and defining a new surface $\Sigma=\td\curve \cup \{ \text{new points}\}$, which is a smooth compact Riemann surface.

$\bullet$ \textbf{Nodal points}\index{nodal point}.
A slightly less (than ramification points) generic type of singular points $(a,b)$, is where both $P'_y(a,b)$ and $P'_x(a,b)$ vanish, to the lowest possible order, i.e. we assume that the second derivative Hessian matrix is invertible $\det\begin{pmatrix}
P''_{xx}(a,b) & P''_{xy}(a,b)\cr P''_{yx}(a,b) & P''_{yy}(a,b)
\end{pmatrix} \neq 0 $.
The intersection of $\tilde\curve$ with a small ball $D((a,b),r)\subset \mathbb C\times  \mathbb C $, is not homeomorphic to a Euclidian disc, instead it is 
homeomorphic to a union of 2 discs, which have a common point $(a,b)$.
This implies that $\tilde\curve$ is in fact not a manifold, it has points whose neighborhoods are not homeomorphic to Euclidian discs.
We say that the surface $\td\curve$ has a \textbf{self intersection}\index{self intersection}, this is called a \textbf{nodal point}.

Nodal points can be desingularized by first removing the point $(a,b)$ from $\tilde\curve$, and adding a 2 new points to $\td\curve$, called $(a,b)_+$ and $(a,b)_-$, and we define the neighborhoods of $(a,b)_\pm$ by one of the 2 punctured discs of $\tilde\curve \cap D((a,b),r)^*$, so that the neighborhoods are now 2 Euclidian discs, as illustrated below:
$$
\includegraphics[scale=0.4]{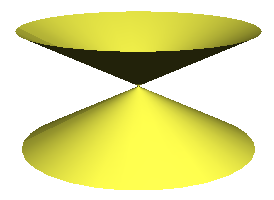}
\qquad  \qquad
\includegraphics[scale=0.4]{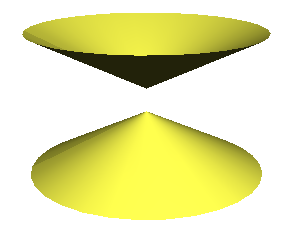}
$$

$\bullet$ In a similar manner, by adding new points to $\td\curve$, all other types of singular points (including neighborhoods of $\infty$, and higher order singular points, at which the Hessian can vanish) can be "desingularized", leading to a Riemann surface $\curve$, which is a smooth compact Riemann surface.

\medskip

$\bullet$ There is a holomorphic map:
\bea
\curve & \to & \td\curve  \cr
p & \mapsto & (x(p),y(p)),
\eea
However this map is not always invertible (it is not invertible at nodal points, since a nodal point is the image of 2 (or more) distinct points of $\curve$).

The map defines 2 holomorphic maps $x:\curve \to \CC P^1$ and $y:\curve \to \CC P^1$.
Since they can reach $\infty\in \CC P^1$, we say that they are meromorphic, i.e. they can have poles.

Eventually this implies that an algebraic equation $P(x,y)=0$ defines a compact Riemann surface, and we have the following theorem

\bt

There exists a smooth compact Riemann surface $\curve$, and 2 meromorphic maps $x:\curve \to \CC P^1$ and $y:\curve \to \CC P^1$, such that

\beq
\td\curve = \{(x,y) \ | \ P(x,y)=0\}  = \{(x(p),y(p)) \ | \ p\in \curve\setminus x^{-1}(\infty)\cup y^{-1}(\infty) \}.
\eeq
The map
\bea
\curve & \to & \td\curve  \cr
p & \mapsto & (x(p),y(p)),
\eea
is meromorphic.
\et

\subsection{Projective algebraic curves}
\label{secprojalgcurve}

\index{projective curve}
Consider a homogeneous polynomial $P(x,y,z)\in \mathbb C[x,y,z]$ of some degree $d$, write its coefficients 
\beq
P(x,y,z)=\sum_{(i,j,k),\ i+j+k=d} P_{i,j,k} \ x^i y^j z^k.
\eeq
Now consider the subset of 
\beq
\mathbb CP^2 = 
\frac{\{(x,y,z)\neq (0,0,0)\}}{ (x,y,z)\equiv (\lambda x,\lambda y,\lambda z) \ \forall \lambda\in\CC^*}
\eeq
annihilated by $P$
\beq
\Sigma= \{ [(x,y,z)]\in \mathbb CP^2 \ | \ P(x,y,z)=0 \}
\eeq
(it is well defined on equivalence classes thanks to the homogeneity of $P$.)
Locally $\mathbb CP^2 \sim \mathbb C^2$, indeed in a neighborhood of a point where at least 1 of the 3 coordinates is not 0 (assume $z\neq 0$), then $(x,y,z)\equiv (x/z,y/z,1)$.
In other words, away from neighborhoods of $z=0$, we can choose $z=1$ and write the equation as
\beq
P(x,y,1)=0.
\eeq
The points of $\Sigma$ where one of the 3 coordinates $x$, $y$ or $z$ vanishes are called \textbf{punctures}\index{puncture}.

Following the same procedure as above, we can find an atlas of $\curve$, by first removing a graph containing all singular points and punctures, with local coordinate $x$, and transition maps are $x\mapsto x$.
Except for charts around singular points, or charts around punctures, where we have to find another local parameter, typically $(x-a)^{1/d_a}$, and possibly desingularize by adding new points to $\curve$.

We shall admit the following:

\bt
Every compact Riemann surface can be algebraically immersed into $\CC P^2$, with at most simple nodal points.

Moreover, every compact Riemann surface can be algebraically embedded into $\CC P^3$ (embedding=bijective, no nodal points).
\et

This is why algebraic  is almost synonymous to compact 
for Riemann surfaces
\begin{center}
\textbf{Algebraic = compact}
\end{center}

\chapter{Functions and forms on Riemann surfaces}
\label{chapforms}

\section{Definitions}

\bd[Functions] An analytic function\index{meromorphic function}\index{holomorphic function}, $f$ on an atlas of a  Riemann surface $\curve$, is the data of a holomorphic function $f_U:U\to \CC P^1$ in each chart, 
satisfying for every transition:
\beq
f_U = f_{U'}\circ \psi_{U\to U'}.
\eeq
This allows to define unambiguously for every point of $\curve$:
\bea
f(\phi_U^{-1}(z)) =f_U(z).
\eea
A holomorphic function that takes values in $\CC P^1$ is called \textbf{meromorphic}\index{meromorphic} if it reaches the value $\infty$ (it has  poles), and \textbf{holomorphic}\index{holomorphic} otherwise.
We shall denote
\begin{itemize}
\item $\mathfrak M^0(\curve)$ the vector space of meromorphic functions on $\curve$,
\item $\mathcal O(\curve)$ (or sometimes $\mathcal O^0(\curve)$) the vector space of holomorphic functions on $\curve$.
\end{itemize}
\ed

\bd[Forms] A meromorphic 1-form\index{meromorphic 1-form}\index{holomorphic 1-form}, $\omega$ on an atlas of a  Riemann surface $\curve$, is the data of a holomorphic function $\omega_U:U\to \CC P^1$ in each chart,
that satisfies for every transition:
\beq
\omega_U(z) = \omega_{U'}(\psi_{U\to U'}(z)) \  \frac{d}{dz}\psi_{U\to U'}(z)  .
\eeq
This allows to define unambiguously
\beq
\omega(\phi_U^{-1}(z)) = \omega_U(z) dz
\eeq
on every point of $\curve$.
A 1-form such that $\omega_U$ takes values in $\CC P^1$ is called meromorphic if $\omega_U$ reaches the value $\infty$ (it has  poles), and holomorphic otherwise.

A meromorphic 1-form $\omega$ is called  exact iff there exists a meromorphic function $f$ such that $\omega=df$, i.e. in each chart $\omega_U(z) = df_U(z)/dz$.

We shall denote
\begin{itemize}
\item $\mathfrak M^1(\curve)$ the vector space of meromorphic forms on $\curve$.
\item $\mathcal O^1(\curve)$ the vector space of holomorphic forms on $\curve$ .
\end{itemize}

\ed

These are in fact special cases of

\bd[Higher order forms]\label{defkthorderform}
\index{meromorphic kth order form}\index{meromorphic kth order form}
A $k^{\rm th}$ order holomorphic (resp. meromorphic) form on an atlas of a  Riemann surface $\curve$, is the data of a holomorphic  function $f_U:U\to \CC P^1$ in each chart,
that satisfies for every transition:
\beq
f_U(z) = f_{U'}(\psi_{U\to U'}(z)) \  \left(\frac{d}{dz}\psi_{U\to U'}(z) \right)^k .
\eeq
This allows to define unambiguously
\beq
f(\phi_U^{-1}(z)) = f_U(z)\ dz^k
\eeq
on every point of $\curve$.

\begin{itemize}
\item If $k=0$ this is called a holomorphic (resp. meromorphic) function.
\item If $k=1$ this is called a holomorphic (resp. meromorphic) 1-form.
\item If $k=2$ this is called a holomorphic (resp. meromorphic) \textbf{quadratic differential}\index{quadratic differential}.
\item It can also be defined for for half--integer $k\in \frac12 \mathbb Z$, and then only $\pm f$ is well defined globally on $\curve$. This is called a \textbf{spinor form}\index{spinor}.

\end{itemize}

We shall denote the vector space of holomorphic and meromorphic order $k$ forms on $\curve$ as
\beq
\mathcal O^k(\curve) \subset \mathfrak M^k(\curve).
\eeq
\ed

\subsection{Examples}

\begin{itemize}

\item On the Riemann sphere, the function $f(z)=z$ is meromorphic, it has a pole at $z=\infty$.
Also on the Riemann sphere, the 1-form $\omega(z)=dz$ is meromorphic, it has a double pole at $\infty$. Indeed in the chart which is the neighborhood of $\infty$, we use the coordinate $z'=1/z$ and we have
\beq
\omega(z) = dz = d(1/z') = \frac{- 1}{z'^2}\ dz'
\eeq
it has a double pole at $z'=0$, i.e. a double pole at $z=\infty$.

\item On the torus $T_\tau=\CC / \mathbb Z+\tau \mathbb Z$, the 1-form
\beq
dz
\eeq
is a holomorphic 1-form.
Indeed it satisfies the transition condition with $\psi(z)=z'=z+a+\tau b$, we have $dz=dz'$.
Moreover, it has no pole.

The following series
\beq
\wp(z) = \frac{1}{z^2} + \sum_{n,m\in \mathbb Z^2-\{(0,0)\} } \frac{1}{(z+n+\tau m)^2} - \frac{1}{(n+\tau m)^2}
\eeq
is absolutely convergent for all $z\notin \mathbb Z+\tau\mathbb Z$, it is clearly bi-periodic $\wp(z+1)=\wp(z+\tau)=\wp(z)$, so that it satisfies the transition conditions, it is thus a meromorphic function on the torus. It has a unique pole at $z=0$, of degee 2. It is called the \textbf{Weierstrass function}\index{Weierstrass function}.


\end{itemize}

\bd[Order]
\index{order}
The $\order_p f=k$ (resp. $\order_p \omega=k$) of a function $f$ (resp. a form) at a point $p\in\curve$ is:
\begin{itemize}
\item the order of vanishing of $f_U$ if $p$ is not a pole, i.e. in any chart $U$, with coordinate $z$, 
$f_U(z) \sim C_U(z-\phi_U(p))^k$. In this case $k>0$.
\item or minus the degree of the pole of $f_U$ if $p$ is a pole, i.e. in any chart $U$, with coordinate $z$, 
$f_U(z) \sim C_U(z-\phi_U(p))^{-|k|}$. In this case $k<0$.
\item For generic points (neither poles nor zeros) we define $\order_p f=0$.
\end{itemize}
The order is independent of a choice of chart and coordinate.
A holomorphic function (resp. form) has non--negative orders at all points.

\ed

\bd[Residue of a form]\index{residue}
Let  $\omega$ a meromorphic 1-form, and $p$ one of its poles.
We define its \textbf{residue}\index{residue} in any chart  $U$ that contains $p$, where $\omega(\phi_U^{-1}(z)) = f_U(z)dz$, as
\beq
\Res_p \omega = c_{-1}
\quad \text{where} \quad f_U(z) =  \sum_{0>j\geq -\order_p \omega} c_j z^j + \text{analytic}
\eeq
It is independent of a choice of chart and coordinate.
\ed
Notice that only the coefficient $c_{-1}$ is independent of a choice of chart and coordinates, all other $c_j$ with $j\neq -1$ do depend on that choice.
Similarly, for a $k^{\rm th}$ order meromorphic form, the residue is defined as the coefficient (independent of chart and coordinate) $c_{-k}$.

\bd[Jordan arcs]
A Jordan arc $\gamma$ on $\curve$, is a continuous map $\gamma:[a,b] \to \curve$, 
such that there exist a finite partition of $[a,b]\subset \RR$
\beq
[a,b] = [a_1,a_2] \cup [a_2,a_3] \cup \dots \cup [a_{n-1},b]
\qquad , \quad
a=a_1<a_2<a_3<\dots <a_n=b,
\eeq
such that each $\gamma([a_i,a_{i+1}])$ is included in a single chart $V_i$, and such that the map
$\gamma_i:[a_i,a_{i+1}] \to U_i$ defined by $\gamma_i(t) = \phi_{V_i}(\gamma(t))$
is a Jordan arc in $U_i\subset \CC$.

There is a notion of homotopic deformations of Jordan arcs on $\curve$ inherited from that in $\CC$, and of concatenation of Jordan arcs.

A closed Jordan arc is called a Jordan curve, or a contour, it is such that $\gamma(b)=\gamma(a)$.

\ed

\bd[Integral of a form]
Let $\omega$ a meromorphic 1-form.
Let $\gamma$ a Jordan arc on $\curve$, not containing any pole of $\omega$, represented by a collection $\gamma_1,\dots ,\gamma_n$ of  Jordan arcs in charts $\gamma_i\subset U_i$.
We define
\beq
\int_\gamma \omega = \sum_i \int_{\gamma_i} \omega_{U_i}(z) dz
\eeq
It is independent of a choice of charts and local coordinates.

Moreover it is invariant under homotopic deformations of $\gamma$.

If $\gamma$ is closed, we write $\oint_\gamma$ rather than $\int_\gamma$, which is then invariant under change of initial point of the Jordan curve.

\ed

\bd[Integral of a form on a chain or cycle]
We define an integer (resp. complex) chain  $\hat\gamma$ as a $\mathbb Z$ (resp. $\CC$) linear combination of homotopy classes of Jordan arcs modulo boundaries of open surfaces.
A chain is a cycle iff its boundary vanishes.

Integration defines the \textbf{Poincar\'e pairing}\index{Poincar\'e pairing} between chains (resp. cycles) and 1-forms
\beq
<\gamma,\omega> = \int_{\hat\gamma} \omega .
\eeq

\ed

\bt[Cauchy]
We have\index{residue}
\beq 
\Res_p \omega = \frac{1}{2\pi\ii} \oint_{\mathcal C_p} \omega
\eeq
with $\mathcal C_p$ a small anti-clockwise contour around $p$.
\et
\proof
It holds in every chart because it holds on $\CC$.
\eproof

\bd[Divisors]
A \textbf{divisor}\index{divisor} is a formal linear combination of points of the surface.
Let $D=\sum_i \alpha_i . p_i$ a divisor.
We define its degree as:
\beq
\deg D = \sum_i \alpha_i.
\eeq
We shall say that the divisor is an integer divisor if $\alpha_i\in \ZZ$ and complex if $\alpha_i\in\CC$.

If $f$ is a meromorphic function, not identically vanishing, we denote the \textbf{divisor of $f$} (it is an integer divisor):
\beq
(f) = \sum_{p\in\curve} \order_p f \ . p.
\eeq
Similarly, if $\omega$ is a meromorphic 1-form not indetically zero, we denote
\beq
(\omega) = \sum_{p\in\curve} \order_p \omega \ . p.
\eeq

If $f=0$ (resp. $\omega=0$) we define
\beq
(0)=0.
\eeq
We define similarly the divisors of any $k^{\rm th}$ order forms.

\ed

\subsection{Classification of 1-forms}

1-forms have been customarily divided into the following classes

\bd[Classification of 1-forms]

A meromorphic 1-form $\omega$ is called 

\begin{itemize}

\item \textbf{1st kind}\index{1st kind form} iff it is holomorphic (it has no poles),

\item \textbf{3rd kind}\index{3rd kind form} iff it has poles of degree at most 1,

\item \textbf{2nd kind}\index{2nd kind form} iff it has some poles of degree $\geq 2$.

\item \textbf{exact}\index{exact form} iff there exists a meromorphic function $f$ such that $\omega=df$, i.e. in each chart $\omega_U(z) = df_U(z)/dz$. In fact all exact forms must be 2nd kind.

\end{itemize}

\ed

\section{Some theorems about forms and functions}

\bt[finite number of poles]
On a compact Riemann surface, each meromorphic function (resp. form) has at most a finite number of poles.
Also each non-vanishing meromorphic function (resp. form) has only finitely many points with non--vanishing order, so the divisor  is a finite sum.
\et 
\proof
Compactness implies that any infinite sequence of points of $\curve$ must have accumulation points.
If the number would be infinite there would be an accumulation point of poles, and the function (resp. form) would not be analytic at the accumulation point.
If there is an accumulation of points of strictly positive orders, i.e. an accumulation of zeros, then the analytic function (resp. form) has to vanish identically in a neighborhood of this point, and thus vanishes identically on $\curve$.
In all cases the divisor is finite.
\eproof

\bt[exact forms]\label{thexact}
A 1-form $\omega$ is exact\index{exact form} if and only if
\beq\label{ointCzeroexact}
\forall \ \mathcal C=\text{cycle} \qquad \oint_{\mathcal C} \omega=0.
\eeq
\et

\proof

Let $o$ a generic point of $\curve$.
The function 
\beq
f(p) = \int_o^p \omega
\eeq
seems to be ill defined on $\curve$ as it seems to depend on a choice of Jordan arc from $o$ to $p$.
However, thanks to \eqref{ointCzeroexact}, its value is independent of the choice of arc and thus depends only on $p$, so it is a well defined function on $\curve$.

$\omega$ may have poles, its integral can have poles and logs, but the condition \eqref{ointCzeroexact} implies that the residues of $\omega$ at all poles vanish.
This implies that $f$,  can't have logartithmic terms, it is thus a meromorphic function on $\curve$, and safisfies
\beq
df=\omega.
\eeq
The converse is obvious.
\eproof

\bt[vanishing total residue]\label{thresidues}
\index{residue}
For a meromorphic form $\omega$ on a compact Riemann surface:
\beq
\sum_{p=\text{poles}} \Res_p \omega = 0.
\eeq

\et

\proof
Let us admit here that $\curve$ can be polygonized, i.e. that there is a graph $\Gamma$ on $\curve$, whose faces are polygons entirely contained in a chart, and such that the poles of $\omega$ are not on on $\Gamma$.
By homotopic deformation, the sum of residues inside each polygon is the integral along edges. Each edge is spanned twice, in each direction, so the sum of integral along edges is zero.

\eproof

\bc[More than 1 simple pole]\label{cor1simplepole}
There is no meromorphic 1-form with only one simple pole.
It has either several simple poles, or poles of higher orders, or no pole at all.
\ec

\bt
If $\genus\geq 1$, there is no meromorphic function with only 1 simple pole.
\et

\proof
If $f$ is a meromorphic function with 1 simple pole $p$.
Then, consider a holomorphic 1-form $\omega$ (it is possible if $\genus>0$, we anticipate on the next section). Assume that $\omega$ either doesn't vanish at $p$ ($k=0$) or has a zero of order $k$ at $p$.
Then $f^{k+1} \omega$ is a meromorphic form, and has only 1 simple pole at $p$, which is impossible.
\eproof

\bt[Functions: $\#$poles = $\#$zeros]\label{thfnumberpoleszeros}
Let $f$ a meromorphic function, not identically vanishing, then
the number of poles (with multiplicity) equals the number of zeros:
\beq
\#\text{zeros}-\#\text{poles} = 
\deg \ (f) = 0.
\eeq

\et

\proof
Use theorem \ref{thresidues} with $\omega = d\log f = \frac{1}{f} df$.
\eproof

\bt[Holomorphic function = constant]\label{thfholconst}
Any holomorphic function is constant.
This implies that
\beq
\mathcal O(\curve) = \mathbb C
\qquad , \qquad
\dim \mathcal O(\curve) = 1.
\eeq
\et

\proof
Let $f$ a holomorphic function. Let $p_0\in\curve$ a given generic point, define the function $g(p)=f(p)-f(p_0)$. This function has no pole and has at least one zero.
This would contradict theorem \ref{thfnumberpoleszeros}, unless $g$ is identically vanishing, i.e. $f$ is constant.
\eproof

\bt[any 2 meromorphic functions are algebraically related]
Let $f$ and $g$ be two meromorphic functions on $\curve$.
Then, there exists a bivariate polynomial $Q$ such that
\beq
Q(f,g)=0.
\eeq

\et

\proof
The proof is nothing but the Lagrange interpolation polynomial.

Let us call $d=\deg f$ the total degree of $f$, i.e. the sum of degrees of all its poles.
Let $x\in\mathbb CP^1$, then $f^{-1}(x)$ has generically a cardinal equal to $d$.
Let us define $Q_0=1$ and for $k=1,\dots,d$:
\beq
Q_k(x) = \sum_{I \subset_k f^{-1}(x)} \prod_{p\in I} g(p),
\eeq
where we sum over all subsets of $f^{-1}(x)$ of cardinal $k$, and we count preimages with multiplicities when $x$ is not generic.
$Q_k$ is clearly a meromorphic function $\mathbb CP^1\to \mathbb CP^1$, therefore it is a rational function $Q_k(x)\in \mathbb C(x)$.
We then define:
\beq
Q(x,y) = \sum_{k=0}^d (-1)^k Q_k(x) y^k \quad \in \mathbb C(x)[y].
\eeq
It is also equal to
\beq
Q(x,y) = \prod_{p\in f^{-1}(x)} (y-g(p)).
\eeq
Therefore, for any $p\in \curve$ we have
\beq
Q(f(p),g(p))=0.
\eeq
\eproof

\bt[Riemann-Hurwitz]\label{thRiemannHurwitz}
\index{Riemann-Hurwitz}
Let $\omega$ a meromorphic 1-form not identically vanishing, then
\beq
\deg \ (\omega) = 2\genus-2
\eeq
where $\genus$ is the genus of $\curve$.
In other words
\beq
\#\text{zeros}-\#\text{poles} = 2\genus-2.
\eeq
\et

\proof

First remark that the ratio of 2 meromorphic 1-forms is a meromorphic function, for which $\#\text{zeros}=\#\text{poles}$, therefore $\#\text{zeros}-\#\text{poles}$ is the same for all 1-forms.

In particular, let us assume that there exist some non--constant meromorphic function $f:\curve\to \CC P^1 $ (we anticipate on the next sections. In the case of an algebraic curve $P(x,y)=0$, one can choose $f=x$), then $\omega=df$ is an exact meromorphic 1-form.

Let $R$ be the set of zeros of $\omega=df$, and $P$ the set of poles.

Choose an arbitrary cellular (all faces are homeomorphic to discs) graph $\Gamma$ on $\CC P^1$, whose vertices are the points of $f(R)\cup \{\infty\}$.

Let $F$ the number of faces, $E$ the number of edges and $V$ the number of vertices.
Let $d$ the degree of $f$, i.e. the sum of degrees of all its poles, which is also $d=\#f^{-1}(x)$ for $x$ in a neighborhood of $\infty$, and thus is the number of preimages of generic $x$.

The Riemann sphere has Euler characteristic $2$:
\beq
\chi(\CC P^1) =2 =F-E+V.
\eeq

Now consider the graph $\Gamma'=f^{-1}(\Gamma)$ on $\curve$.
Since the faces of $\Gamma$ contain no zero of $df$, then the preimages of each face is homeomorphic to a disc too, so that $\Gamma'$ is a cellular graph on $\curve$.
Its Euler characteristic is
\beq
\chi(\curve) = 2-2\genus = F'-E'+V'.
\eeq
The number of faces $F'=d F$ and edges $E'=dE$ because they are made of generic points.
The points of $R$ and $P$ are by definition not generic.
For $x\in f(R)$ we have
\beq
\#f^{-1}(x) = d- \sum_{r\in f^{-1}(x)} \order_{r}df.
\eeq
Their sum is
\beq
\sum_{x\in f(R)} \#f^{-1}(x) = d(V-1)- \deg (df)_+
\eeq
where $(df)_+$ is the divisor of zeros of $df$.
Similarly for $x=\infty$ we have
\beq
\#f^{-1}(\infty) =  \sum_{p\in P} (-1-\order_{p}df).
\eeq
This implies
\beq
V'=d(V-1)-d-\deg (df).
\eeq
Putting all together we get
\beq
\deg (df)=2\genus-2.
\eeq

Since $\deg (\omega)$ is the same for all 1-forms, it must be the same as for the exact form $df$ and the theorem is proved.

In fact all what remains to prove is the existence of at least one non--constant meromorphic function.
For Riemann surfaces coming from an algebraic equation $P(x,y)=0$, the function $(x,y)\mapsto x$ can play this role.
More generally the existence of non--constant meromorphic functions will be established below (and won't use this theorem).
\eproof

\section{Existence of meromorphic forms}

Before going further, we need to make sure that meromorphic functions and forms do actually exist.
First let us define the Hodge star:

\bd[Hodge star, harmonic forms]
Let $\omega$ a $C^\infty$ 1-form (real or complex) on $\curve$ viewed as a smooth manifold of dimension $2$ rather than a complex manifold. 
In a local coordinate $z=x+iy$ it can be written
\beq
\omega = pdx+qdy = \frac{p-iq}{2} dz + \frac{p+iq}{2} d \bar z.
\eeq
$p$ and $q$ can be viewed as real valued or complex valued $C^\infty$ functions on $\curve$, they are not assumed to be analytic.

We define the \textbf{Hodge star} \index{Hodge star} of a differential 1-form
\beq
*\omega = -qdx+pdy = -i \ \frac{p-iq}{2} dz + i\ \frac{p+iq}{2} d \bar z.
\eeq
We have
\beq
\omega \wedge * \bar\omega
= (|p|^2+|q|^2) \ dx\wedge dy.
\eeq

The Hilbert space $L^2(\curve)$ of real (resp. $L^2(\curve,\CC)$\index{$L^2$} of complex)  square integrable 1-forms, is equipped with the norm (it is positive definite)
\beq
||\omega||^2 = \int_{\curve} \omega \wedge *\bar\omega.
\eeq

A 1-form $\omega$ is called \textbf{closed}\index{closed} (resp. \textbf{co--closed}\index{co-closed}) iff $d\omega=0$ (resp. $d*\omega=0$).
A 1-form $\omega$ is called \textbf{harmonic}\index{harmonic} iff it is closed and co--closed.
Let $\mathcal H$ the set of real harmonic forms.

If $f$ is a function, then its Laplacian\index{Laplacian} is
\beq
\Delta f = \frac{\partial^2 f}{\partial x^2}+\frac{\partial^2 f}{\partial y^2} = d *d f = -2i\ \bar\partial \partial f.
\eeq

Let $E$ (resp. $E^*$) the closure of the set of exact (resp. co-exact) 1-forms, i.e. the set of differentials $df$ (resp. $*df$) and limits (with respect to the topology induced by the $L^2(\curve)$ norm) of sequences $\lim_{n\to\infty} df_n$ (resp. $\lim_{n\to\infty} *df_n$).

\ed

There is the following theorem (that we shall not use in these lectures, in fact it will be reproved case by case corresponding to our needs):

\bt[Hodge decomposition theorem]
\index{Hodge decomposition theorem}
Every real square integrable 1-form can be uniquely decomposed as a harmonic+exact+co-exact:
\beq
L^2(\curve) = \mathcal H \oplus E \oplus E^*.
\eeq

\et
\proof
Admitted, and in fact not needed in these lectures.
\eproof

The following theorem is the key to existence of holomorphic and meromorphic 1-forms:
\bt[Harmonic forms]\label{thexistharmonicforms}
Let $\curve$ a Riemann surface of genus $\genus \geq 1$, and let $\acycle_1,\dots, \acycle_{2\genus}$ a basis of $H_1(\curve,\mathbb Z)$ (non--contractible cycles).

Given $(\epsilon_1,\dots,\epsilon_{2\genus}) \in \mathbb R^{2\genus}$, there exists a unique real harmonic form $\nu$ on $\curve$ such that
\beq\label{condthharmforms}
\forall\ i=1,\dots,2\genus \ , \quad \oint_{\acycle_i} \nu = \epsilon_i.
\eeq
\et
Moreover, $\mathcal H$ is a vector space over $\RR$ of dimension
\beq
\dim \mathcal H = 2\genus.
\eeq

\proof
Let us prove it for $(\epsilon_1,\dots,\epsilon_{2\genus})=(1,0,\dots,0)$, the general case will then hold by linearity and relabelling.

The constraints  \eqref{condthharmforms} define an affine subspace $V\subset L^2(\curve)$.
Let us first show that this affine subspace is non--empty.

Let us choose some Jordan arcs representative of the $\acycle_i$s, and let $\bcycle$ a closed Jordan arc that intersects $\acycle_1$ once and no other $\acycle_j$: $\acycle_j\cap \bcycle=\delta_{1,j}$.
Choose a  tubular neighborhood $U$  of $\bcycle$.
Choose a real function $\theta$, not continuous, but such that $\theta$ is $C^\infty$ on $\curve\setminus \bcycle$, is constant equal to 1 in a left neighborhood of $\bcycle$, is constant equal to 0 in a right neighborhood of $\bcycle$, and is identically zero outside $U$ (left and right refer to the orientation of the arc $\acycle_1$). Such a function exists in the annulus of $\RR^2$, and is easy to define with the help of the $C^\infty(\RR)$ function $x\mapsto e^{-1/x}$ if $x>0$ and $x\mapsto 0$ if $x\leq 0$.
$$
\includegraphics[scale=0.35]{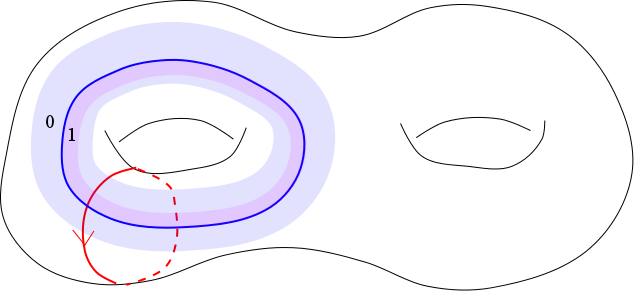}
$$
Then $d\theta$ is a $C^\infty$ 1-form on $\curve$, it belongs to $L^2(\curve)$, and it  satisfies \eqref{condthharmforms}, so it belongs to $V$.

In order to satisfy \eqref{condthharmforms}, one can only add exact forms (and their limits), and thus
\beq
V=d\theta+E ,
\eeq
which is a closed set since $E$ is closed. 
The Hilbert projection theorem ensures that there exists at least one element $\nu\in V$ whose norm $||\nu||$ is minimal.

This means that for any $C^\infty$ function $f$, one has $||\nu+df||^2\geq ||\nu||^2$, and this implies that $(\nu,df)=0$, i.e. $\int_\curve f d *\nu=0$. 
Since this has to hold for all $f$, this implies that  $d*\nu=0$ and thus $\nu$ is a harmonic form $\nu\in V\cap \mathcal H$.

\smallskip

The space $\mathcal H$ is clearly a real vector space, and we have just seen that 
the morphism
\bea
\mathcal H & \to & \RR^{2\genus} \cr
\nu & \mapsto & \left( \oint_{\acycle_1}\nu,\oint_{\acycle_2}\nu,\dots,\oint_{\acycle_{2\genus}}\nu\right)
\eea
is surjective.
It is also injective, because if $\nu$ is in the kernel, then all its cycle integrals vanish, implying that it is an exact form,  $\nu=df$, and since $d*\nu=0$, $f$ must be a harmonic function $\Delta f=0$.
Stokes theorem implies
\beq
||\nu||^2 = -\int_\curve f \Delta f = 0,
\eeq
which implies $\nu=0$.
This proves that there is an isomorphism $\mathcal H\sim \RR^{2\genus}$.

\eproof

In other words, for $\genus\geq 1$, there exists harmonic forms.
Notice that if $\nu$ is a real harmonic form, then $\omega=\nu+i*\nu$ is a holomorphic form on $\curve$.

Remark that we have obtained a harmonic form by an "extremization" procedure, this is often called "\textbf{Dirichlet Principle}\index{Dirichlet principle}".

This method can be generalized to get "meromorphic forms" (and then this works also for genus 0), with some adaptations, and we get

\bt[Harmonic forms with 2 simple poles]
Given $q_+,q_-$ distinct in $\curve$, there exists a unique real harmonic form $\nu$ on $\curve\setminus \{q_+,q_-\}$, such that in a neighborhood $U_+$ (resp. $U_-$) of $q_+$ (resp. $q_-$):
\beq
\nu(p) \sim_{p\to q_{\pm}} \pm d\text{Arg}{|\phi_{U_{\pm}}(p)-\phi_{U_\pm}(q_\pm)|} 
+ \text{harmonic at } \ q_\pm \ 
\eeq
and such that all its cycle integrals vanish
\beq
\forall \ i=1,\dots,2\genus \ , \qquad  \oint_{\acycle_i} \nu=0
\eeq
In other words it satisfies
\beq
\oint_{\mathcal C} \nu = \left\{\begin{array}{l}
\pm 2\pi \quad \text{if} \ \mathcal C\ \text{surrounds} \ q_\pm \cr
0 \quad \text{otherwise}\end{array}\right.
\eeq
\et

\proof
We shall denote $z$ (resp. $z'$) the coordinate of a point $p$ in a neighborhood chart $U_+$ (resp. $U_-$) of $q_+$ (resp. $q_-$):
\beq
z=\phi_{U_+}(p)-\phi_{U_+}(q_+)
\qquad , \ \text{resp.} \quad z'=\phi_{U_-}(p)-\phi_{U_-}(q_-) \ ,
\eeq
vanishing at $p=q_+$ (resp $p=q_-$).

Let us choose a Jordan arc $\gamma$ with boundary $\partial\gamma=q_+-q_-$, and let $U$ a tubular neighborhood of $\gamma$.
Let us choose a function $\theta\in\mathcal C^\infty(\curve\setminus \gamma)$, such that:
\bea
\theta &=& \text{Arg} z \quad \text{in} \ U_+ \cr
\theta &=& 2\pi-\text{Arg} (-z') \quad \text{in} \ U_- \cr
\theta &=&0 \quad \text{in} \ U_{\text{left}} - U_+-U_- \cr
\theta &=&2\pi \quad \text{in} \ U_{\text{right}} - U_+-U_- \cr
\eea
$d\theta$ is then a $\mathcal C^\infty$ 1-form on $\curve-\{q_+\}-\{q_-\}$.
$$
\includegraphics[scale=0.45]{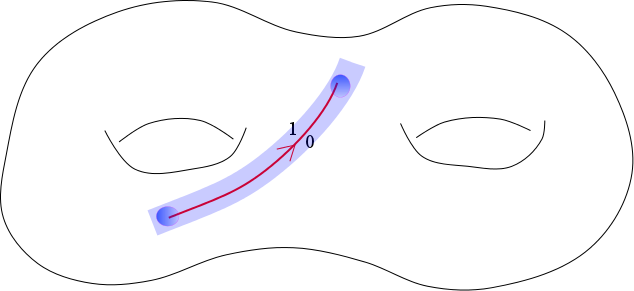}
$$
However $d\theta\notin L^2(\curve)$ because it has poles at $q_+$ and $q_-$, so its norm is infinite.
Let us define the "regularized norm" on $d\theta+E$ as
\beq
||d\theta+df||^2-||d\theta||^2 \stackrel{\text{def}}{=} ||df||^2 - 2 \int_{\curve\setminus \gamma} \theta \Delta f + 2\pi (f(q_+)-f(q_-)),
\eeq
which is a strictly convex functional on the closed space $E$.
The Hilbert projection theorem says that there is an element $\nu$ of $d\theta+E$ with minimal norm, i.e. for any smooth $f$ compactly supported on $\curve-\{q_+\}-\{q_-\}$, we have
\beq
||\nu+df||^2-||\nu||^2 = -2\int_\curve f  d * \nu + ||df||^2 \geq 0.
\eeq
This having to be  positive for all $f$ implies that $d*\nu=0$, and thus $\nu$ is harmonic on $\curve-\{q_+\}-\{q_-\}$.
Moreover, it has the correct  behaviors near $q_+$ and $q_-$, and the correct cycle integrals.

$\nu$ is unique because if there would exist another $\td\nu$, then the difference $\nu-\td\nu$ would be harmonic, with all cycle integrals vanishing, so it would be exact $\nu-\td\nu=df$ with $f$ a harmonic function, i.e. it would have to vanish.
\eproof

\bc[Green function]\label{cordefGreen}
\index{Green function}
Given $q_+,q_-$ distinct in $\curve$, there exists a real function $\mathcal G_{q_+,q_-}$ harmonic on $\curve\setminus \{q_+,q_-\}$, such that in a neighborhood $U_+$ (resp. $U_-$) of $q_+$ (resp. $q_-$):
\beq
\mathcal G_{q_+,q_-}(p) \sim_{p\to q_\pm } \pm \log{|\phi_{U_{\pm}}(p)-\phi_{U_\pm}(q_\pm)|} + \text{harmonic}.
\eeq
It is unique up to adding a constant.

Moreover it satisfies the Poisson equation\index{Poisson equation}
\beq
\Delta \mathcal G_{q_+,q_-}(p) = 4\pi \left( \delta^{(2)}(p,q_+) - \delta^{(2)}(p,q_-) \right).
\eeq
\ec

\proof
We choose $\mathcal G_{q_+,q_-}$ an integral of $*\nu$, to which we add the unique linear combination of harmonic forms that ensure that all its cycle integrals vanish.
\eproof

\bc[Third kind forms]\label{corexistthridkindforms}
Given any 2 distinct points $q_+,q_-$ on $\curve$, there exists a unique meromorphic 1-form $\omega_{q_+,q_-}$ that has a simple pole at $q_+$ and a simple pole at $q_-$, and such that
\beq
\Res_{q_+} \omega_{q_+,q_-} = 1 = -\Res_{q_-} \omega_{q_+,q_-}.
\eeq
and such that, for any non--contractible cycle $\mathcal C$ one has
\beq
\Re \oint_{\mathcal C} \omega_{q_+,q_-} =0.
\eeq
\ec

\proof
Choose
$\omega_{q_+,q_-} = i \nu-*\nu $, 
and add the unique linear combination of holomorphic forms, that cancels the real parts of all cycle integrals.
\eproof

\bc[Higher order poles]\label{cor:highorderpoleompk}
Let $U$ a chart, and let $\phi$ a coordinate in $U$.
Let $q\in U$ and let $k\geq 1$.
There exists a unique meromorphic 1-form $\omega_{q,k}$ that has a pole of order $k+1$ at $q$ and no other pole, and that behaves near $q$ like
\beq
\omega_{q,k}(z) \sim \frac{d\phi(z)}{(\phi(z)-\phi(q))^{k+1}} + \text{analytic at}\ q, 
\eeq
and such that, for any non--contractible cycle $\mathcal C$ one has
\beq
\Re \oint_{\mathcal C} \omega_{q,k} =0.
\eeq
\ec

\proof
Choose $q'\neq q$ generic in $U$, and choose
$\omega_{q,k} = \frac{1}{k!}\ \frac{d^{k} \omega_{q,q'}}{dq^{k}}  $.
It is unique (and in particular independent of $q'$), because the difference of 2 such forms would have no pole and all real parts of its cycle integrals vanishing so it would vanish.
\eproof

\subsubsection{Fundamental second kind form}

However, the form $\omega_{q,k}$, and in particular $\omega_{q,1}$ depends on the choice of chart neighborhood of $q$, and on the choice of coordinate in that chart.
As we shall see now, $\omega_{q,1}$ in fact transforms as a 1-form of $q$ under chart transitions, i.e. $B_S(z,q) = \omega_{q,1}(z) d\phi(q)$ is a well defined 1-form (chart independent) of both variables $z$ and $q$.
We call it a bilinear differential, the product is a tensor product, that we write \beq
B_S(z,q) = \omega_{q,1}(z) d\phi(q) = \omega_{q,1}(z) \otimes d\phi(q) = \omega_{q,1}(z) \boxtimes d\phi(q) .
\eeq
The tensor product notation $\otimes$ just means that this is a linear combination of 1-forms of $z$, whose coefficients are themselves 1-forms of $q$.
The box tensor product notation $\boxtimes$\index{$\boxtimes$} means that this is a differential form on the product $\curve\times \curve$, which is a 1-form on the first factor $\curve$ of the product, tensored by a 1-form of the second factor $\curve$ of the product.

We have the following theorem:

\bp[Schiffer kernel]\label{propSchifferBS}
\index{Schiffer kernel}
Define the following bi-differential of the Green function
\beq
B_S(p,q) =  \partial_p \partial_q \ \mathcal G_{q,q'}(p) \ \ dp \otimes dq
\eeq
where, if $z$ is a coordinate in a chart,  $\partial_z  = \frac12 \left( \frac{\partial }{\partial \Re z} - \ii \ \frac{\partial }{\partial \Im z}  \right)$, and $dz = d\Re z + \ii d\Im z$.
It is independent of a choice of chart and coordinate, and it has the following properties:

\begin{itemize}

\item it is a meromorphic bilinear differential (meromorphic 1-form of $p$, tensored by a meromorphic 1-form of $q$),

\item it is independent of $q'$,

\item it has a double pole on the diagonal, such that in any chart and coordinate
\beq
B_S(p,q) \mathop{\sim}_{p\to q} \frac{d\phi_U(p) \boxtimes d \phi_U(q)}{(\phi_U(p)-\phi_U(q))^2} + \text{analytic at }p=q,
\eeq

\item for any pair of non--contractible cycles $\mathcal C, \mathcal C' $, one has
\beq
\Re \oint_{p\in \mathcal C} \oint_{q\in \mathcal C'} B_S(p,q) =0.
\eeq
\end{itemize}

\ep

Moreover we shall see later in section \ref{secfundformgen} of chap.~\ref{chapAbel} that $B_S$ is in fact symmetric $B_S(p,q)=B_S(q,p)$.

The Schiffer kernel is a special example of the following notion:

\bd[Fundamental form of second kind]\label{deffundform2kind}
A fundamental form of second kind\index{fundamental form} $B(z,z')$ is a symmetric $1\boxtimes 1$ form on $\curve\times \curve$ that has a double pole at $z=z'$ and no other pole, and such that in any chart
\beq
B(z,z') \sim \frac{d\phi_U(z) \ d\phi_U(z')}{(\phi_U(z)-\phi_U(z'))^2} + \text{analytic at} \ z=z'.
\eeq
\ed
If the genus is $\geq 1$, a fundamental form of second kind is not unique since we can add to it any symmetric bilinear combination of holomorphic 1-forms:
\beq
B(z,z') \to B(z,z') + \sum_{i,j=1}^{\genus} \kappa_{i,j} \ \omega_i(z) \otimes \omega_j(z').
\eeq
Therefore, and due to the existence of at least one such form (the Schiffer kernel),  we have
\bt
The space of fundamental second kind differentials is not empty, and is an affine space, with linear space $\mathcal O^1(\curve)\stackrel{\text{sym}}{\otimes} \mathcal O^1(\curve)$.

\et

\subsection{Uniqueness of Riemann surfaces of genus 0}

As a corollary of corollary \ref{cor:highorderpoleompk}, we have

\bt[Genus 0 = Riemann sphere]\label{thg0RS}
Every compact simply connected (genus zero) Riemann surface is isomorphic to the Riemann sphere.
\et

\proof
Let $U$ a chart with its coordinate, and $p\neq o$ two distinct points in $U$. 
The 1-form $\omega_{p,1}$ as defined in corollary \ref{cor:highorderpoleompk} has a double pole at $p$ and no other pole, so it has vanishing residue.
The function $f(z)=\int_o^z \omega_{p,1}$ is meromorphic on $\curve$, with a simple pole at $p$ and no other pole, and a zero at $o$ (an no other zeros since $\#$zeros=$\#$poles).
The map $f:\curve\to \CC P^1$ is injective, indeed if there would exist $q\neq q'$ with $f(q)=f(q')$, then the function $z\mapsto f(z)-f(q)$ would have 2 zeros $q$ and $q'$, which is impossible.
$f$ is continuous, and since $\curve$ is open and compact, its image must be $\CC P^1$, it is thus surjective. $f$ is an isomorphism.
\eproof

\section{Newton's polygon}

So far we have been building 1-forms by Dirichlet principle, which is not explicit and not algebraic.

Newton's polygon's method allows to build and classify functions and forms on algebraic Riemann surfaces, by combinatorial and algebraic methods.

Let $P\in \mathbb C[x,y] $ a bivariate polynomial, and
\beq
\td\curve = \{(x,y) \ | \ P(x,y)=0 \} \subset \mathbb C\times \mathbb C,
\eeq
and the compact Riemann surface $\curve$ its desingularization.

\bd[Newton's polytope]
Let $\mathcal N\subset \mathbb Z\times\mathbb Z$ be the finite set of pairs $(i,j)$ such that $P_{i,j}\neq 0$:
\beq
P(x,y) = \sum_{(i,j)\in \mathcal N} P_{i,j} \ x^i \ y^j.
\eeq
The set of integer points $\mathcal N\subset \mathbb Z\times\mathbb Z$, is called the \textbf{Newton's polytope}\index{Newton's polytope} of $P$.
\ed

$$
\includegraphics[scale=0.25]{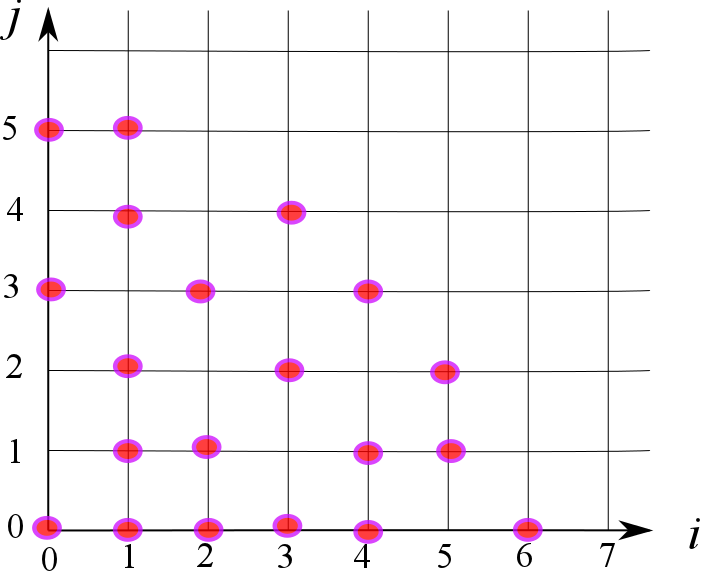}
$$

\subsection{Meromorphic functions and forms}

The maps $(x,y)\stackrel{x}{\mapsto} x$ and $(x,y)\stackrel{y}{\mapsto} y$ are meromorphic functions on $\curve$.
In fact, they provide an algebraic basis for all meromorphic functions:

\bp
Any meromorphic function $f$ on $\curve$ is a rational function of $x$ and $y$, there exists $R\in \mathbb C(x,y)$ such that
\beq
f = R(x,y).
\eeq
More precisely, if $d=\deg_y P$, there exists rational fractions $Q_0,Q_1,\dots,Q_{d-1} \in \mathbb C(x)$ such that
\beq
f =  \sum_{k=0}^{d-1}  Q_k(x) \ y^k.
\eeq
\ep

\proof

This is  Lagrange interpolating polynomial.
Let us define
\beq
Q(x,y) = \sum_{Y  \ | \ P(x,Y)=0} f(x,Y) \ \frac{P(x,y)}{P'_y(x,Y) \ (y-Y)} 
\eeq
it clearly satisfies $f(x,y) = Q(x,y)$ whenever $P(x,y)=0$, i.e. on $\curve$.
Moreover it is clearly a polynomial of $y$ of degree at most $d-1$, it can be written
\beq
Q(x,y) = \sum_{k=0}^{d-1} Q_k(x) \ y^k.
\eeq
Its coefficients $Q_k(x)$ are analytic and meromorphic functions of $x$, defined on $\mathbb CP^1\to \mathbb CP^1$.
Therefore they are rational fractions: $Q_k(x)\in \mathbb C(x)$.

\eproof

Since $dx$ is a 1-form, it immediately follows that
\bc
Any meromorphic 1-form $\omega$ on $\curve$ can be written as
\beq
\omega(x,y) = R(x,y) dx
\eeq
with $R(x,y)\in \mathbb C(x,y)$.

Up to changing $R\to R/P'_y$, it is more usual to write it in Poincar\'e form\index{Poincar\'e residue}:
\beq
\omega(x,y) = R(x,y) \frac{dx}{P'_y(x,y)}
\eeq
with $R(x,y)\in \mathbb C(x,y)$.

\ec

\subsection{Poles and slopes of the convex envelope}

\index{convex envelope}
The meromorphic functions $x$ and $y$ have poles and zeros.
Let $p$ a pole or zero of $x$ and/or $y$, and let $z$ a coordinate in a chart containing $p$, and such that $p$ has coordinate $z=0$.

If $p$ is a pole or zero of $x$ (resp. $y$) of order $-\alpha$ (resp. $-\beta$) we have
\beq
x(z) \sim  c\ z^{-\alpha}
\qquad ( \text{resp.} \quad
y(z) \sim  \tilde c\ z^{-\beta} \ ) .
\eeq
We assume that $(\alpha,\beta)\neq (0,0)$.

Let $D_{\alpha,\beta,m}$ the line of equation
\beq
D_{\alpha,\beta,m} = \{ (i,j) \ | \ \alpha i+\beta j=m \}.
\eeq
There exists $m_{\alpha,\beta}\in\mathbb Z$ such that
\beq
m_{\alpha,\beta} = \max \{ m \ | \ D_{\alpha,\beta,m}\cap \mathcal N\neq \emptyset\}.
\eeq
In other words, the line $D_{\alpha,\beta,m_{\alpha,\beta}}$ is the leftmost line parallel to the vector $(\beta,-\alpha)$ (left with respect to the orientation of  this vector), i.e. such that all the Newton's polytope lies to its right.

\bt[Poles and convex envelope]
The line $D_{\alpha,\beta,m_{\alpha,\beta}}$ is tangent to the convex envelope of the Newton's polytope, i.e. it contains an edge of the convex envelope of the polytope, or equivalently it contains at least 2 distinct vertices.
There is a 1-1 correspondance between poles/zeros of $x$ and/or $y$, and tangent segments to the convex envelope.
\et

\proof
By definition, the line $D_{\alpha,\beta,m_{\alpha,\beta}}$ intersects the Newton's polytope, and is such that all points of the Newton's polytope lie on the right side of that line.
It remains to prove that it intersects the Newton's polytope in at least 2 points.
Near $p$ we have:
\bea
0 = P(x,y) 
&=& \sum_{(i,j)\in \mathcal N} P_{i,j} \ x^i y^j \cr
&=& \sum_m \sum_{(i,j)\in \mathcal N\cap D_{\alpha,\beta,m} } P_{i,j} \ x^i y^j \cr
&\mathop{\sim}_{z\to 0} & \sum_{m\leq m_{\alpha,\beta}} z^{-m} \sum_{(i,j)\in \mathcal N\cap D_{\alpha,\beta,m} }  P_{i,j} \ c^i \tilde c^j \  (1+O(z)) \cr
&\mathop{\sim}_{z\to 0} &  z^{-m_{\alpha,\beta}} \sum_{(i,j)\in \mathcal N\cap D_{\alpha,\beta,m_{\alpha,\beta}} }  P_{i,j} \ c^i \tilde c^j  + O(z^{1-m_{\alpha,\beta}}) \cr
\eea
If $D_{\alpha,\beta,m_{\alpha,\beta}}\cap \mathcal N =\{(i_{p},j_p)\}$ would be a single point, we would have
\bea
0 \mathop{\sim}_{z\to 0} & z^{-m_{\alpha,\beta}}  P_{i_p,j_p} \ c^{i_p} \tilde c^{j_p}   (1+O(z))
\eea
which can't be zero, leading to a contradiction.

\eproof

$$
\includegraphics[scale=0.2]{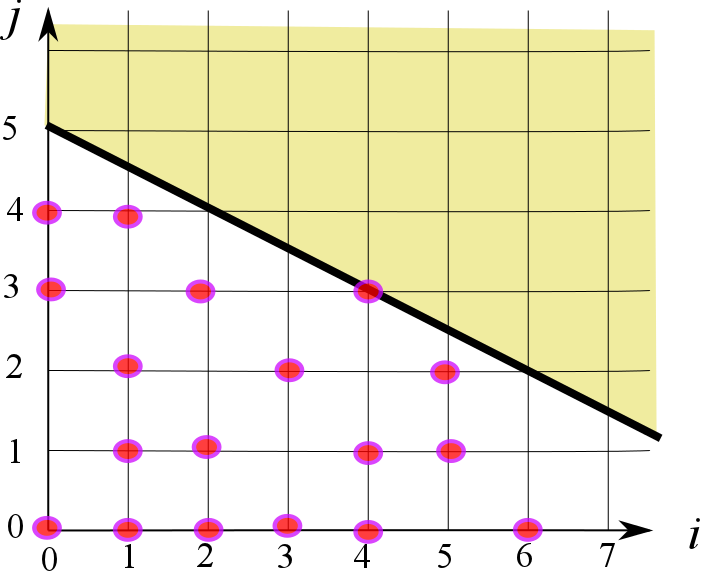}
\quad
\includegraphics[scale=0.2]{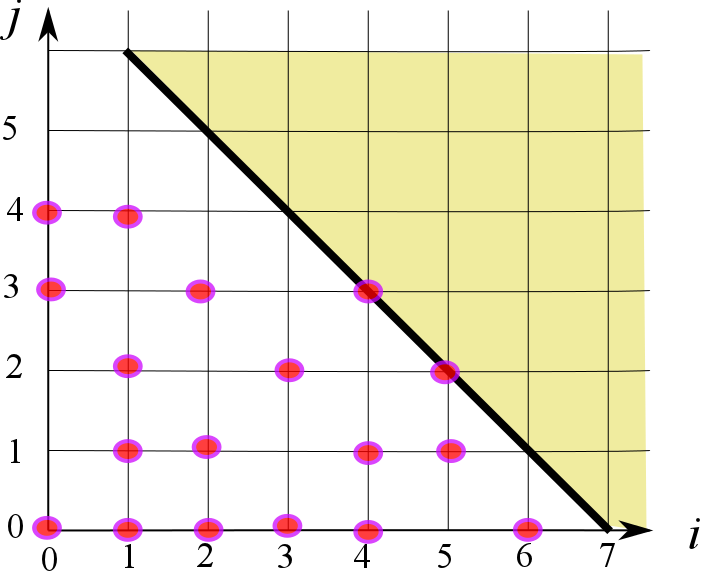}
\quad
\includegraphics[scale=0.2]{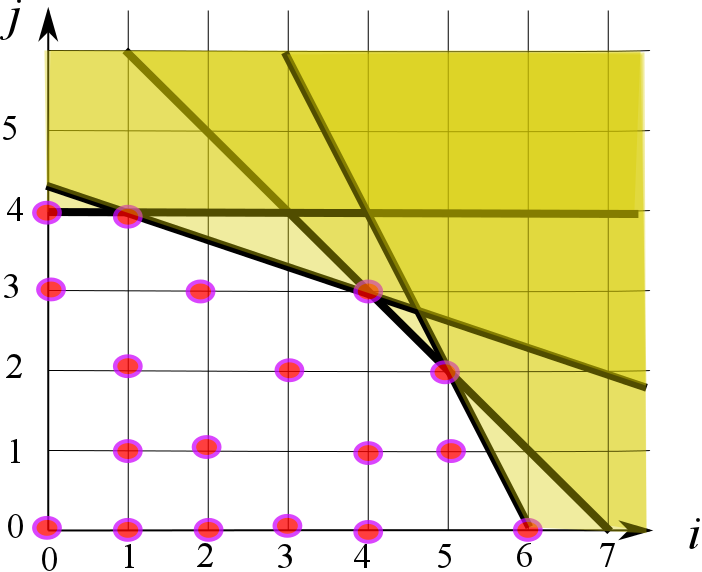}
$$

\bd[Amoeba]
The \textbf{Amoeba}\index{amoeba}, is the set $\mathcal A\subset \mathbb R\times \mathbb R$ defined by:
\beq
\mathcal A = \{ (\log |x|,\log |y|) \ | \ P(x,y)=0 \}.
\eeq
\ed
The amoeba has asymptotic lines in directions where $x$ and/or $y$ have poles or zeros, these lines have slope $\frac{\beta}{\alpha}$, and are orthogonal to the tangents to the convex envelope.
In fact, the amoeba looks like a thickening of a graph dual to a complete triangulation (with triangles whose vertices are on $\ZZ^2$ and of area $\frac12$) of the Newton's polytope.

Example: for the equation $P(x,y) = 1+c\ xy+x^2y+xy^2$, the amoeba looks like
$$
\includegraphics[scale=0.3]{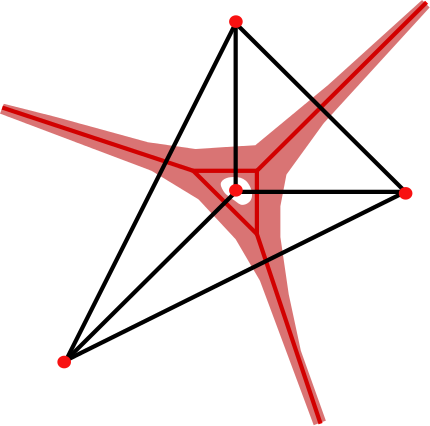}
$$

\subsection{Holomorphic forms}

Since $x$ is a meromorphic function, then $dx$ is a meromorphic 1-form.
It has poles and zeros.
Near a generic branchpoint $x=a, y=b$, a local coordinate is $z=y-b$, and we have
\beq
x \sim a -\frac12 \ \frac{P''_{yy}(a,b)}{P'_x(a,b)} (y-b)^2 + O(y-b)^3,
\eeq
so that
\beq
dx \sim  - \ \frac{P''_{yy}(a,b)}{P'_x(a,b)} (y-b) dy + O(y-b)^2dy.
\eeq
And we have
\beq
P'_y(x,y) \sim (y-b) P''_{yy}(a,b) + O(y-b)^2,
\eeq
therefore the 1-form
\beq
\frac{dx}{P'_y(x,y)} = - \ \frac{dy}{P'_x(x,y)}
\eeq
is analytic at generic branchpoints.
Can it be a holomorphic 1-form ?
\bp[Holomorphic forms and Newton's polytope]
We denote $\stackrel{\circ}{\mathcal N}$ the \textbf{interior}\index{interior of Newton's polygon} of the convex envelop of the Newton's polytope, i.e. the set of all integer points in $\ZZ\times \ZZ $ strictly inside the convex envelope of $\mathcal N$.

Assume that the coefficients $P_{i,j}$ are generic so that $\tilde\curve=\{(x,y) \ | \ P(x,y)=0\}$ has only generic branchpoints and no nodal points.

For every $(k,l)\in \ZZ\times \ZZ$,
let
\beq
\omega_{(k,l)} = \frac{x^{k-1} y^{l-1}}{P'_y(x,y)}\ dx
\eeq

Then
\beq
\omega_{(k,l)} \in \mathcal O^1(\curve) 
\quad \Leftrightarrow \quad
(k,l) \in \ \stackrel{\circ}{\mathcal N}
\eeq
It follows that
\beq
\genus=\dim \mathcal O^1(\curve) = \#\stackrel{\circ}{\mathcal N}.
\eeq

\ep
$$
\includegraphics[scale=0.25]{Newtonex1.png}
\qquad\qquad
\includegraphics[scale=0.25]{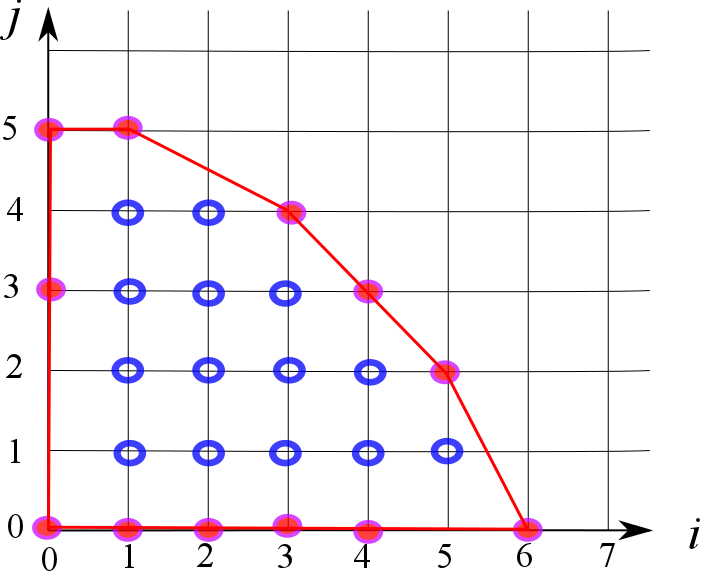}
$$

Remark that if $P$ is not generic, there can be nodal points,  and thus some zeros of the denominator $P'_y(x,y)$ are not ramification points, and are not zeros of $dx$, which means that $\omega_{(k,l)}$ can have poles at nodal points. 
However, by taking linear combinations $\sum_{k,l} c_{k,l}\omega_{(k,l)}$, one can cancel the poles at nodal points, and thus $\mathcal O^1(\curve)$ is generated by a linear subspace of $\CC^{\stackrel{\circ}{\mathcal N}}$, of codimension equal to the number of nodal points, i.e.:
\beq
\dim \mathcal O^1(\curve) = \#\stackrel{\circ}{\mathcal N}- \#\text{nodal points}.
\eeq
On the other hand, if $P$ is not generic, nodal points can be seen as degeneracies of cycles that have been pinched, therefore the genus has been decreased compared to the generic case, in such a way that
\beq
\genus = \#\stackrel{\circ}{\mathcal N}- \#\text{nodal points}.
\eeq
Eventually we always have
\beq
\dim \mathcal O^1(\curve) = \genus.
\eeq

\proof
Let $\omega$ a holomorphic 1-form on $\curve$.
$\omega/dx$ is a function, it may have poles at the zeros of $dx$. If we assume $P$ generic, then $dx$ has simple zeros, and thus $\omega/dx$ can have at most simple poles at the zeros of $dx$.
Moreover, the function $P'_y(x,y)$ vanishes at the ramification points, therefore $P'_y(x,y) \omega/dx$ is a meromorphic function on $\curve$ with no poles at ramification points. Its only poles can be at points where $P'_y(x,y)=\infty$, i.e. at poles of $x$ and/or $y$.
It can thus be written (the sum of non--vanishing terms is finite)
\beq
P'_y(x,y)\frac{\omega}{dx} = \sum_{(k,l) \in \ZZ\times \ZZ} c_{k,l} x^{k-1} y^{l-1}.
\eeq
Let us see how it behaves at a pole $p$ of $x$ and/or $y$, corresponding to a tangent  $D_{\alpha,\beta,m_{\alpha,\beta}}$ of the convex envelope.
We have
\beq
P'_y(x,y) \sim z^{\beta-m_{\alpha,\beta}} \sum_{(i,j)\in D_{\alpha,\beta,m_{\alpha,\beta}}\cap \mathcal N} j P_{i,j} c^i \tilde c^{j-1} .
\eeq
We also have $dx \sim -\alpha c z^{-\alpha-1} dz$ and therefore
\beq
\omega \sim \sum_{k,l} c_{k,l}\  z^{m_{\alpha,\beta}-k\alpha-l\beta-1} dz \left(\frac{-1}{\alpha}\sum_{(i,j)\in D_{\alpha,\beta,m_{\alpha,\beta}}\cap \mathcal N} j P_{i,j} c^{i-1} \tilde c^{j-1}\right)^{-1}.
\eeq
Since $\omega$ has no pole at $p$, i.e. at $z=0$, we must have
\beq
k\alpha+l\beta \leq m_{\alpha\beta}-1 < m_{\alpha,\beta}.
\eeq
In other words, the point $(k,l)$ must be strictly below the line $D_{\alpha,\beta,m_{\alpha,\beta}}$.
Since this must be true for all poles, i.e. all tangents to the convex envelope, we deduce that 
\beq
(k,l) \in \ \stackrel{\circ}{\mathcal N}.
\eeq

\eproof

As an immediate corollary of the proof we have

\bt[classification]
Let $(k,l)\in \ZZ^2$ two integers.
The 1-form 
\beq
\omega_{k,l} = \frac{ x^{k-1} y^{l-1} \ dx}{P'_y(x,y)}
\eeq
is 
\begin{itemize}

\item 1st kind\index{1st kind form} iff $(k,l)$ is strictly interior,

\item 3rd kind\index{3rd kind form} iff $(k,l)$ belongs to the boundary, it is thus at the intersection of 2 tangent segments, corresponding to 2 poles $p_1,p_2$. It is then a 3rd kind form with simple poles only at $p_1$ an $p_2$.

\item 2nd kind\index{2nd kind form} if $(k,l)$ is strictly exterior. It has poles at all tangents that are to its right (it is to the left of). 
The degree of the pole of $\omega_{k,l}$ at a pole $p=(\alpha,\beta)$ of corresponding tangent line $D_{\alpha,\beta,m_{\alpha,\beta}}$ is 
\beq
-\order_{p} \omega_{k,l} = \alpha k+\beta l -m_{\alpha,\beta}+1 \geq 2.
\eeq

\end{itemize}

\et

\subsection{Fundamental second kind form}

We defined fundamental second kind forms in def. \ref{deffundform2kind}.
We recall that  $B(z,z')$ is a symmetric $1\otimes 1$ form on $\curve\times \curve$ that has a double pole at $z=z'$ and no other pole, and such that
\beq\label{Bbehavdiag}
B(z,z') \sim \frac{dz \ dz'}{(z-z')^2} + \text{analytic at} \ z=z'.
\eeq
If the genus is $\geq 1$, it is not unique since we can add to it any symmetric bilinear combination of holomorphic 1-forms:
\beq
B(z,z') \to B(z,z') + \sum_{i,j=1}^{\genus} \kappa_{i,j} \ \omega_i(z) \otimes \omega_j(z').
\eeq

Remark that the $1\otimes 1$ form
\beq
 - \ \frac{  \frac{P(x',y)P(x,y')}{(x-x')^2(y-y')^2}     }{  P'_y(x,y) \ P'_y(x',y')} \  dx \ dx'
\eeq
is a symmetric $1\otimes 1$ form, it has a double pole at $(x,y)=(x',y')$, with the behavior $\frac{dxdx'}{(x-x')^2}$, and it has no pole if $x=x'$ and $y\neq y'$ nor if $y=y'$ but $x\neq x'$, so it is a good candidate for $B$.
However, it can have poles where $P$ has poles, i.e. at $x$ or $x'$ or $y$ or $y'$  $=\infty$.
Therefore the form
\beq
\td B((x,y),(x',y')) = B((x,y),(x',y'))+\frac{  \frac{P(x',y)P(x,y')}{(x-x')^2(y-y')^2}     }{  P'_y(x,y) \ P'_y(x',y')} \  dx \ dx'
\eeq
must be a symmetric 1-form $\otimes$ 1-form, whose poles can be only at $x,x',y,y'=\infty$.
There must exist a polynomial
\beq
 P'_y(x,y) \ P'_y(x',y') \ \frac{\td B((x,y),(x',y'))}{dx dx'}
=  Q(x,y;x',y') \in \CC[x,y;x',y'].
\eeq
At fixed $(x',y')$, this polynomial must be such that
\beq
\left( \frac{P(x',y)P(x,y')}{(x-x')^2(y-y')^2} \right)_{+} -  Q(x,y;x',y') 
\eeq
has monomials $x^{u-1}y^{v-1}$ only inside the Newton's polygon, i.e. only if $(u,v)\in \stackrel{\circ}{\mathcal N}$.

\subsubsection{General case}

\bp[Fundamental 2nd kind form]
the following $1\otimes 1$ form
\beq\label{defBarithm}
B_0((x,y);(x',y')) = - \ \frac{\frac{P(x,y')P(x',y)}{(x-x')^2(y-y')^2} - Q(x,y;x',y')}{P_y(x,y) P_y(x',y')} \ dx \ dx' 
\eeq
where $Q\in \CC[x,y,x',y'] $ is a polynomial
\bea
Q(x,y;x',y')
&=& \sum_{(i,j)\in \mathcal N} \sum_{(i',j')\in \mathcal N} P_{i,j} P_{i',j'}
\sum_{(u,v)\in \ZZ^2\cap \text{ triangle }(i,j),(i',j'),(i,j')}
|u-i| \ | v-j'| \cr
&& \Big( \delta_{(u,v)\notin {\stackrel{\circ}{\mathcal N}} \cup [(i,j),(i',j')] } \ \ x^{u-1}y^{v-1} x'^{i+i'-u-1}y'^{j+j'-v-1} \cr
&& +\delta_{(u,v)\notin {\stackrel{\circ}{\mathcal N}} \text{ and } (i+i'-u,j+j'-v)\in {\stackrel{\circ}{\mathcal N}} } \ \ x'^{u-1}y'^{v-1} x^{i+i'-u-1}y^{j+j'-v-1}  \cr
&& + \frac12\ \delta_{(u,v)\in  [(i,j),(i',j')]} \ \ x^{u-1}y^{v-1} x'^{i+i'-u-1}y'^{j+j'-v-1}  \Big) \ .
\eea
has a double pole at $(x,y)=(x',y')$ with behaviour \eqref{Bbehavdiag}, it has no pole if $x=x'$ and $y\neq y'$, it has no pole if $y=y'$ and $x\neq x'$, and it has no pole at the poles/zeros of $x$ and $y$.

\ep

Its only possible poles could be at  zeros of $P'_y(x,y)$ that are not zeros of $dx$, if these exist, i.e. these are common zeros of $P'_y(x,y)$ and $P'_x(x,y)$, and these are nodal points.

Generically, there is no nodal point, and the expression above is the fundamental second kind differential.
If nodal points exist, one can add to $Q$ a symmetric bilinear combination of monomials belonging to the interior of Newton's polygon, that would exactly cancel these unwanted poles.

\proof
When $x\to x'$ and $y\to y'$, we have $P(x,y')\sim (y'-y)P'_y(x,y) $ and $P(x',y)\sim (y-y')P'_y(x',y') $, so that  expression \eqref{defBarithm} has a pole of the type \eqref{Bbehavdiag}.

When $y\to y'$ and $x\neq x'$, we have $P(x,y')\sim (y'-y)P'_y(x,y) $ and $P(x',y)\sim (y-y')P'_y(x',y) $, so that  expression \eqref{defBarithm} behaves as
\beq
\frac{-P'_y(x,y) P'_y(x',y)}{(x-x')^2}
\eeq
which has no pole at $x\neq x'$. Same thing for $x\to x'$ with $y\neq y'$.

It remains to study the behaviors at poles/zeros of $x$ and/or $y$. 
Let us consider a point where both $x$ and $y$ have a pole (the other cases, can be obtained by changing $x\to 1/x$ and/or $y\to 1/y$, and remarking that expression \eqref{defBarithm} is unchanged under these changes).
At a pole $x\to\infty$ and $y\to\infty$, we have
\bea
\frac{P(x,y')P(x',y)}{(x-x')^2(y-y')^2} 
&\sim & \sum_{(i,j)\in \mathcal N} \sum_{(i',j')\in\mathcal N} \sum_{k\geq 1} \sum_{l\geq 1} P_{i,j} P_{i',j'} \ k l \ x^{i-k-1}y'^{j+l-1} x'^{i'+k-1}y^{j'-l-1} \cr
&\sim & \sum_{(u,v)\in \ZZ_+^2} x^{u-1} y^{v-1} \ \Big( \sum_{(i,j)\in \mathcal N} \sum_{(i',j')\in\mathcal N} \sum_{k\geq 1} \sum_{l\geq 1} P_{i,j} P_{i',j'} \ k l \cr
&& \ \delta_{u,i-k}\delta_{v,j'-l} x'^{i'+k-1}y'^{j+l-1} \Big) 
\eea
where the last bracket contains in fact a finite sum.
All the monomials such that $(u,v)\notin \stackrel{\circ}{\mathcal N} $ and $(u,v)$ is at the NE of the Newton's polygon, would yield a pole, and must be compensated by a term in $Q$. 

Let us consider such an $(u,v)$ monomial that enters in $Q$.
Notice that $(u,v)$ at the NE of the Newton's polygon implies that
$u=i-k\geq i'$ and $v=j'-l\geq j$, which implies in particular that this can occur only if $i>i'$ and $j'>j$.
Moreover, since all the line $[(i,j),(i',j')]$ is contained in the Newton's polygon, we see that $(u,v)$ must belong to the triangle $((i,j),(i',j'),(i',j))$.

Consider the point $(u',v')=(i'+k,j+l)=(i+i'-u,j+j'-v)$, which is the symmetric of $(u,v)$ with respect to the middle of $[(i,j),(i',j')]$.

Let us thus assume that $>i'$ and $j'>j$ and $(u,v)$ belongs to the triangle $((i,j),(i',j'),(i',j))$, and let us consider different cases:

$\bullet$ $(u,v)\notin [(i,j),(i',j')] $.
If $(u,v)\notin \stackrel{\circ}{\mathcal N}$, then the monomial $P_{i,j} P_{i',j'} k l x^{u-1}y^{v-1} x'^{u'-1} y'^{v'-1}$ should appear in $Q$ and is indeed the first term in \eqref{defBarithm}.
Notice that in that case the point $(u',v')$ can't be at the NE of Newton's polygon.
There are then 2 sub-cases:

$\bullet\bullet$ $(u',v')\in \stackrel{\circ}{\mathcal N}$, then we can add to $Q$ a monomial proportional to $x^{u'-1}y^{v'-1}$ without changing the pole property of $B$. In particular we can add
\beq
P_{i,j}P_{i',j'} \ k l \ x^{u'-1}y^{v'-1} x'^{u-1} y'^{v-1}
\eeq
which is the second term in \eqref{defBarithm}.
It allows to make $Q$ symmetric under the exchange $(x,y)\leftrightarrow (x',y')$.

$\bullet\bullet$ $(u',v')\notin \stackrel{\circ}{\mathcal N}$.
Notice that since $(u,v)\notin [(i,j),(i',j')] $, we also have $(u',v')\notin [(i,j),(i',j')] $.
Moreover, if $(u',v')\notin \stackrel{\circ}{\mathcal N}$, this implies that $(u',v')$ is ae SW of Newton's polygon.
This means that the monomial $P_{i,j}P_{i',j'} kl x^{u'-1}y^{v'-1} x^{u-1}y^{v-1}$ will appear in $Q$ in the contribution with $(i,j)\leftrightarrow(i',j')$.

$\bullet$ $(u,v)\in [(i,j),(i',j')] $.
This implies that $(u',v')\in [(i,j),(i',j')] $ as well.
Remarking that if $(u,v)\in [(i,j),(i',j')] $, we have $kl=(i-u)(j'-v)=(u-i')(v-j) $, we have
\beq
P_{i,j} P_{i',j'} (i-u) (j'-v) x^{u-1}y^{v-1} x'^{u'-1} y'^{v'-1}
=
P_{i',j'} P_{i,j} (i'-u) (j-v) x^{u-1}y^{v-1} x'^{u'-1} y'^{v'-1}
\eeq
i.e. this monomial appears twice in the sum \eqref{defBarithm} because it also appears in the term $(i,j)\leftrightarrow(i',j')$, and this is why it has to be multiplied by $\frac12$.

Also, if $(u,v)\in [(i,j),(i',j')] $, we have $kl=(i-u)(j'-v)=(i-u')(j'-v')$, the monomial $P_{i,j}P_{i',j'} kl x^{u'-1}y^{v'-1} x^{u-1}y^{v-1}$ also appears in \eqref{defBarithm}.

Also, if $(u,v)\in \stackrel{\circ}{\mathcal N}\cap [(i,j),(i',j')] $, this implies that $(u',v')\in \stackrel{\circ}{\mathcal N}\cap [(i,j),(i',j')] $, and thus this monomial and its symmetric under $(x,y)\leftrightarrow(x',y')$ are both inside Newton's polygon, so don't contribute to poles of $B$.

\smallskip

Eventually we have shown that the polynomial of \eqref{defBarithm} is symmetric under $(x,y)\leftrightarrow(x',y')$, and up to monomials inside $\stackrel{\circ}{\mathcal N}$, it compensates all the terms of $\frac{P(x,y')P(x',y)}{(x-x')^2(y-y')^2} $ that could possibly diverge.

This concludes the proof.
\eproof

\subsubsection{Hyperelliptical case}

\bp[Hyperellitical curves]
Consider the case $P(x,y)=y^2-Q(x)$, with $Q(x)\in\mathbb C[x]$ a polynomial of even degree, whose zeros are all distinct.
Let $U(x)=(\sqrt{Q(x)})_+$ be the polynmial part near $\infty$ of its square-root, and let $V(x)=Q(x)-U(x)^2$.
We then have
\beq
B((x,y);(x',y')) = \frac{y y' + U(x) U(x')+\frac12 V(x)+\frac12 V(x')}{2y y' (x-x')^2} \ dx \ dx'.
\eeq
It is a 1-form of $z=(x,y)\in\curve $, with a double pole at $(x,y)=(x',y')$, and no other pole, in particular no pole at $(x,y)=(x',-y')$. 
\ep

\chapter{Abel map, Jacobian and Theta function}
\label{chapAbel}

\section{Holomorphic forms}

We recall that we called $\mathcal O^1(\curve)$ the vector space of holomorphic 1-forms (no poles) on $\curve$.
We also call $H_1(\curve,\mathbb Z)$ the $\ZZ$--module (resp. $H_1(\curve,\CC)$ the $\CC$--vector space) of cycles, and for a surface of genus $\genus$ we have
\beq
\dim H_1(\curve,\mathbb Z) = \dim H_1(\curve,\CC) = 2\genus .
\eeq

\subsection{Symplectic basis of cycles}

We admit that, if $\genus\geq 1$, it is always possible to choose a 

\bd[symplectic basis of cycles]
\index{symplectic basis} of $H_1(\curve,\mathbb Z)$:
\beq
\acycle_i\cap \bcycle_j=\delta_{i,j}
\quad , \quad
\acycle_i\cap \acycle_j=0
\quad , \quad
\bcycle_i\cap \bcycle_j=0
\ .
\eeq
A choice of symplectic basis of cycles, is called a \textbf{marking}\index{marking} or \textbf{Torelli marking}\index{Torelli marking} of $\curve$.

\ed
In this definition, the \textbf{intersection numbers}\index{intersection numbers} are counted algebraically (taking the orientation into account): 
\beq
\gamma\cap \gamma' = \sum_{p\in \gamma \cap \gamma'} \pm 1
\eeq
where at a crossing point $p$, $\pm 1$ is $+1$ if the oriented $\gamma'$ crosses $\gamma$ from its right to its left, and $-1$ otherwise. 
The intersection number is invariant under homotopic deformations, is compatible with addition by concatenation, and thus descends to the homology classes by linearity.

We insist that a choice of symplectic basis is {\em not unique}.

$$
\includegraphics[scale=0.35]{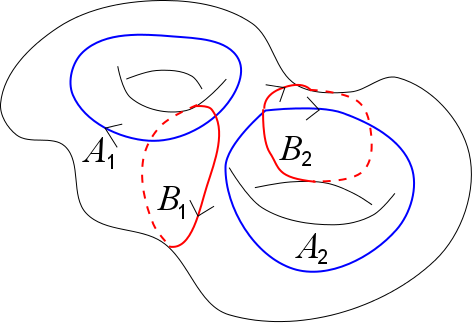}
$$

\subsection{Small genus}

\bt[Riemann sphere]
There is no non--identically--vanishing holomorphic 1-form on the Riemann sphere:
\beq
\mathcal O^1(\mathbb CP^1) = \{0\}
\qquad , \qquad
\dim \mathcal O^1(\mathbb CP^1) = 0.
\eeq

\et

\proof
write $\omega(z)=f(z)dz=-z'^{-2} f(1/z') dz'$ with $z'=1/z$. We want $f(z)$ to have no pole in the chart $\mathbb C$, so $f(z)$ could only be a polynomial, and we want $z'^{-2}f(1/z')$ to have no pole at $z'=0$, which implies the polynomial should be of degree $\leq -2$ which is not possible.
\eproof

In fact this applies to every genus zero curve (but as we shall see later, every genus zero Riemann surface is isomorphic to the Riemann sphere):
\bt[Genus zero]
There is no non--identically--vanishing holomorphic 1-form on a curve $\curve$ of genus $0$:
\beq
\mathcal O^1(\curve) = \{0\}
\qquad , \qquad
\dim \mathcal O^1(\curve) = 0.
\eeq

\et
\proof
Let $\omega$ a holomorphic 1-form on $\curve$.
Choose a base point $o\in\curve$, and define the function
\beq
f(p) = \int_o^p \omega.
\eeq
The function $f$ is well defined, in particular is independent of the integration path choosen to go from $o$ to $p$, since $\curve$ is simply connected.
The function $f$ is then a holomorphic function on $\curve$, and from theorem II-\ref{thfholconst}, it must be constant, which implies $\omega=df=0$.
\eproof

\bt[Torus]
On the torus $T_\tau=\CC / (\mathbb Z+\tau\mathbb Z)$
\beq
\mathcal O^1(\curve) \sim \mathbb C
\qquad , \qquad
\dim \mathcal O^1(\curve) = 1.
\eeq

\et

\proof
We know that $dz$ is a holomorphic 1-form.
If $\omega(z)=f(z)dz$ is another holomorphic 1-form, we must have $f(z)=f(z+1)=f(z+\tau)$, and $f$ can have no pole, from theorem II-\ref{thfholconst} it must be a constant, and thus $\mathcal O^1(\curve) \sim \mathbb C$.
\eproof

\subsection{Higher genus $\geq 1$}

Let $\acycle_1,\dots,\acycle_{2\genus}$ be a basis of $H_1(\curve,\mathbb Z)$.

\bt
The space of real harmonic 1-forms $\mathcal H(\curve)$ has real dimension:
\beq
\dim_{\mathbb R} \mathcal H(\curve) = 2\genus.
\eeq
The space of complex holomorphic 1-forms $\mathcal O^1(\curve)$ has complex dimension:
\beq
\dim_{\CC} \mathcal O^1(\curve) =  \genus.
\eeq
\et

\proof
We have already proved that the dimension of the space of real harmonic forms is $2\genus$. 
If $\nu$ is a real harmonic form, then $\omega=\nu+\ii*\nu$ is a complex holomorphic 1-form, such that $\Re\omega = \nu$. 
Therefore the map
\bea
\mathcal H(\curve) & \to & \mathcal O^1(\curve) \cr
\nu & \mapsto & \nu+\ii *\nu \cr
\Re \omega & \leftarrow & \omega
\eea
is an invertible isomorphism of real vector spaces. This implies that 
$\dim_{\mathbb R} \mathcal O^1(\curve) = 2\genus$.
Moreover, $\mathcal O^1(\curve)$ is clearly a complex vector space, and thus its dimension over $\CC$ is half of its dimension over $\RR$, therefore it is
$\genus$.
\eproof

\subsection{Riemann bilinear identity}

The \textbf{Riemann bilinear identity}\index{Riemann bilinear identity} is the key to many theorems, let us state it and prove it here.

Let $\omega$ and $\tilde \omega$ be 2 meromorphic forms.

Let $\acycle_i,\bcycle_i$ be Jordan arcs representative of a symplectic basis of cycles, chosen in a way that they all intersect (transversally) at the same unique point. We admit that it is always possible. 
Also up to homotopic deformations, we chose them to avoid all singularities of $\omega$ and $\tilde \omega$.
Let
\beq
\curve_0 = \curve \setminus \cup_i \acycle_i \cup_i \bcycle_i.
\eeq
By definition of a basis of non--contractible cycles, $\curve_0$ is a simply connected domain of $\curve$, called a \textbf{fundamental domain}\index{fundamental domain}, it is bounded by the cycles $\acycle_i,\bcycle_i$, its boundary is
\beq\label{RBIbordsumAB}
\partial\curve_0 = \sum_i {\acycle_i}_{\, \text{left}}-{\acycle_i}_{\, \text{right}} + \sum_i {\bcycle_i}_{\, \text{left}}-{\bcycle_i}_{\, \text{right}}.
\eeq

$$
\includegraphics[scale=0.35]{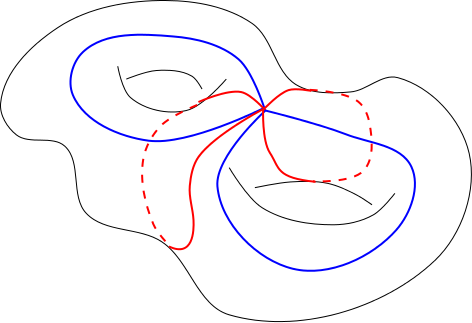}
$$

\bl
Let $U$ a tubular neighborhood of $\partial\curve_0$ in $\curve_0$, that contains no pole of $\omega$, and let $o\in U$ a generic point.
Then $f(p)=\int_o^p \omega$ is independent of a choice of integration path from $o$ to $p$ in $U$. 
$f(p)$ is a  holomorphic function on $U$, such that
\beq
df=\omega 
\qquad \text{on} \quad U.
\eeq
Moreover $f$ can be analytically continued to the boundary of $U$.
\el

\proof
A priori the integral $\int_o^p \omega$ depends on the path from $o$ to $p$ in $U$. 
Topologically $\curve_0$ is a disc, its boundary is a circle and its tubular neighborhood $U$ is an annulus.
There are 2 homotopically independent paths $\gamma_+,\gamma_-$  from $o$ to $p$ in $U$.
The difference between the integrals along the 2 independent paths, is
\bea
\int_{\gamma_+} \omega - \int_{\gamma_-} \omega 
&=& \oint_{\gamma_+-\gamma_-} \omega \cr
&=& 2\pi\ii \sum_{q=\text{poles of}\ \omega} \Res_q \omega \cr
&=& 0,
\eea
thanks to theorem II-\ref{thresidues}.
Therefore $f(p)$ is independent of the path chosen, it defines a function on $U$. 
It clearly satisfies $df=\omega$ and is thus holomorphic on $U$.
\eproof

\bt[Riemann bilinear identity]

Let $\omega$ and $\td\omega$ be 2 meromorphic forms on $\curve$, and let $f$, $\tilde f$ be 2  functions, holomorphic on a tubular neighborhood of $\partial\curve_0$ in $\curve_0-\text{poles}$, such that
\beq
df = \omega.
\quad , \quad 
d\tilde f = \tilde\omega.
\eeq
Then we have
\beq
\oint_{\partial \curve_0} f \td \omega = -\oint_{\partial \curve_0} \td f \omega  = \sum_{i=1}^\genus \oint_{\acycle_i} \omega \oint_{\bcycle_i} \td\omega - \oint_{\bcycle_i} \omega \oint_{\acycle_i} \td\omega.
\eeq
\et
\textbf{Remark:}
Depending on our choice of $\omega$ and $\td\omega$, the contour integral on the left hand side can often be contracted to surround only singularities of $f$ or of $\tilde \omega$, and eventually reduced to a sum of residues.
\smallskip

\proof
Observe that $\tilde \omega$ is continuous across any cycles, it takes the same value on left and right, whereas $f$ can have a discontinuity:
\bea\label{RBIdiscf}
\text{on} \ \acycle_i \qquad f_{\text left} - f_{\text right} = - \oint_{\bcycle_i} \omega \cr
\text{on} \ \bcycle_i \qquad f_{\text left} - f_{\text right} = \oint_{\acycle_i} \omega .
\eea
Inserted into \eqref{RBIbordsumAB} this immediately yields the theorem.
\eproof

\bc
If $\omega$ and $\td\omega$ are both holomorphic, then the left hand side can be contracted to 0, and 
\beq\label{eqRBI1}
\sum_{i} \oint_{\acycle_i}\omega \oint_{\bcycle_i}\tilde \omega - \oint_{\bcycle_i}\omega \oint_{\acycle_i}\tilde \omega = 0.
\eeq
\ec

\bt[Riemann bilinear inequality]\label{RBIineq}
\index{Riemann bilinear inequality}
If $\omega$ is a non--identically--vanishing holomorphic 1-form we have
\beq\label{eqRBI2}
2\ii \left(\sum_{i} \oint_{\acycle_i}\omega \oint_{\bcycle_i} \bar \omega - \oint_{\bcycle_i}\omega \oint_{\acycle_i}\bar\omega \right)>0.
\eeq
\et

\proof
Observe that the $L^2(\curve)$ norm of $\omega$ is
\beq
||\omega||^2 = 2\ii \int_\curve \omega \wedge \bar\omega >0 .
\eeq
Use Stokes theorem on the fundamental domain $\curve_0$:
\beq
\int_{\curve} \omega \wedge \bar\omega = \int_{\curve_0} \omega \wedge \bar\omega
= - \int_{\partial \curve_0} \bar f \omega 
\eeq
with $\omega=df$ on $\curve_0$.
Using \eqref{RBIbordsumAB} and \eqref{RBIdiscf} gives
\beq
\int_{\partial \curve_0} \bar f \omega 
= \sum_i \oint_{\bcycle_i} \omega \oint_{\acycle_i} \bar \omega - \oint_{\acycle_i} \omega \oint_{\bcycle_i} \bar \omega .
\eeq

\eproof

\section{Normalized basis}

\bt[Normalized basis of holomorphic forms]
\index{normalized basis of forms}
Given a symplectic basis of cycles, there exists a unique basis $\omega_1,\dots,\omega_\genus$ of $\mathcal O^1(\curve)$ such that
\beq
\forall\ i=1,\dots,\genus \ , \quad \oint_{\acycle_i} \omega_j = \delta_{i,j}.
\eeq
\et

\proof
Define the map
\bea
\epsilon : \quad
\mathcal O^1(\curve) &\to & \mathbb C^\genus \cr
\omega & \mapsto & \epsilon_i = \oint_{\acycle_i} \omega .
\eea
We shall prove that it is an isomorphism.
Since the dimension of the spaces on both sides are the same, it suffices to prove that the kernel vanishes.
Let us also denote 
\beq
\td\epsilon_i = \oint_{\bcycle_i} \omega.
\eeq
The Riemann bilinear inequality of theorem \ref{RBIineq} implies that if $\omega\neq 0$ we have
\beq
\Im (\sum_{i=1}^\genus \epsilon_i \bar{\td\epsilon}_i)<0,
\eeq
and therefore the vector $(\epsilon_1,\dots,\epsilon_{\genus})$ can't vanish.
This implies that $\operatorname{Ker}\epsilon =0$, and thus $\epsilon$ is invertible.

The normalized basis is:
\beq
\omega_i = \epsilon^{-1}( \{\delta_{i,j}\}_{j=1,\dots,\genus} ).
\eeq

\eproof

Then we define

\bd[Riemann matrix of periods]
The $\genus\times \genus$ matrix 
\beq
\tau_{i,j} = \oint_{\bcycle_i} \omega_j
\eeq
is called the \textbf{Riemann matrix of periods}\index{Riemann matrix of periods}.
\ed

We shall now prove that the matrix $\tau$ is a \textbf{Siegel matrix}\index{Siegel}: $\tau$ is symmetric and $\Im \tau$ is positive definite.
The proof relies on the Riemann bilinear identity.

\bc[Period $\implies$ Siegel matrix]
The $\genus\times \genus$ matrix of periods $\tau_{i,j}$ is symmetric and its imaginary part is positive definite.
\ec

\br
The converse is not true, not all Siegel matrices are Riemann periods of Riemann surfaces.
The subset of Siegel matrices that are periods of Riemann surfaces is characterized by the Krichever--Novikov conjecture, later proved by T. Shiota \cite{Shiota}.
For genus $\genus=1$, every $1\times 1$ Siegel matrix $\tau$ (i.e. a complex number whose imaginary part is $>0$) is a Riemann period, namely the Riemann period of the torus $T_\tau$. This starts being wrong for $\genus>2$.
\er

\smallskip

\proof
Indeed
Choosing $\omega=\omega_i$ and $\tilde\omega=\omega_j$ in \eqref{eqRBI1} gives
\beq
\tau_{i,j}-\tau_{j,i}=0
\eeq
and thus $\tau$ is a symmetric matrix.
Choosing $\omega=\omega_i$ in \eqref{eqRBI2} gives
\beq
2 \ii (
\bar\tau_{i,i} - \tau_{i,i} )>0.
\eeq
More generally, let $c\in \RR^\genus-\{0\}$, then choosing $\omega=\sum_i c_i \omega_i$ in \eqref{eqRBI2} yields
\beq
\sum_{i,j} c_i \ \Im\tau_{i,j} \ c_j >0
\eeq
i.e.
\beq
\Im\tau>0.
\eeq
\eproof

\section{Abel map and Theta functions}

Let $\curve^{\text{univ}}$ a universal cover of $\curve$, and $\curve_0$ a fundamental domain.
\bd
We define  the map
\bea
\mathbf u  : & \curve^{\text{univ}} &\to \CC^{\genus} \cr
& p & \mapsto
\mathbf u(p) = (u_1(p),\dots,u_\genus(p))
\quad , \quad
u_i(p) = \int_o^p \omega_i .
\eea
We also denote, by the same name $\mathbf u$, the quotient modulo $\mathbb Z^{\genus}+\tau \mathbb Z^{\genus}$, which is then defined on $\curve$ rather than $\curve^{\text{univ}}$
\bea
\mathbf u : &  \curve & \to \mathbb J = \CC^{\genus}/(\mathbb Z^{\genus}+\tau \mathbb Z^{\genus}) \cr
& p & \mapsto
\mathbf u(p) = (u_1(p),\dots,u_\genus(p)) \mod \mathbb Z^{\genus}+\tau \mathbb Z^{\genus}.
\eea
It is called the \textbf{Abel map}\index{Abel map}.
The $2\genus$-dimensional torus $\mathbb J = \CC^{\genus}/(\mathbb Z^{\genus}+\tau \mathbb Z^{\genus})$ is called the \textbf{Jacobian}\index{Jacobian} of $\curve$.

\ed

\bd[Abel map for divisors]
If $D=\sum_i \alpha_i.p_i$ is a divisor, we define by linearity
\beq
\mathbf u(D) = \sum_i \alpha_i \mathbf u(p_i).
\eeq
\ed

\bd[Riemann Theta function]
\index{Theta function}
If $\tau$ belongs to the $\genus$-dimensional \textbf{Siegel} space\index{Siegel} (symmetric complex matrices whose imaginary part is positive definite), then the map
\bea
\Theta: \CC^\genus & \to & \CC \cr
\mathbf u & \mapsto & \Theta(\mathbf u,\tau) \stackrel{\text{def}}{=} \sum_{\mathbf n\in \mathbb Z^{\genus}} e^{2\pi\ii (\mathbf n,\mathbf u)} \ e^{\pi\ii (\mathbf n,\tau\mathbf n)}
\eea
is analytic on $\CC^{\genus}$ (the sum is absolutely convergent for all $u\in \CC^\genus$).

\ed

\bl\label{lemThetaproperties}
It satifies
\bea\label{eqThetaphase}
&& \Theta(-\mathbf u)= \Theta(\mathbf u) \cr
&& \mathbf n \in \mathbb Z^{\genus} \ , \quad \Theta(\mathbf u+\mathbf n)= \Theta(\mathbf u) \cr
&& \mathbf n \in \mathbb Z^{\genus} \ , \quad \Theta(\mathbf u+\tau\mathbf n)= \Theta(\mathbf u) \  e^{-2\pi\ii (\mathbf n,\mathbf u)} \ e^{-\pi\ii (\mathbf n,\tau\mathbf n)}\cr
&& \mathbf a,\mathbf b \in \mathbb Z^{\genus}\times \mathbb Z^{\genus} \, , \quad (\mathbf a,\mathbf b)\in 2\mathbb Z+1 \quad \implies \quad \Theta(\frac12 \mathbf a+\frac12 \tau\mathbf b)= 0 \cr
\eea
\el

\proof
Easy computations.
\eproof

By composing the Abel map $\curve^{\text{univ}}\to \CC^{\genus}$ together with the Theta function $\CC^{\genus}\to \CC$, we can build complex functions on $\curve^{\text{univ}}$, and if we take "good" combinations, they can sometimes be defined on $\curve$ rather than $\curve^{\text{univ}}$. 

\begin{center}
{\em Theta functions will serve as building blocks for any meromorphic functions.}
\end{center}

As we shall see, Theta functions will be to meromorphic functions, what linear functions are to rational fractions (ratios of products of linear functions) for genus $0$.

We define
\bd
Let $q$ a generic point of $\curve$, and $\zeta\in \CC^\genus $.
Define the map $\curve_0\to \CC$:
\beq
p \mapsto 
g_{\zeta,q}(p) = \Theta(\mathbf u(p)-\mathbf u(q)+\zeta).
\eeq


\ed

\bl
Let $\zeta$ a zero of the function $\Theta$, i.e. $\Theta(\zeta)=0$.
If $(\Theta'_1(\zeta),\dots,\Theta'_\genus(\zeta))=0$, then $\zeta$ is called a singular zero.

If $\zeta$ is a singular zero, then the function $g_{\zeta,q}$ vanishes identically on $\curve_0$ for any $q$.
If $\zeta$ is not singular, then the map $\curve_0\to\CC$, $p\mapsto g_{\zeta,q}(p) $  has $\genus$ zeros on $\curve_0$.
One of these zeros is $q$, and let $P_1,\dots,P_{\genus-1}$ the other zeros. 
The Abel map of the divisor $P_1+\dots+P_{\genus-1}$ is
\beq
\mathbf u(P_1)+\dots +\mathbf u(P_{\genus-1}) = K-\zeta
\eeq
where $K$ is the \textbf{Riemann's constant}\index{Riemann's constant}:
\beq
K_k= \frac{-\tau_{k,k}}{2} - \sum_i \oint_{\acycle_i} u_i du_k.
\eeq
\el

\proof
It again uses Riemann bilinear identity.
Let $\curve_0$ a fundamental domain.
If $g_{\zeta,a}$ is not identically vanishing, we have
\bea
\frac{1}{2\pi\ii} 
\int_{\partial\curve_0} d\log g_{\zeta,q}(p) = \#\text{zeros of }\ g_{\zeta,q}(p).
\eea
Then, as in the Riemann bilinear identity, we decompose the boundary as \eqref{RBIbordsumAB}, and use the fact that $g_{\zeta,q}(p)$ is continuous across the cycle $\bcycle_i$ and $\log g_{\zeta,q}(p)$ has a discontinuity across the $\acycle_i$ boundary given by \eqref{eqThetaphase}, thus equal to
\beq
\log g_{\zeta,q}(p)_{\acycle_i \, \text{left}}-\log g_{\zeta,q}(p)_{\acycle_i \, \text{right}}
= 2\pi\ii ( u_i(p)-u_i(q)+\zeta_i+\frac12 \tau_{i,i} )
\eeq
and thus taking the differential:
\beq
d\log g_{\zeta,q}(p)_{\acycle_i \, \text{left}}-d\log g_{\zeta,q}(p)_{\acycle_i\,\text{right}}
= 2\pi\ii d u_i(p).
\eeq
It follows that
\beq
\#\text{zeros of }\ g_{\zeta,q} = \sum_{i=1}^\genus \oint_{\acycle_i} du_i = \genus.
\eeq
Clearly, since $\Theta(\zeta)=0$ we have $g_{\zeta,q}(q)=0$.
We call $P_{\genus}=q$ and $P_1,\dots,P_{\genus-1}$ the other zeros.

Moreover, we have
\bea
\frac{1}{2\pi\ii} \int_{\partial\curve_0} u_k(p) d\log g_{\zeta,q}(p) =  \sum_{j=1}^{\genus} u_k(P_j) 
= u_k(q) + u_k(\sum_{j=1}^{\genus-1} P_j).
\eea
Integrating by parts
\bea
\frac{1}{2\pi\ii} \int_{\partial\curve_0} u_k(p) d\log g_{\zeta,q}(p) =  
- \frac{1}{2\pi\ii} \int_{\partial\curve_0} \log g_{\zeta,q}(p) \ du_k(p)
\eea
and using the discontinuity of $\log g_{\zeta,q}$ across $\acycle_i$, we have
\bea
u_k(\sum_{j=1}^\genus P_j) 
&=& - \sum_i \oint_{p\in\acycle_i} du_k(p) ( u_i(p)-u_i(q)+\zeta_i+\frac12 \tau_{i,i} ) \cr
&=& u_k(q) - \zeta_k - \frac{\tau_{k,k}}{2} - \sum_i \oint_{p\in\acycle_i}  u_i \ du_k  \cr
&=& u_k(q) +K_k -\zeta_k
\eea
where $K$ is the Riemann's constant.
\eproof

\bl[Theta divisor]
The set $(\Theta)$ of zeros of $\Theta$, called the \textbf{Theta divisor}\index{Theta divisor} is a submanifold of  $\CC^{\genus}$ of dimension $\genus-1$.
Let $W_{\genus-1}$ be the set of divisors sums of $\genus-1$ points.
The map
\bea
W_{\genus-1} & \to & (\Theta) \cr
D & \mapsto & K-\mathbf u(D)
\eea
is an isomorphism.
In other words, every zero $\zeta$ of $\Theta$ can be uniquely written (modulo $\ZZ^\genus+\tau \ZZ^\genus$) as (minus) the Abel map of a sum of $\genus-1$ points shifted by $K$, and vice-versa, the Abel map of any sum of $\genus-1$ points, shifted by $K$ is a zero of $\Theta$.
\beq
\forall \ \zeta \ | \ \Theta(\zeta)=0
\qquad \exists ! \ D=P_1+\dots+P_{\genus-1}  \ | \quad
\zeta = K-\mathbf u(D).
\eeq
\beq
\forall \ D=P_1+\dots+P_{\genus-1}\in W_{\genus-1} 
\qquad , \quad
\zeta = K-\mathbf u(D) \in (\Theta).
\eeq

\el

\proof
Choose $q$ a generic point of $\curve$.
Let $\zeta$ a non-singular zero of $\Theta$.
Since one of the zeros of $g_{\zeta,q}$ is $q$, 
the others are $P_1,\dots,P_{\genus-1}$, and we have
\beq
\zeta = K - \sum_{i=1}^{\genus-1} \mathbf u(P_i).
\eeq
A priori, the $P_i$s are functions of $\zeta$ and $q$.

At fixed $q$, we thus have a map
\bea
(\Theta) & \to & W_{\genus-1} \cr
\zeta & \mapsto & P_1+\dots +P_{\genus-1}
\eea
where $W_{\genus-1}$ is the set of divisors sums of $\genus-1$ points.
The inverse  map is:
\bea
 W_{\genus-1} & \to & (\Theta) \cr
D & \mapsto & K-\mathbf u(D)
\eea
which is clearly independent of $q$.
So the map is invertible, and independent of $q$.
We thus have
\beq
(\Theta) = K - \mathbf u(W_{\genus-1}).
\eeq

\eproof

\bt[Divisors of meromorphic functions]\label{thprincdiv}
If $f\neq 0$ is a meromorphic function, with divisor $(f)=\sum_i \alpha_i . p_i$, then
\beq
\deg (f)= 0
\qquad , \qquad 
\mathbf u((f))=0
\eeq
and, for any non--singular choice of $\zeta\in (\Theta)$, there exists $C\in \CC^*$, such that
\beq\label{eqfprodgq}
f(p) = C \prod_{i} g_{\zeta,p_i}(p)^{\alpha_i}.
\eeq
Reciprocally, if $D=\sum_i \alpha_i .p_i$, is a divisor such that
\beq
\deg D= 0
\qquad , \qquad 
\mathbf u(D)=0
\eeq
then $D$ is the divisor of a meromorphic function.
\et

\proof
If $D$ is a divisor of degree $0$ with $\mathbf u(D)=0$, then \eqref{eqfprodgq} clearly defines a meromorphic function on $\curve$, proving the last part of the theorem.

Vice versa, let $f$ a meromorphic function, let $\zeta$ a regular zero of $\Theta$, and $D=(f)=\sum_i \alpha_i.p_i$.
We already know that $\deg D=\sum_i \alpha_i=0$.
Define on $\curve_0$:
\beq
g(p) = f(p) \ \prod_{i} g_{\zeta,p_i}(p)^{-\alpha_i}.
\eeq
It has no pole nor zeros in $\curve_0$, therefore $\log g$ is holomorphic on $\curve_0$ and $d\log g$ is a holomorphic 1-form on $\curve_0$.

$g$ has no monodromy around $\acycle_i$, and around $\bcycle_i$ it gets multiplied by a phase independent of $p$:
\beq
g(p+\acycle_i)=g(p)
\qquad , \qquad
g(p+\bcycle_i)=g(p) \ e^{2\pi\ii u_i(D)},
\eeq
which implies that $d\log g$ is analytic across the boundaries of $\curve_0$, it thus defines a holomorphic 1-form on $\curve$.
Therefore there exists $\lambda_1,\dots,\lambda_{\genus}$ such that
\beq
d\log g = \sum_{i=1}^\genus \lambda_i du_i
\eeq
and thus there exists $C\in \CC^*$ such that
\beq
g(p) = C e^{\sum_{i=1}^\genus \lambda_i u_i(p)}.
\eeq
This last expression has monodromy $e^{\lambda_i}=1$ around $\acycle_i$  implying $\lambda_i \in 2\pi\ii \mathbb Z$, and it has monodromy $e^{\sum_k \tau_{i,k}\lambda_k}=e^{2\pi\ii u_i(D)}$ around $\bcycle_i$, implying 
\beq
\mathbf u(D) \in \mathbb Z^{\genus}+\tau \mathbb Z^{\genus} \equiv 0 \ \text{in} \ \mathbb J.
\eeq

This implies the theorem.
\eproof

\bc
The Abel map $\curve \to \mathbb J$ is injective.
\ec
\proof
If there would be $p_1,p_2$ distinct and such that $\mathbf u(p_1)-\mathbf u(p_2)=0$ in $\mathbb J$, this would imply that $D=p_1-p_2$ would be a divisor of degree 0, and such that $\mathbf u(D)=0$, and there would thus exist a meromorphic function $f$ with only one simple pole at $p_2$ and no other pole.
First assume that there exists a holomorphic 1-form $\omega$ that doesn't vanish at $p_2$, then $f\omega$ would contradict corollary II-\ref{cor1simplepole}.
Therefore assume that every holomorphic 1-form in $\mathcal O^1(\curve)$ vanishes at $p_2$. Let
\beq
k=\min \{ \operatorname{order}_{p_2} \omega \ | \ \omega\in H_1(\curve)\} \geq 1.
\eeq
Then if $\omega\in \mathcal O^1(\curve)$ is a holomorphic 1-form that vanishes at order $k$ at $p_2$, then $f\omega$ would vanish at order $k-1$, that would contradict the minimality of $k$.
Therefore this is impossible, proving the corollary.
\eproof

\bt[Jacobi inversion theorem]
There is a bijection
\bea
W_{\genus} & \to & \mathbb J \cr
D & \mapsto & \mathbf u(D)-K
\eea
where $W_\genus$ is the set of divisors that are sums of $\genus$ points.
In other words any point $v\in \mathbb J$, can be uniquely written as
\beq
v=-K+\mathbf u(P_1)+\dots+\mathbf u(P_\genus).
\eeq

\et

\proof
Let $v\in \mathbb C^\genus$.
By the same-as-usual Riemann bilinear identity argument, the function $p\mapsto g(p)=\Theta(\mathbf u(p)-v)$ on $\curve_0$, has $\genus$ zeros in $\curve_0$.
Let us call them $q_1,\dots,q_{\genus}$.
By definition $\mathbf u(q_1)-v=\zeta$ is a zero of $\Theta$, and therefore there exists a unique divisor $D'=P_1+\dots+P_{\genus-1}$ in $W_{\genus-1}$ such that
\beq
\mathbf u(q_1)-v = K-\mathbf u(P_1+\dots+P_{\genus-1})
\eeq
This implies, after defining $D=q_1+D'\in W_{\genus} $ that
\beq
v=-K+\mathbf u(D).
\eeq
The map $v\mapsto D$ seems to depend on the choice $q_1$ of zero of $g$, but we can see that the map is invertible, with inverse map $W_{\genus}\to \mathbb J$, $D\mapsto -K+\mathbf u(D)$ independent of this choice.
This ends the proof.
\eproof

\bl
Let $\zeta\in (\Theta)$. Since $\Theta$ is an even function then  $-\zeta\in(\Theta)$, and 
\beq
\omega_\zeta = \sum_{i=1}^\genus \Theta'_i(\zeta) \ \omega_i = -\omega_{-\zeta}
\eeq
is a holomorphic 1-form with divisor
\beq
(\omega_\zeta) = \sum_{i=1}^\genus P_i + \sum_{i=1}^\genus \td P_i
\eeq
where $D=P_1+\dots+P_{\genus-1}$ is the unique divisor such that $\mathbf u(D)=K-\zeta$, and $\td D=\td P_1+\dots+\td P_{\genus-1}$ is the unique divisor such that $\mathbf u(\td D)=K+\zeta$.
\el

\proof
The function $g_{\zeta,q}(p)= \Theta(\mathbf u(p)-\mathbf u(q)+\zeta)$ has a zero at $p=q$, and at $P_1,\dots ,P_{\genus-1}$.
Near $q=p$ it behaves, in a chart coordinate, as
\beq
g_{\zeta,q}(p) \sim_{p\to q} (p-q) \  \frac{\omega_\zeta(q)}{dq} .
\eeq
We can choose $q=P_i$, and then $g_{\zeta,P_i}(p)$ has a zero of order at least 2 at $P_i$, implying that $\omega_\zeta(P_i)$ must vanish (in fact it must vanish at the same order as the multiplicity of $P_i$ in $D$).
This also holds for $\omega_{-\zeta}(\td P_i)=-\omega_\zeta(\td P_i)$ by the same reason.
Therefore $\omega_\zeta$ vanishes at $D+\tilde D$.
Since a holomorphic 1-form has $2\genus-2$ zeros, these are the only zeros, and thus
\beq
(\omega_\zeta) = D+\td D.
\eeq

\eproof

\bt
Let $\omega$ a meromorphic 1-form, and $D=(\omega)=\sum_i \alpha_i .p_i$ its divisor, then:
\beq\label{eqdivom}
\deg D=2\genus-2
\qquad , \qquad
\mathbf u(D) = 2K.
\eeq
\et

\proof
We have already proved (Riemann-Hurwitz theorem) that a meromorphic 1-form is such that $\deg D=2\genus-2$.
Moreover, if $\omega_1$ and $\omega_2$ are meromorphic forms, then $\omega_1/\omega_2$ is a meromorphic function, its divisor is $(\omega_1)-(\omega_2)$ and it satisfies $\mathbf u((\omega_1)-(\omega_2)) = 0$, therefore $\mathbf u(D)$ is the same for all  1-forms.

The previous lemma shows that $\mathbf u((\omega_\zeta))=2K$ for $\zeta$ a zero of $\Theta$.
\eproof

\bt\label{thmoneformdivisor}
Let $\omega$ a holomorphic 1-form, and $D=(\omega)=\sum_i \alpha_i .p_i$ its divisor, then:
\beq\label{eqdivom1}
\deg D=2\genus-2
\qquad , \qquad
\mathbf u(D) = 2K.
\eeq
Vice--versa, if $D$ is a positive divisor satifying \eqref{eqdivom1}, then there exists a unique (up to scalar multiplication) holomorphic 1-form whose divisor is $D$.
\et

\proof
The first part of the theorem is already proved.
Now let $D$ a positive divisor of degree $2\genus-2$ that satisfies \eqref{eqdivom1}:
\beq
D=\sum_{i=1}^{2\genus-2} P_i.
\eeq
Let $D'=\sum_{i=1}^{\genus-1} P_i$, then $\zeta=K-\mathbf u(D')\in (\Theta)$, which implies that $\omega_\zeta$ is a holomorphic 1-form whose divisor is $D'+\td D'$, where $\td D'$ is the unique positive divisor of degree $\genus-1$ such that $K-\mathbf u(\td D')=-\zeta=-(K-\mathbf u(D'))= K-\mathbf u(D-D') $, i.e. $\td D'=D-D'$.
This implies that $(\omega_\zeta)=D$. 

To prove uniqueness (up to scalar multiplication), observe that if two holomorphic forms have the same divisor of zeros, their ratio is a meromorphic function without poles, therefore it is a constant.
\eproof

\subsection{Divisors, classes, Picard group}

\bd
A divisor $D$ is called principal, iff there exists a meromorphic function $f$ such that $(f)=D$.
From theorem \ref{thprincdiv}, principal divisors are those such that:
\beq
\deg D=0 \qquad , \qquad \mathbf u(D)=0.
\eeq

The $\ZZ$--module of divisors, quotiented by principal divisors is called the Picard group:
\beq
\text{Pic}(\curve) = \text{Div}(\curve)/\text{Principal divisors}(\curve).
\eeq
The degree and Abel maps are morphisms (they can be pushed to the quotient)
\beq
\deg: \text{Pic}(\curve) \to \ZZ,
\eeq
\beq
\mathbf u: \text{Pic}(\curve) \to \mathbb J(\curve).
\eeq
\ed

\bd[Canonical divisor]
The canonical divisor class $\mathfrak K\in \text{Pic}(\curve)$ is the divisor class (modulo principal divisors) of any meromorphic 1-form.
(It is well defined since the ratio of 2 meromorphic forms is a meromorphic function).
We have
\beq
\deg \mathfrak K = 2\genus-2
\qquad , \qquad
\mathbf u( \mathfrak K) = 2K.
\eeq

\ed

\section{Prime form}

Let us consider special zeros of $\Theta$ as follows:
Let $\mathbf c  = \frac12 \mathbf a + \frac12 \tau \mathbf b$ be an \textbf{half--integer characteristic}\index{half--integer characteristic}, i.e. $\mathbf c=-\mathbf c$. 
We say it is odd iff the scalar product $(\mathbf a,\mathbf b) $ is odd 
\beq
\mathbf c  = \frac12 \mathbf a + \frac12 \tau \mathbf b \ \ \text{odd}
\qquad \Leftrightarrow \qquad
(\mathbf a,\mathbf b) = \sum_{i=1}^\genus a_i b_i \in 2\mathbb Z+1 ,
\eeq
it is then a zero of $\Theta$.
Let us admit that there exists non--singular half-integer odd characteristics.

\bl
Let $\mathbf c$ a non--singular half-integer odd characteristics.
The 1-form
\beq
h_{\mathbf c} = \sum_{i=1}^{\genus}  \Theta'_i(\mathbf c)\ \omega_i
\eeq
has $\genus-1$ double zeros, located at the unique positive integer divisor  of degree $\genus-1$ such that $\mathbf u(D)=K-\mathbf c$.

The square root $\sqrt{h_{\mathbf c}}$ is well defined and analytic on a fundamental domain $\curve_0$, it is a $\frac12$ spinor form.
\el

\proof
We have seen that the zeros of $h_c$ have divisor $D+\tilde D$, with $D$ and $\td D$ the unique divisors such that $u(D)=K-\mathbf c$ and $u(\tilde D)=K+\mathbf c$, but since $\mathbf c=-\mathbf c$ in $\mathbb J$, we have $D=\tilde D$, and thus all zeros are double zeros.
\eproof

\bd[Prime form]
\index{prime form}
\beq
E_{\mathbf c}(p,q) = \frac{\Theta(\mathbf u(p)-\mathbf u(q)+\mathbf c)}{\sqrt{h_{\mathbf c}(p) \ h_{\mathbf c}(q)} }
\eeq
It is a $\frac{-1}{2} \otimes \frac{-1}{2}$ bi-spinor defined on a fundamental domain $\curve_0\times \curve_0$, it vanishes at $p=q$ and nowhere else, it behaves near $p=q$, in any local coordinate $\phi_U$, as
\beq
E_{\mathbf c}(p,q) \sim \frac{\phi_U(p)-\phi_U(q)}{\sqrt{d\phi_U(p) \ d\phi_U(q)}} \ (1+O(\phi_U(p)-\phi_U(q))).
\eeq
It has monodromies (up to a sign):
\bea
E_{\mathbf c}(p+\acycle_i,q) &=& \pm \ E_{\mathbf c}(p,q) \cr
E_{\mathbf c}(p+\bcycle_i,q) &=& \pm \ E_{\mathbf c}(p,q) \ e^{-2\pi\ii(u_i(p-q)+c_i + \frac12 \tau_{i,i})}
\eea

Under a change of $\mathbf c$, we have
\beq
E_{\mathbf c'}(p,q) = \pm \ 
E_{\mathbf c}(p,q) \ e^{\pi\ii (\mathbf b-\mathbf b',\mathbf u(p)-\mathbf u(q))} . 
\eeq

\ed

\proof
Notice that $g_{\mathbf c,q}(p)$ has $\genus$ zeros, one of them is $q$, and the others $P_1,\dots P_{\genus-1}$ are such that $u(\sum_i P_i)=K-D$, therefore they are the same as the zeros of $\sqrt{h_{\mathbf c}}$.
This shows that as a function of $p$, $E_{\mathbf c}(p,q)$ vanishes only at $p=q$ and nowhere else.
The other properties are rather obvious.

\eproof

\bt[Green function]
The Green function defined in cor II-\ref{cordefGreen}, is (up to an additive constant $C$ independent of $p$):
\beq
\mathcal G_{q_+,q_-}(p) = C+ \log{\left|\frac{\Theta(\mathbf u(p)-\mathbf u(q_+)+\mathbf c)}{\Theta(\mathbf u(p)-\mathbf u(q_-)+\mathbf c)}\right|} -\pi \  \Im \mathbf u(p)^T \ (\Im\tau)^{-1} \ \Im (\mathbf u(q_+)-\mathbf u(q_-)).
\eeq

\et

This can also be written with the prime form
\beq
\mathcal G_{q_+,q_-}(p) = C+ \log{\left|\frac{E_c(p,q_+)}{E_c(p,q_-)} \ \frac{\nu(q_+)}{\nu(q_-)}\right|} -\pi \  \sum_{i,j=1}^\genus \Im \mathbf u_i(p) \ ((\Im\tau)^{-1})_{i,j} \ \Im (\mathbf u_j(q_+)-\mathbf u_j(q_-)).
\eeq
where $\nu$ is any meromorphic 1-form.
Observe that changing the 1-form $\nu$, or the odd characteristic $c$, or the origin $o$ for the definition of the Abel map, or the basis of symplectic cycles, amount to changing $\mathcal G$ by an additive constant $C$ independent of $p$.

\section{Fundamental form}
\label{secfundformgen}

The following object is probably the most useful form that can be defined on a Riemann surface, it allows to reconstruct everything.
Riemann already introduced it, it then received several names, and we shall call it the  \textbf{Fundamental second kind differential}.

Although it differs from another object named Bergman kernel in operator theory, it is sometimes called also \textbf{Bergman kernel}\index{Bergman kernel}, because it was extensively studied by Bergman and Schiffer \cite{BergSchif}, and this name was used in a series of Korotkin-Kokotov seminal articles \cite{KoKo,KoKo2}.
It is also very close to other objects, called Schiffer kernel, or Klein kernel, that were introduced, and that we shall see below.

\bd[Fundamental second kind differential]\label{defBergman}
\index{fundamental form}
\index{fundamental second kind differential}
\beq
B(p,q) = d_p d_q \left(\log \Theta(\mathbf u(p)-\mathbf u(q)+\mathbf c)\right) .
\eeq
It is independent of a choice of $\mathbf c$.
$B$ is a meromorphic symmetric bilinear $1\boxtimes 1$ form on $\curve\times \curve$, it has a double pole at $p=q$ and no other pole, and near $p\to q$ it behaves (in any choice of local coordinate $\phi_U$) as
\beq\label{Bbehavesdiag}
B(p,q) = \frac{d\phi_U(p) \ d\phi_U(q)}{(\phi_U(p)-\phi_U(q))^2} + \text{analytic at }\ q.
\eeq
It also satisfies
\beq
B(q,p)=B(p,q)
\eeq
\beq
\oint_{q\in \acycle_i} B(p,q)=0
\eeq
\beq
\oint_{q\in \bcycle_i} B(p,q)=2\pi\ii \ \omega_i(p) .
\eeq
It is sometimes called \textbf{Bergman kernel}.
\ed

This definition contains assertions that need to be proved.

\proof
It is clearly symmetric because $\Theta$ is even.

It has a double pole at $p=q$ because it is the second derivative of the log of something that vanishes linearly.
There is no simple pole contribution at $p=q$ because of parity, or because of the same reason, it is the second derivative of a log. 

It is globally meromorphic on $\curve$, because when we go around a cycle, $\Theta$ gets multiplied by a phase, $\log\Theta$ receives an additive contribution which is linear in $\mathbf u(p)-\mathbf u(q)+c$, and thus is killed by taking the second derivative  $d_p d_q$. 

$\Theta$ also has zeros at $P_1,\dots,P_{\genus-1}$, but since those points are independent of $p$ and $q$, taking derivatives with respect to $p$ and $q$ kills these poles.

The values of the $\acycle$-cycle and $\bcycle$-cycle integrals are easy from the quasiperiodicity properties of $\Theta$, indeed integrating a derivative just gives the difference of values of $\log\Theta$ between the end and start of the integration path, i.e. the phase shift of $\log\Theta$.

The fact that it is independent of $\mathbf c$ follows from uniqueness. 
Indeed If $B$ and $B'$ are 2 fundamental forms of the second kind, for instance corresponding to $\mathbf c$ and $\mathbf c'$, their difference has no pole at all, and has vanishing $\acycle$-cycle integrals, so it is a vanishing holomorphic 1-form, showing that $B$ is unique.
\eproof

\smallskip

As a corollary we get

\bd[Third kind forms]\label{defthirdkindfromB}
For any distinct points $q_1,q_2 \in \curve_0\times \curve_0$, the following 1-form of $p\in\curve$
\beq
\omega_{q_1,q_2}(p) = d_p \left( \log \frac{\Theta(\mathbf u(p)-\mathbf u(q_1)+\mathbf c)}{\Theta(\mathbf u(p)-\mathbf u(q_2)+\mathbf c)} \right)
= \int_{q_2}^{q_1} B(p,.)
\eeq
is independent of a choice of $\mathbf c$, it is meromorphic on $\curve$, it has a simple pole at $p=q_1$ with residue $+1$ and a simple pole at $p=q_2$ with residue $-1$ and no other pole, and it is normalized on $\acycle$--cycles
\bea
\oint_{p\in\acycle_i} \omega_{q_1,q_2}(p)=0 \cr
\Res_{q_1} \omega_{q_1,q_2}=1 \cr
\Res_{q_2} \omega_{q_1,q_2} = -1.
\eea
\ed

\subsubsection{Other kernels}

There exist  other classical bilinear differentials, slightly different from $B$.
First notice that for every symmetric $\genus\times \genus$ matrix $\kappa$, then
\beq
B_\kappa(p,q) = B(p,q) + 2\pi\ii \sum_{i,j} \kappa_{i,j} \ \omega_i(p) \omega_j(q)
\eeq
is also a meromorphic symmetric bilinear $1\otimes 1$ form on $\curve\times \curve$, with a double pole at $p=q$ and no other pole with behavior \eqref{Bbehavesdiag}, the only difference is that it satisfies
\bea
\oint_{q\in\acycle_i} B_\kappa(p,q) &=& 2\pi\ii \sum_j \kappa_{i,j} \omega_j(p) \cr
\oint_{q\in\bcycle_i} B_\kappa(p,q) &=& 2\pi\ii \left( \omega_i(p) +  \sum_{j,l}  \tau_{i,j}\kappa_{j,l} \omega_l(p) \right) .
\eea

In particular we have special choices of $\kappa$ as follows:

\bd[Klein kernel]
Let $\zeta\in \CC^\genus$, and define the \textbf{Klein kernel}\index{Klein kernel}
\beq
B_\zeta(p,q) = B_{\frac{1}{\pi\ii}\log\Theta''(\zeta)}(p,q) = B(p,q) + 2 \sum_{i,j} 
(\log\Theta)''(\zeta)_{i,j} \ \omega_i(p) \omega_j(q)
\eeq
where
\beq
(\log\Theta)''(\zeta)_{i,j} = \left(\frac{\partial^2}{\partial u_i \ \partial u_j} \ \log\Theta(u) \right)_{u=\zeta}.
\eeq

\ed

\bd[Schiffer kernel]
\index{Schiffer kernel}
We have already defined the Schiffer kernel, it corresponds to the choice $\kappa=\frac\ii 2 \ \Im \tau^{-1}$:
\beq
B_S(p,q) = B_{\frac\ii 2 \ \Im \tau^{-1}}(p,q) 
\eeq

\ed

\subsection{Modular transformations}

A choice of symplectic basis of cycles (marking) is not unique, and some notions, like the normalized basis of holomorphic forms, the Riemann matrix of periods, and the fundamental second kind differential depend on the marking.
The group of changes of symplectic basis, that keeps the intersection matrix constant, is the \textbf{symplectic group}\index{symplectic group} $Sp_{2\genus}(\ZZ)$, of $2\genus\times 2\genus$ matrices (written as 4 block matrices of size $\genus\times \genus$)
\beq
\begin{pmatrix}
\alpha & \beta \cr
\gamma & \delta
\end{pmatrix} \
\eeq
satisfying
\beq
\begin{pmatrix}
\alpha & \beta \cr
\gamma & \delta
\end{pmatrix} \
\begin{pmatrix}
0 & \text{Id}_{\genus} \cr -\text{Id}_{\genus} & 0
\end{pmatrix} \
\begin{pmatrix}
\alpha & \beta \cr
\gamma & \delta
\end{pmatrix}^T
= 
\begin{pmatrix}
0 & \text{Id}_{\genus} \cr -\text{Id}_{\genus} & 0
\end{pmatrix}.
\eeq

Now consider 2 markings, related by a $Sp_{2\genus}(\ZZ)$ symplectic transformation:
\beq
\begin{pmatrix}
\acycle \cr
\bcycle 
\end{pmatrix}
= \begin{pmatrix}
\alpha & \beta \cr
\gamma & \delta
\end{pmatrix}
\
\begin{pmatrix}
\td\acycle \cr
\td\bcycle 
\end{pmatrix}.
\eeq

\bt[Modular transformations]
\index{modular transformation}
Under a symplectic transformation, the Riemann matrix of periods changes as
\beq
\td \tau = (\delta-\tau\beta)^{-1} ( \tau\alpha-\gamma),
\eeq
and the normalized basis of holomorphic 1-forms as
\beq
\td \omega_i = \sum_j J_{i,j} \omega_j
\qquad, \qquad
J= (\delta-\tau\beta)^{-1} = \alpha^T + \td\tau \beta^T.
\eeq

The Klein kernel and the Schiffer kernel are $Sp_{2\genus}(\ZZ)$ modular invariant.

The kernel $B_\kappa$ of def. \ref{defBergman} changes as
\beq
B_\kappa = \td B_{\td\kappa}
\quad , \quad
\td\kappa = (\alpha+\td\tau\beta)^{-1}\kappa (\delta-\tau\beta) -
(\alpha+\td\tau\beta)^{-1} \beta.
\eeq
In particular for $\td\kappa=0$:
\beq
\td B_0 = B_{ \beta(\delta-\tau\beta)^{-1}},
\eeq
i.e.
\beq
\td B_0(p,q) = B_0(p,q) + 2\pi\ii \sum_{i,j=1}^\genus (\beta(\delta-\tau\beta)^{-1})_{i,j} \ \omega_i(p) \ \omega_j(q).
\eeq

\et

Remark that if $\genus=1$, the matrices $\alpha,\beta,\gamma,\delta$ are scalar elements of $\ZZ$, and we have
\beq
\td\tau = \frac{\alpha\tau-\gamma}{\delta-\beta\tau}.
\eeq

\subsection{Meromorphic forms and generalized cycles}

\bt[All meromorphic forms are generated by integrating $B$]\label{BtoM1}
A basis of $\mathfrak M^1(\curve)$ can be constructed by integrating $B$ as follows:
\begin{itemize}
\item 1st kind forms
\beq
\omega_i(p) = \frac{1}{2\pi\ii} \oint_{q\in\bcycle_i} B(p,q)
\eeq
It is the unique holomorphic 1-form normalized on $\acycle$--cycles as
\bea
\oint_{\acycle_j} \omega_{i}=\delta_{i,j} .
\eea
We call $\acycle_i$ and $\bcycle_i$ \textbf{1st kind cycles}\index{1st kind cycle}.

\item 3rd kind forms
\beq\label{eqdefB3rd}
\omega_{q_1,q_2}(p) = \int_{q=q_2}^{q_1} B(p,q)
\eeq
where the integration chain from $q_2$ to $q_1$ is the unique one that doesn't intersect any $\acycle_i$-cycles nor $\bcycle_i$-cycles.
It is the unique 1-form normalized on $\acycle$--cycles that has a simple pole with residue $1$ at $q_1$ and a simple pole with residue $-1$ at $q_2$ and no other poles, such that
\bea
\Res_{q_1} \omega_{q_1,q_2}=1 \cr
\cr
\Res_{q_2} \omega_{q_1,q_2}=-1 \cr
\cr
\oint_{\acycle_i} \omega_{q_1,q_2}=0 .
\eea
We call a chain $q_2\to q_1$ a  \textbf{3rd kind cycle}\index{3rd kind cycle} (we shall call it a generalized cycle below).

\item 2nd kind forms. 
For $q\in\curve$, choose $\phi_U$ a local coordinate in a neighborhood of $q$, and let $f$ a function, meromorphic in the neighborhood $U$ of $q$, and 
let $d=\operatorname{order}_q f-f(q)$.

For $k\in \ZZ_+$, if $d>0$, define
\beq\label{eqdefBqk}
\omega_{f,q,k}(p) = \frac{1}{2\pi\ii} \ \frac{d}{k}\oint_{q'\in\mathcal C_q} (f(q')-f(q))^{\frac{-k}{d}} \ B(p,q').
\eeq
It is the unique 1-form normalized on $\acycle$--cycles that has a pole of degree $k+1$ at $q$, that behaves as
\beq
\omega_{f,q,k}(p) \sim \frac{df(p)}{(f(p)-f(q))^{1+\frac{k}{d}}} + \text{analytic at} \ q \qquad , \qquad
\oint_{\acycle_i} \omega_{f,q,k}=0 .
\eeq

If $d<0$, i.e. $q$ is a pole of $f$, of degree $-d$ we define 
\beq\label{eqdefBqkpole}
\omega_{f,q,k}(p) = \frac{1}{2\pi\ii} \ \frac{d}{k}\oint_{q'\in\mathcal C_q} f(q')^{\frac{-k}{d}} \ B(p,q').
\eeq
It is the unique 1-form normalized on $\acycle$--cycles that has a pole of degree $k+1$ at $q$, that behaves as
\beq
\omega_{f,q,k}(p) \sim \frac{df(p)}{(f(p))^{1+\frac{k}{d}}} + \text{analytic at} \ q 
\qquad , \qquad
\oint_{\acycle_i} \omega_{f,q,k}=0 .
\eeq

We call a pair $(\mathcal C_q.f)$, that we denote $\mathcal C_q.f$,  made of a small cycle $\mathcal C_q$ around $q$, together with a function $f$ meromorphic in a neighborhood of $\mathcal C_q$ with poles only at $q$, a  \textbf{2nd kind cycle}\index{2nd kind cycle} (we shall call it a generalized cycle below).

\end{itemize}

\et

In this theorem, we have implicitely defined a notion of "\textbf{generalized cycles}". 
$B$ allows to define a form--cycle duality as follows:

through integration (Poincar\'e pairing), cycles can be viewed as acting linearly on the space of forms, and are thus elements of the dual of the space of 1-forms:
\beq
<\gamma,\omega> = \oint_\gamma \omega \in \CC.
\eeq
In other words
\beq
H_1(\curve) \subset \mathfrak M^1(\curve)^*.
\eeq
However, the dual $\mathfrak M^1(\curve)^*$ is much larger than $H_1(\curve)$, it may contain for instance integrals on a cycle, together with multiplication by a function (as in \eqref{eqdefBqk}) or a distribution, or integrals on open chains with boundaries as in \eqref{eqdefB3rd}, and also 2-dimensional integrals, and many other things.
Since $B$ is a $1\boxtimes 1$ form, acting on its second variable by an element of the dual, yields a 1-form of the first variable, i.e. a 1-form.
However, if we chose an arbitrary element of $\mathfrak M^1(\curve)^*$, this 1-form is not necessarily meromorphic.
Let us define:

\bd[Generalized cycles]
We define the space of generalized cycles\index{generalized cycle} $\mathfrak M_1(\curve)$, as the subspace of $\mathfrak M^1(\curve)^*$, whose integral of $B$ is a meromorphic 1-form.
$B$ defines a map $\hat B$ from generalized cycles to meromorphic 1-forms:
\bea
\hat B: \mathfrak M_1(\curve) & \to & \mathfrak M^1(\curve) \cr
\gamma & \mapsto & \hat B(\gamma)=<\gamma,B> \ , \ \hat B(\gamma)(p) = \oint_{q\in \gamma} B(p,q)
\eea
From theorem \ref{BtoM1}, this map is surjective.
It is not injective.

\ed

\bd
The intersection of generalized cycles is the symplectic form on $\mathfrak M_1(\curve)\times \mathfrak M_1(\curve)$ defined by
\beq
\gamma_1 \cap \gamma_2 = - \gamma_2\cap \gamma_1 = \frac{1}{2\pi\ii} \left( \oint_{\gamma_1}\oint_{\gamma_2} B -  \oint_{\gamma_2}\oint_{\gamma_1} B \right)
\eeq
One can check that on $H_1(\curve)\times H_1(\curve)$ this is indeed the usual intersection.

\ed

Notice that $\Ker \hat B$ is obviously a Lagrangian submanifold.

\subsubsection{To go further}

To learn more about generalized cycles, their sumplectic structure, their Lagrangian submanifolds, and Hodge structures, see for instance \cite{EynISlectures}.

\chapter{Riemann-Roch}

The question of interest is: given a set of points $p_i$ and orders $\alpha_i\in\ZZ $ (i.e. an integer divisor $D=\sum_i \alpha_i.p_i$), is it possible to find a meromorhic function (resp. 1-form) with poles at $p_i$ of degree at most $-\alpha_i$ if $\alpha_i<0$, and zeros of order at least $\alpha_i$ if $\alpha_i>0$ ?
And how many linearly independent such functions (resp. 1-forms) is there ? 

\section{Spaces and dimensions}

\bd[Positive divisors]
A divisor  $D=\sum_i \alpha_i.p_i$ is said positive\index{positive divisor} 
\beq
\sum_i \alpha_i.p_i \geq 0 \qquad \text{iff} \, \ \forall\ i \ , \ \alpha_i\geq 0.
\eeq
\beq
D>0  \qquad \text{iff} \, \ D\geq 0 \ \text{and} \ D\neq 0 .
\eeq
This induces a (partial) order relation in the set of divisors:
\beq
D_1\geq D_2 \qquad \text{iff} \,  \  \ D_1-D_2\geq 0.
\eeq
\beq
D_1> D_2 \qquad \text{iff} \,  \  \ D_1-D_2> 0.
\eeq
\ed

\bd
Given a divisor $D$, we define the vector spaces (over $\mathbb C$)
\beq
\mathcal L(D) = \{ f\in \mathfrak M^0(\curve) \ | \ (f)\geq D\}
\qquad , \quad
r(D) = \dim \mathcal L(D),
\eeq
\beq
\Omega(D) = \{ \omega\in \mathfrak M^1(\curve) \ | \ (\omega)\geq D\} .
\qquad , \quad
i(D) = \dim \Omega (D).
\eeq
\ed

\bt
$r(D)$ and $i(D)$ depend only on the divisor class (modulo principal divisors):
\beq
r(D)=r([D])
\qquad , \qquad
i(D)=i([D]).
\eeq
\et
\proof
If $D_2-D_1=(f)$ is a principal divisor, then the maps
\bea
\Omega(D_1) &\to & \Omega(D_2) \cr
\omega_1 & \mapsto & f\omega_1
\eea
\bea
\mathcal L(D_1) &\to & \mathcal L(D_2) \cr
f_1 & \mapsto & f f_1
\eea
are isomorphisms (they are obviously invertible: the inverse is dividing by $f$), therefore $i(D_1)=i(D_2)=i([D_1])$ and $r(D_1)=r(D_2)=r([D_1])$.
\eproof

The Riemann-Roch theorem\index{Riemann-Roch}, that we shall prove in this chapter, is:
\bt[Riemann-Roch]
For every divisor $D$ we have
\beq
r(-D) = \deg D +1-\genus+i(D).
\eeq
\et

This theorem is extremely useful and powerful. One reason is that most often one of the two indices $r(-D)$ or $i(D)$ is easy to compute and the other is more difficult, so the theorem gives it without effort.

Also, using positivity $i(D)\geq 0$ (resp. $r(-D)\geq 0$) we easily get a lower bound for $r(-D)$ (resp. $i(D)$), and thus the Riemann-Roch theorem is often used to prove the existence of functions or forms with given poles and degrees.

\section{Special cases}\label{sec:RRspecial}

Let us first prove it in special cases:

$\bullet$ If $D=0$, then from theorem II-\ref{thfholconst} we have $\mathcal L(0)=\mathcal O(\curve)=\CC$ and $r(0)=1$.
On the other hand, $\Omega(0)=\mathcal O^1(\curve)$ and $i(0)=\genus$.
The Riemann-Roch theorem is satisfied.

$\bullet$ If $-D> 0$, then $\mathcal L(-D)=\{0\}$ and $r(-D)=0$, and $\deg D<0$.
We have (using the basis of theorem III-\ref{BtoM1})
\beq
\Omega(D) = \{\sum_i \sum_{k=2}^{-\alpha_i} t_{p_i,k} \omega_{p_i,k} + \sum_i t_{p_i,1} \omega_{p_i,o} + \sum_{i=1}^\genus \epsilon_i \omega_i \ | \ \sum_i t_{p_i,1}=0 \ \}
\eeq
Therefore
\beq
i(D) = \sum_i (-\alpha_i) +\genus-1=-\deg D+\genus-1.
\eeq
The Riemann-Roch theorem is satisfied.

$\bullet$ $\genus\geq 2$ and $D> 0$ and $\deg D=\genus-1$.
$\Omega(D)$ is then the set of holomorphic forms $\omega$
with zeros at the $\genus-1$ point of $D$. 
$\zeta=K-u(D)$ is a zero of $\Theta$, and by parity, $-\zeta=u(D)-K$ is also a zero of $\Theta$, therefore there is a unique positive divisor $\td D> 0$ of degree $\deg \td D=\genus-1$ with Abel map $\mathbf u(\td D) =K+\zeta  = 2K-\mathbf u(D)$.
From theorem III-\ref{thmoneformdivisor}, there exists a unique (up to scalar multiplication) holomorphic one form $\omega$ whose divisor of zeros is $D+\td D$, which implies that $i(D)=1$.
It also implies that $\Omega(D)=\Omega(D+\td D) = \Omega(\td D)$.

For all $f\in \mathcal L(-D)$, the 1-form $f\omega$ has no pole, and its divisor of zeros is $\geq \td D$, so it belongs to $\Omega(\td D)$.
Reciprocally, if $\td\omega\in \Omega(\td D)= \Omega(D+\td D)$, then $\td \omega$ must be proportional to $\omega$, and
$f=\td \omega/\omega$ is a constant, in particular it belongs to $\mathcal L(-D)$.
Therefore the map 
\bea
\mathcal L(-D) & \to & \Omega(\td D) \cr
f & \mapsto & f \omega \cr
\frac{\td \omega}{\omega} & \leftarrow & \td \omega
\eea
is an isomorphism.
This implies $r(-D)=i(\td D)=1$.
The Riemann-Roch theorem is satisfied.

\section{Genus 0}

On the Riemann sphere we have
\beq 
\mathcal L(-D) = \left\{ C \frac{P(z)}{\prod_{i} (z-p_i)^{\alpha_i}}  \ | \ P\in \CC[z] \quad \deg P\leq \deg D \right\}
\eeq
\beq
\implies \qquad
r(-D) = \max(0,1+\deg D)
\eeq

\beq 
\Omega(D) = \left\{ C \frac{P(z)}{\prod_{i} (z-p_i)^{-\alpha_i}}  \ dz \ | \ P\in \CC[z] \quad  \deg P\leq -2-\deg D \right\}
\eeq
\beq
\implies \qquad
i(D) = \max(0,-1-\deg D)
\eeq
We easily see that the Riemann-Roch theorem is satisfied for all divisors.

\section{Genus 1}

On a genus 1 curve, there is a unique --up to a scalar factor-- holomorphic 1-form $\omega_0$ without poles nor zeros.
Choosing a symplectic basis of cycles $\acycle\cap \bcycle=1$, we choose $\omega_0$ normalized on $\acycle$, so that $\oint_{\acycle}\omega_0=1$, and we define $\tau=\oint_{\bcycle}\omega_0$ and we define the Abel map $p\mapsto u(p)$ such that $du=\omega_0$.

The map $f\mapsto \omega=f \omega_0$ is an isomorphism 
\beq
\mathcal L(D) \to \Omega(D),
\eeq
and thus
\beq
r(D)=i(D).
\eeq

If $f\in \mathcal L(D)$, we have $(f)=D+\td D$ with $\td D\geq 0$,  $\deg \td D=-\deg D$ and $\mathbf u(\td D)=-\mathbf u(D)$.
Let us denote
\beq
W(D) = \{ \td D \ | \ \td D\geq 0  \ \text{and} \ \deg \td D=-\deg D \ \text{and} \ \mathbf u(\td D)=-\mathbf u(D)\} .
\eeq
There is an isomorphism $W(D)\oplus \CC \to \mathcal L(D) $, given by
\beq
(\td D,C) \mapsto f(p) = C \ \prod_{i} \Theta(u(p)-u(p_i)+\frac{1+\tau}{2})^{\alpha_i} \ \prod_{i} \Theta(u(p)-u(p'_i)+\frac{1+\tau}{2})^{\alpha'_i},
\eeq
and thus
\beq
r(D) = \dim W(D) = \max(0,-\deg D).
\eeq
We have
\beq
r(-D)-r(D)=\deg D
\eeq
so that the Riemann-Roch theorem is satisfied.

\bt
Every genus 1 Riemann surface is isomorphic to a torus $\CC/\mathbb Z+\tau\mathbb Z$.
\et

\proof
The map $p\mapsto u(p)$ is an isomorphism $\curve \to \mathbb J(\curve) = \CC/\mathbb Z+\tau\mathbb Z$. 
Indeed it is injective, of maximal rank, and continuous, so it must be surjective.
\eproof

\section{Higher genus $\geq 2$}

Let us first make some observations.

\subsection{General remarks}

\bt
\beq
i([D])=r([D]-\mathfrak K).
\eeq
\et
\proof
Let $\omega$ an arbitrary meromorphic 1-form.
The map
\bea
\Omega(D) & \to & \mathcal L(D-(\omega))  \cr
\omega' & \mapsto & \omega'/\omega 
\eea
is an isomorphism.
\eproof

Write $D=D_+-D_-$, where $D_+\geq 0$ and $D_-\geq 0$.
We clearly have
\beq
\Omega(D_+) \subset \Omega(D) \subset \Omega(-D_-)
\qquad \implies \qquad
0\leq i(D_+) \leq i(D) \leq i(-D_-) .
\eeq
\beq
\mathcal L(-D_-) \subset \mathcal L(-D) \subset \mathcal L(-D_+)
\qquad \implies \qquad
0\leq r(-D_-) \leq r(-D) \leq r(-D_+) .
\eeq

\subsection{Proof of the Riemann--Roch theorem}

\subsubsection{Negative divisors}

\bt
Let $D$ a divisor.
If there exists a negative divisor $\td D\leq 0$ whose divisor class is $[\td D]=[D]$ or if $[\td D]=\mathfrak K-[D]$, then the Riemann--Roch theorem holds
\et

\proof
The case $\td D=0$, and $\td D<0$ were already done in section \ref{sec:RRspecial}.

\eproof

\subsubsection{Other cases}

The following proposition is very useful:

%
%
%
%

\bp[Riemann inequality]
Let $D> 0$, we have
\beq
r(-D)\geq \deg D-\genus+1.
\eeq
\ep
\proof
Writing $D=\sum_{i=1}^k \alpha_i p_i$, define $D' = \sum_i (\alpha_i+1) p_i$.

The map
\bea
\mathcal L(-D) & \to & \Omega(-D') \cr
f & \mapsto & df
\eea
is a morphism whose kernel consists of constant functions, i.e. whose kernel has dimension 1.
Its image consists of exact forms, those for which all cycle integrals and residues vanish.
It is a space of codimension at most $2\genus+k-1$ (because there are at most $2\genus+k$ independent non contractible cycles, and since the sum of all residues is $0$, at most $k-1$ of them are actually independent), therefore
\beq
r(-D)-1 \geq i(-D')-2\genus+1-k.
\eeq
Moreover, we have already proved Riemann--Roch theorem for negative divisors, and we have $r(D')=0$ and
\beq 
i(-D') = \deg D' +\genus-1= \deg D+k+\genus-1.
\eeq
Therefore:
\beq
r(-D)-1 \geq \deg D-\genus.
\eeq

\eproof

\bp\label{propRR1}
If $r(-D)>0$, then there exists $\td D\geq 0$ such that $[\td D]=[D]$.
\ep
\proof
Choose $0\neq f\in \mathcal L(-D)$, then $\td D=(f)+D\geq 0$ and $[\td D]=[D]$.
\eproof

\bp\label{propRR2}
If $i(D)>0$, then there exists $\td D\geq 0$ such that $[\td D]=\mathfrak K-[D]$.
\ep
\proof
Use $i([D])=r(\mathfrak K-[D])$.
\eproof

\bt
If $D$ is a divisor such that there exists no positive divisor $\td D\geq 0$ such that $[D]=[\td D]$ nor $[\td D]=\mathfrak K-[D]$, then $i(D)=r(-D)=0$ and
\beq
\deg D=\genus-1.
\eeq
So that the Riemann-Roch theorem holds in this case too.
\et
\proof
Write $D=D_+-D_-$ with $D_+> 0$ and $D_-\geq 0$, with $D_+$ and $D_-$ having no points in common.
We have from the Riemann inequality
\beq
r(-D_+)\geq \deg D_+ -\genus+1 = \deg D+\deg D_- - \genus+1
\eeq
Assume that $\deg D\geq \genus$, this implies that
\beq
r(-D_+)\geq \deg D_- +1.
\eeq
The subspace of  $\{ f \ | \ (f)\geq D_--D_+\} \subset \mathcal L(-D_+)$, is of codimension $\deg D_-$, and thus it is non--vanishing, showing that there exists some $f\neq 0$ such that $(f)+D\geq 0$. This contradicts our hypothesis that $D$ is not equivalent to a positive divisor.
Therefore we must have $\deg D\leq \genus-1$.
By the same reasoning we have
\beq
\genus-1 \geq \deg (\mathfrak K-[D]) =2\genus-2-\deg D
\eeq
which implies that $\genus-1\leq \deg D\leq \genus-1$ and thus $\deg D=\genus-1$.
The fact that $i(D)=0$ and $r(-D)=0$ follow from prop \ref{propRR1} and prop \ref{propRR2}.
The riemann--Roch theorem thus holds.
\eproof

\chapter{Moduli spaces}

Throughough this section we shall denote
\beq\label{defdgnchign}
d_{\genus,n} = 3\genus-3+n
\qquad , \qquad
\chi_{\genus,n}=2-2\genus-n.
\eeq

\bd
Let $(\curve,p_1,\dots,p_n)$ and $(\curve',p'_1,\dots,p'_n)$ be two compact Riemann surfaces of genus $\genus$, with $n$ distinct and labeled marked points $p_i\in \curve$, $p'_i\in\curve'$.
They are called isomorphic iff there exists a holomorphic map $\phi:\curve \to \curve'$, invertible and whose inverse is holomorphic, such that $\phi(p_i)=p'_i$ for all $i=1,\dots,n$.

Automorphisms of $(\curve,p_1,\dots,p_n)$ form a group. We say that $(\curve,p_1,\dots,p_n)$ is stable iff its automorphism group is a finite group.

We define the \textbf{moduli space}\index{moduli space}
\beq
\mathcal M_{\genus,n} = \{ (\curve,p_1,\dots,p_n) \ \text{of genus} \ \genus \}/\text{isomorphisms} .
\eeq

\ed

\section{Genus 0}

From corollary II-\ref{thg0RS}, every Riemann surface of genus 0 is isomorphic to the Riemann sphere.
Automorphisms of the Riemann sphere are M\"obius\index{M\"obius transformation} transformations $z\mapsto \frac{az+b}{cz+d}$ with $ad-bc=1$.

\bt
\beq
\mathcal M_{0,0} = \{(\CC P^1)\}
\qquad , \quad \text{Aut}=PSL(2,\CC).
\eeq
\beq
\mathcal M_{0,1} = \{(\CC P^1,\infty) \}
\qquad , \quad \text{Aut}= \{z\mapsto az+b\} \sim \CC^*\times \CC.
\eeq
\beq
\mathcal M_{0,2} = \{(\CC P^1,0,\infty) \}
\qquad , \quad \text{Aut}= \{z\mapsto az\} \sim \CC^* .
\eeq
\beq
\mathcal M_{0,3} = \{(\CC P^1,0,1,\infty) \}
\qquad , \quad \text{Aut}= \{\text{Id}\} .
\eeq
And if $n\geq 4$
\beq
\mathcal M_{0,n} = \{(\CC P^1,0,1,\infty,p_4,\dots,p_n) \ | \ p_4,\dots,p_n \ \text{distinct and } \ \neq 0,1,\infty \}
\quad , \quad \text{Aut}= \{\text{Id}\} ,
\eeq
\beq
\mathcal M_{0,n} \sim (\CC P^1\setminus \{0,1,\infty\})^{n-3} \setminus \{ \text{coinciding points} \} .
\eeq

$\mathcal M_{0,n}$ is a smooth complex connected manifold  of dimension 
\beq
\dim \mathcal M_{0,n} = \max(0,n-3).
\eeq
$\mathcal M_{0,n}$ is stable iff $n\geq 3$.
If $n\geq 4$, it is not compact.

\et

\subsection{$\mathcal M_{0,3}$}

$\mathcal M_{0,3}$ is a single point, with trivial automorphism, this is the simplest possible manifold: the point.
\beq
\mathcal M_{0,3} = \{(\CC P^1,0,1,\infty) \}
\qquad , \quad \dim \mathcal M_{0,3} = d_{0,3}=0
\qquad , \quad \text{Aut}= \{\text{Id}\} .
\eeq
There is a unique topology on it, and it is compact.

Its Euler characteristic is
\beq
\chi(\mathcal M_{0,3}) = 1.
\eeq

\subsection{$\mathcal M_{0,4}$}

$\mathcal M_{0,4}$ is of dimension 1, it is a sphere with 3 points removed:
\begin{multline}
\mathcal M_{0,4} = \{(\CC P^1,0,1,\infty,p) \ | \ p\neq 0,1,\infty \} = \CC P^1-\{0,1,\infty\}
\qquad , \quad \dim \mathcal M_{0,4} = d_{0,4}=1 \cr
\text{Aut}= \{\text{Id}\} .
\end{multline}
We put on it the induced topology from $\CC P^1$. It is not compact.
We also put on it the complex structure inherited from $\CC P^1$, it is thus a non compact Riemann surface.

Topologically it is $\text{sphere--less--3--points}$, its Euler characteristic is
\beq
\chi(\mathcal M_{0,4}) = -1.
\eeq

\subsubsection{Boundary}

The boundary is reached when we consider a sequence in $\mathcal M_{0,4}$, or equivalently a sequence of points $p$ in $\CC P^1-\{0,1,\infty\}$, which has no adherence value (and thus no limit). 
This means a sequence that tends to $0$ or $1$ or $\infty$.
In other words, a boundary corresponds to 2 of the marked points colliding.

Consider the limit $p_4=p\to 0=p_1$, and $p_2=1, p_3=\infty$.
The chart $\{ z \ | \ |z|> 2|p|\}$ is a neighborhood that contains $p_2$ and $p_3$ but not $p_1$ neither $p_4$. In the limit $p\to 0$, this neighborhood becomes  $\CC P^1-\{0\}$. In this chart, a whole basis of neighborhoods of $p_1,p_4$ becomes contracted to the point $\{0\}$.
By using a M\"obius transformation and the coordinate $z'=z/p$, the chart $\{ z' \ | \ |z'|<2\}$  is a neighborhood that contains $p_1$ and $p_4$ but not $p_2$ neither $p_3$. In the limit $p\to 0$, this neighborhood becomes  $\CC P^1-\{\infty\}$, and a whole basis of neighborhoods of $p_2,p_3$ becomes contracted to the point $\{\infty\}$.

In other words, in the limit $p\to 0$, we have 2 charts, entirely disconnected, that touch each other only by one point.

This can be described by a notion of \textbf{Nodal surface}\index{nodal surface}.

\bd[Nodal  Riemann surface]
A nodal Riemann surface $\curve$, is a finite union of compact surfaces $\curve_i$, together with a set of disjoint nodal points. 
A nodal point is a pair of distinct points on the union.
The nodal surface is
\beq
\curve = \cup_i \curve_i / \equiv
\eeq
with the quotient by the equivalence relation $p \equiv q $ iff $p=q$ or if $(p,q)$ is a nodal point.
The topology of $\curve$ is made of neighborhoods of non--nodal points in the $\curve_i$s and a neighborhood of a nodal point is the union of 2 neighborhoods of each of the 2 points.
With this topology, $\cup_i \curve_i$ (before taking the quotient) is not a separated space, and the quotient $\curve$ is separated but is not a manifold because there is no neighborhood of nodal points homeomorphic to Euclidian discs.

Connectivity is well defined,  nodal surfaces can be connected or not, also Jordan arcs and their homotopy classes are well defined, and a nodal surface can be simply connected or not.

The Euler characteristic is:
\beq
\chi(\curve) = \sum_i \chi(\curve_i-\{\text{nodal points}\}).
\eeq
$$
\includegraphics[scale=0.45]{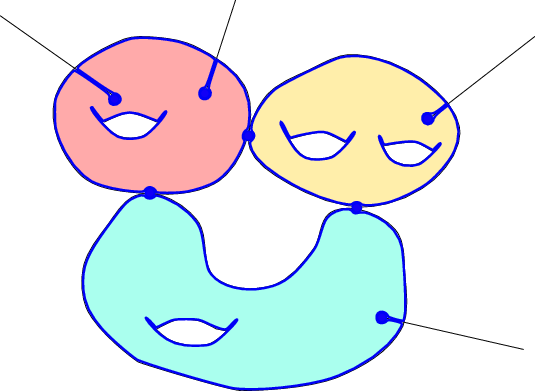}
$$
\ed

We see that the limit $p\to 0$ in $\mathcal M_{0,4}$ is described by a nodal surface, with 2 components, and one nodal point:
$$
\includegraphics[scale=0.75]{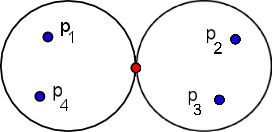}
$$
\begin{itemize}
\item The first component is a Riemann sphere containing $p_2=1$ and $p_3=\infty$ and one side of the nodal point (the point $z=0$). 
It is thus $(\CC P^1 ,1,\infty,0)$, which is an element of $\mathcal M_{0,3}$.

\item The second component is a Riemann sphere containing $p_1=0$ and $p_4=1$ (in the coordinate $z'=z/p$) and one side of the nodal point (the point $z'=\infty$).
It is thus $(\CC P^1 ,0,1,\infty)$, which is an element of $\mathcal M_{0,3}$.
\end{itemize}
Notice that the Euler characteristic of a nodal surface with 2 sphere components having each 3 points removed is:
\bea
\chi(\curve_1\cup\curve_2-\{p_1,p_2,p_3,p_4,\text{nodal points}\})
&=& \chi(\curve_1-\{p_2,p_3,\text{nodal point}\}) \cr
&& + \chi(\curve_2-\{p_1,p_4,\text{nodal point}\}) \cr
&=& -1 -1 =-2,
\eea
and agrees with the Euler characteristic $2-2\genus-n=-2$ of a surface of genus $\genus=0$ with $n=4$ points removed.

Eventually this boundary of $\mathcal M_{0,4}$ can be viewed as an element of 
$\mathcal M_{0,3}\times \mathcal M_{0,3} $.

Notice that the boundary $p_4\to p_1$, is by a M\"obius transform, the same as the boundary $p_2\to p_3$.

There are $3$ possibilities to choose a pair of colliding points among 4.
Therefore there are 3 boundaries, and we have
\beq
\partial \mathcal M_{0,4} \sim 
   \mathcal M_{0,3} \times \mathcal M_{0,3}  \ \cup  \  \mathcal M_{0,3} \times \mathcal M_{0,3} \ \cup \   \mathcal M_{0,3} \times \mathcal M_{0,3} \ ,
\eeq
Each of these boundaries is a point, and $\mathcal M_{0,4}$ is a sphere with 3 points missing.
We can compactify $\mathcal M_{0,4}$ by adding its boundary, and then we get the full sphere:
\beq
\overline{\mathcal M_{0,4}}
= \mathcal M_{0,4} \cup \partial \mathcal M_{0,4} \sim \CC P^1.
\eeq
We equip it with the topology and complex structure of the Riemann sphere $\CC P^1$.
It is then a manifold, and in fact a complex manifold, a compact Riemann surface.
We have
\beq
\chi(\overline{\mathcal M_{0,4}}) = 2.
\eeq

\subsubsection{Deligne--Mumford compactification}

The general case works similarly. 
Moduli spaces $\mathcal M_{g,n}$  are not compact, they have boundaries whenever some marked points collapse, or whenever some cycles get pinched.
The limit can always be described by nodal surfaces.

We thus define:
\bd[Deligne Mumford compactified moduli space]
Let $(\genus,n)$ such that $2-2\genus-n<0$.

The Deligne--Mumford\index{Deligne Mumford} compactified moduli space $\overline{\mathcal M_{\genus,n}}$ is defined as the set (modulo isomorphisms) of connected nodal Riemann surfaces with $n$ smooth labelled marked points (smooth means they are distinct from nodal points and from each other), and \textbf{stable}\index{stable} (the Euler characteristic of each component with all marked and nodal points removed is $<0$), and of total Euler characteristic $\chi=2-2\genus-n$:
\beq
\overline{\mathcal M_{\genus,n}}
= \{ (\curve,p_1,\dots,p_n) \ | \ \chi=2-2\genus-n=\sum_i \chi_i \ , \ \forall \ i \ \chi_i <0 \}/\text{isomorphisms} .
\eeq
Isomorphisms are the holomorphic maps whose inverse is analytic, that conserve labeled points and that conserve (up to possible permutations) the nodal points.

\ed

We shall not describe here the topology of this moduli space, but just mention that with the appropriate topology it is indeed compact.
Also it can be equipped with a differentiable structure, and a complex structure, that we do not describe here. It shall be explained in section \ref{secStrebel} below, by providing an explicit atlas of charts and coordinates.

However, it is not a manifold (some neighborhoods are not homeomorphic to Euclidian neighborhoods), it is an orbifold (a manifold quotiented by a group: each neighborhood is homeomorphic to a Euclidian neighborhood quotiented by a group), and it is a \textbf{stack}\index{stack}.
As we shall see now it has a rather non--trivial topology, in particular, it can have pieces of different dimensions.

Let us first study some simple examples.

\subsubsection{$\bullet \ (\genus,n)=(0,5)$}

We have
\beq
\mathcal M_{0,5} = \{ (\CC P^1, 0,1,\infty, p ,q) \ | p\neq 0,1,\infty, \ q\neq 0,1,\infty , \ p\neq q \}.
\eeq
It has no non--trivial automorphisms, because there is a unique M\"obius map that fixes $0,1,\infty$.

Topologically, this is:
\beq
\mathcal M_{0,5} = (\text{sphere--less--3--points})\times (\text{sphere--less--3--points}) - (\text{sphere--less--3--points})
\eeq
where the last sphere--less--3--points is the diagonal of the product.

We have
\beq
\dim \mathcal M_{0,5} =2 .
\eeq
\beq
\chi(\mathcal M_{0,5})=(-1)\times (-1) - (-1) = 2.
\eeq

Naively, one could think that to get a smooth differentiable (and complex) compact manifold, one could add to it the missing pieces to complete the product into 2 full spheres.
We would need to add 7 missing sphere--less--3--points, and 9 missing points.

This is wrong, let us study the boundary.
There are boundaries of codimension 1 when exactly 2 points collide, and boundaries of codimension 2, when 2 pairs collide.

The number of boundaries of codimension 1 is the numbers of pairs of 2 points chosen among 5, i.e.
\beq
\begin{pmatrix}
5 \cr 2
\end{pmatrix} = 10.
\eeq
Each such boundary is described by a nodal surface with 1 nodal point and 2 components, one component carrying the 2 colliding points + nodal point, and one component carrying the other 3 points and the nodal point, each component is a sphere.
Therefore
\beq
\partial_{\text{codim 1}} \ \mathcal M_{0,5} = 10 \times \ (\mathcal M_{0,3} \times \mathcal M_{0,4} ).
\eeq
Topologically each boundary of codim 1, is a sphere--less--3--points ($\mathcal M_{0,4}$ times a point).
$$
\includegraphics[scale=0.75]{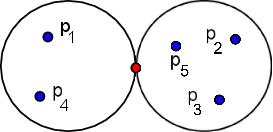}
$$

Similarly  boundaries of codimension 2 are obtained by choosing 2 pairs of colliding points among 5 points, i.e. choose the non colliding point ($5$ choices), then split the 4 remainings into 2 pairs ($3$ choices),
\beq 
5\times 3 =  15.
\eeq
Each such boundary is described by a nodal surface with 2 nodal points and 3 components, two components carrying the 2 pairs of colliding points + 1 nodal point, and one component carrying the  5th marked point and 2 nodal points, each component is a sphere.
Therefore
\beq
\partial_{\text{codim 2}} \ \mathcal M_{0,5} = 15 \times \ \mathcal M_{0,3} \times \mathcal M_{0,3}\times \mathcal M_{0,3}.
\eeq
Topologically each boundary of codim 2, is a point.
$$
\includegraphics[scale=0.55]{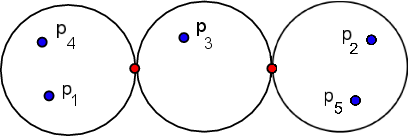}
$$
$$
\includegraphics[scale=0.55]{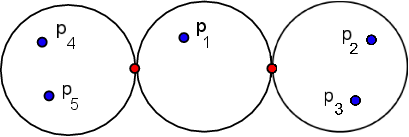}
$$

Finally we have
\beq
\overline{\mathcal M_{0,5}} = 
\mathcal M_{0,5} \ \cup \ 10 \times \ \mathcal M_{0,3} \times \mathcal M_{0,4}
\ \cup \ 15 \times \ \mathcal M_{0,3} \times \mathcal M_{0,3}\times \mathcal M_{0,3}.
\eeq
Topologically, it is a product of 2 spheres, with 7 sphere--less--3--points removed, and 9 points removed, and with 10 sphere--less--3--points added and 15 points added.
It is definitely not a smooth manifold.

Some 7 among 10 of the sphere--less--3--points should be glued to the missing ones, namely the colliding $p\to 0$ should be glued to the missing $p=0$ in the product, and the same for $p\to 1$, $p\to \infty$, $q\to 0$, $q\to 1$, $q\to\infty$, and the colliding $p\to q$ should be glued to the diagonal $p=q$.
The 3 remaining, namely $0\to 1$, $0\to\infty$ and $1\to\infty$ should be glued to just a point respectively to $p=q=\infty$, $p=q=1$, $p=q=0$.

9 among 15 of the codimension 2 boundaries, for example the boundary $(p\to 0 , q\to 1 , \infty)$ should be glued to the corresponding missing points in the product.
The 6 remaining codimension 2 boundaries, for example $(0\to 1, p, q\to\infty)$ should be glued to the missing point in one of the 3 sphere--less--3--points, the one glued to the point $p=q=\infty$.

In the end we find a union of pieces of different dimensions.
Many points have neighborhoods that are not homeomorphic to Euclidian subspaces.

This is a \textbf{stack}\index{stack}.

\subsection{$n>5$}

The same holds for higher $n$:
\beq
\dim \mathcal M_{0,n} =n-3 
\eeq
it is a product of $n-3$ sphere--less--3--points, and to which we remove all submanifolds of coinciding points.
One can compute that
\beq
\chi(\mathcal M_{0,n})=(-1)^{n-1} (n-3)!.
\eeq

Boundaries are nodal surfaces.
For $n\geq 5$ there are $n(n-1)/2$ codimension 1 boundaries (choose 2 points among $n$).
All boundaries can be described by a graph whose vertices are the components of the nodal surface, and whose edges are nodal points.
Since the genus is 0, we must have
\beq
2-2g-n=2-n = \sum_i (2-2g_i-n_i-k_i)
\eeq
where $\sum_i n_i$ is the number of marked points and $\frac12 \sum_i k_i$ is the number of nodal points.
Observe that the connected components can't exceed 1+ number of nodal points, otherwise the surface would not be connected, which implies that
\beq
\sum_i (k_i-2) \geq -2.
\eeq
The relationship $2-n = -n+\sum_i (k_i-2) - 2\sum_i g_i$
implies 
\beq
\sum_i g_i =1-\frac12 \sum_i (k_i-2) \leq 0,
\eeq
therefore all connected components must have genus 0, and
the number of nodal points (edges in the dual graph) is equal to the number of connected components (vertices in the dual graph)-1, i.e. the graph must be a tree.

\section{Genus 1}

By the Abel map, every Riemann surface of genus 1 is isomorphic to a standard torus $T_\tau = \CC/\mathbb Z+\tau\mathbb Z$ for some $\tau$ with $\Im\tau>0$.

\bt
$T_{\tau}$ and $T_{\tau'}$ are isomorphic iff 
\beq
\tau =\frac{a\tau'+b}{c\tau'+d}
\qquad , \ (a,b,c,d)\in \mathbb Z^4 \ , \ ad-bc=1.
\eeq
\et

\proof
Assume that there exists an isomorphism $f:T_\tau \to T_{\tau'}$.
It must satisfy, in each charts:
\beq\label{eqftorusmapcond}
f(z+n+\tau m) = f(z)+n'+\tau' m'
\eeq
Its differential $df$ must therefore satisfy
\beq
df(z+n+\tau m) =df(z)
\eeq
showing that it is globally  a holomorphic 1-form on $T_\tau$, 
it must therefore be proportional to $dz$, i.e.
\beq
df(z) =\alpha dz.
\eeq
This implies that $f$ must be a chart-wise affine function:
\beq
f(z) = \alpha z + \beta.
\eeq
A priori this function is defined on the fundamental domain.
It must satisfy \eqref{eqftorusmapcond}, in particular we must have
\beq
\alpha = f(1)-f(0) = c\tau'+d
\qquad , \qquad
\alpha\tau =  f(\tau)-f(0) = a\tau'+b,
\eeq
and therefore
\beq
\tau = \frac{a\tau'+b}{c\tau'+d}.
\eeq
Saying that this transformation is invertible in the same form implies that $ad-bc=1$.

Vice-versa, if $\tau$ and $\tau'$ are related by such a transformation, the map:
\beq
f:z\mapsto (c\tau'+d) z \ \text{mod}\ \mathbb Z+\tau'\mathbb Z
\eeq
is a holomorphic map $T_\tau\to T_{\tau'}$ since it satisfies the transition condition that $f(z+n+\tau m)\equiv f(z)$.
\eproof

\bc[Moduli space $\mathcal M_{1,0}$]
\beq
\mathcal M_{1,0}
= \{ T_\tau \ | \ \tau\in \CC_+ \}/\text{isomorphisms} =   \CC_+ / PSL(2,\mathbb Z) .
\eeq
However, each $T_\tau$ has an infinite group of automorphisms, indeed every translation $z\mapsto z+\beta$ for $\beta\in \CC$ is an automorphism.

$\mathcal M_{1,0}$ is unstable.
\ec

$$
\includegraphics[scale=0.75]{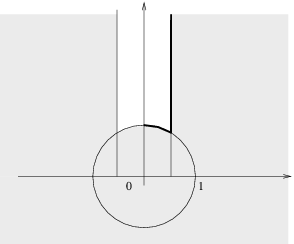}
$$

If we have a marked point $p_1$, up to performing a translation, we choose it to be the origin $z=0$. Therefore:

\bt[Moduli space $\mathcal M_{1,1}$]
\beq
\mathcal M_{1,1}
= \{ (T_\tau,0) \ | \ \tau\in \CC_+ \}/\text{isomorphisms} =   \CC_+ / PSL(2,\mathbb Z) .
\eeq
Since the modular group $PSL(2,\mathbb Z)$ is generated by $\tau\mapsto \tau+1$ and $\tau \mapsto -1/\tau$, a fundamental domain is
\bea
\mathcal M_{1,1} 
&=& \{ z \ | \ \Im\, z>0, \ \frac{-1}{2}<\Re\, z< \frac{1}{2} , \ |z|>1\} \cr
&& \cup \{ z \ | \ \Re\, z=\frac12 , \ \Im\, z > \frac{\sqrt{3}}{2} \} 
 \cup \{ z \ | \ |z|=1 , \ \frac\pi{3}< \text{Arg}\, z < \frac{\pi}{2} \} \cr
&& \cup \{\ii\} \cup \{e^{\frac{2\pi\ii}{3}}\}.
\eea
$\mathcal M_{1,1}$ is an orbifold of dimension $1$.

\smallskip
\textbf{Automorphisms:}

The map $z\mapsto -z$ is always an automorphism.
\begin{itemize}
\item For generic $(\curve,p)\in \mathcal M_{1,1}$, i.e. generic $\tau$ we have $\operatorname{Aut}=\ZZ_2$.

\item For $\tau=\ii$, the map $z\mapsto \ii z$ is an automorphism, and we have $\operatorname{Aut}=\ZZ_4$.

\item For $\tau=e^{\ii\pi/3}$, the map $z\mapsto e^{\ii\pi/3} z$ is an automorphism, and we have $\operatorname{Aut}=\ZZ_6$.

\end{itemize}

In all cases the number of automorphisms is finite, $\mathcal M_{1,n}$ is stable iff $n\geq 1$.

\et

\textbf{Euler characteristic of $\mathcal M_{1,1}$:}

$\mathcal M_{1,1}$ is made of a 2-cell (with $\ZZ_2$ automorphism), two 1-cells (with $\ZZ_2$ automorphism), and 2 points with automorphisms $\ZZ_4$  and $\ZZ_6$. 
The Euler characteristics is thus
\beq
\chi(\mathcal M_{1,1}) = \frac{1}2-2\times \frac12+\frac14+\frac16 = \frac{-1}{12}.
\eeq

%
\medskip
What is interesting, is to see that for an orbifold, the Euler characteristic is a rational number rather than an integer.

\subsection{Boundary of $\mathcal M_{1,1}$}

The boundary is reached when a cycle gets pinched into a nodal point, and this corresponds to $\tau\to\infty$ (or in fact $(a\tau+b)/(c\tau+d)\to \mathbb Q\cup\{\infty\}$ at the boundary of the hyperbolic plane $\CC_+$).
We can identify a torus with a pinched cycle (a nodal point) and a marked point, with a sphere with 3 marked points, 2 of the marked points  when glued together provide a nodal point (and they can be exchanged by a $\ZZ_2$ symmetry), and the 3rd marked point is the initial marked point of the torus.
In other words
\beq
\partial \mathcal M_{1,1}  \sim \mathcal M_{0,3}/\ZZ_2 .
\eeq

$$
\includegraphics[scale=0.45]{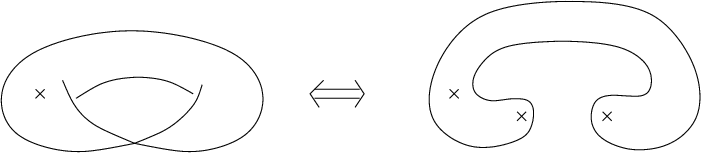}
$$

We have
\beq
\overline{\mathcal M_{1,1}} = \mathcal M_{1,1} \cup \mathcal M_{0,3}/\ZZ_2.
\eeq
The fundamental domain of $\mathcal M_{1,1}$ is a hyperbolic triangle in the upper complex plane $\CC_+$ equipped with the hyperbolic metric.
In this metric,  geodesics are lines or circles orthogonal to the real axis, so that indeed the boundaries of the fundamental domain are geodesics, it is an hyperbolic triangle.
This triangle has 3 vertices: 2 of them ($e^{\\ii\pi/3}$ and $e^{2\ii\pi/3}$) have angle $\pi/3$ and one of them ($\infty$) has angle $0$.

In hyperbolic geometry, the area of a triangle is its deficit angle:
\beq
\operatorname{Volume}_{\text{Hyperbolic}}(\overline{\mathcal M_{1,1}}) = \pi - \frac{\pi}{3}-\frac{\pi}{3}-0 = \frac{\pi}{3}.
\eeq

\section{Higher genus}

For $\genus\geq 2$, the Abel map embeds the curve into a submanifold of its Jacobian, which is a compact torus of dimension $2\genus$.

Given a symplectic basis of cycles, and an origin point to define the Abel map $p\mapsto\mathbf u(p)$, define
\bea
\forall \ i=1,\dots, \genus \quad , \qquad
v_i(p) = \sum_{j} (\Im\tau)^{-1}_{i,j} \Im u_j(p) \cr
\forall \ i=1,\dots, \genus \quad , \qquad
\td v_i(p) = \Re u_i(p) - \sum_{j} (\Re\tau)_{i,j} v_j(p)
\eea
By definition we have
\beq
u_i(p) = v_i(p) + \sum_j \tau_{i,j} \td v_j(p),
\eeq
which we write
\beq
\mathbf u(p) = v(p) + \tau \td v(p).
\eeq

The following sets of points of $\curve$ can be chosen as arcs representing the cycles $\acycle_i$ or $\bcycle_i$s:
\bea
\{ p \ | \ \td v_i(p)=0\} \sim \acycle_i \cr
\{ p \ | \ \td v_i(p)=1\} \sim \acycle_i \cr
\{ p \ | \ v_i(p)=0\} \sim \bcycle_i \cr
\{ p \ | \ v_i(p)=1\} \sim \bcycle_i
\eea
If we remove them from $\curve$, we get the fundamental domain $\curve_0\subset \curve$.
The map
\bea
\curve & \to & ]0,1[^{2\genus} \cr
p & \mapsto & \{ v_i(p),\td v_i(p)\}
\eea
embedds the curve into a polygon with $4\genus$ sides in $]0,1[^{2\genus}$.
The surface $\curve$ is obtained by gluing corresponding sides together.

It is equipped with a K\"ahler metric corresponding to a symplectic form (which is nothing but the  restriction of the canonical symplectic form of $[0,1]^{2\genus}\subset \RR^{2\genus}$ to the image of the curve)
\beq
 \sum_{i=1}^\genus d \td v_i\wedge d v_i = \sum_{i,j} (\Im \tau)^{-1}_{i,j} \ d\Re u_i \wedge d\Im u_i
 = \frac{\ii}{2} \sum_{i,j} (\Im \tau)^{-1}_{i,j} \ d u_i \wedge d\bar u_i.
\eeq
Notice that this metric is positive definite, and is modular invariant.

With this metric, the total area is (this is proved by Riemann bilinear identity)
\beq
\text{Area}(\curve) = \genus.
\eeq

\textbf{Example: Torus.}
The torus is not only embedded in the Jacobian, it is isomorphic to its Jacobian. Writing $u=v+\tau \td v$ where $v$ and $\td v$ are real, the lines $v=0$, $v=1$ are the 2 sides of the cycle $\bcycle$, the lines $\td v=0$ and $\td v=1$ are the 2 sides of the cycle $\acycle$, and the fundamental domain comprised between them is the parallelogram with summits $0,1,1+\tau,\tau$.
This is a parallelogram whose basis has length 1, and whose height is $\Im\tau$, therefore with the metric 
\beq
\frac{1}{\Im\tau} \ d\Re u \wedge d\Im u  = d\td v \wedge dv,
\eeq
its area is $1$, which is equal to the genus.

\section{Coordinates in the moduli space}
\label{secStrebel}

We shall find an explicit atlas of the moduli space $\mathcal M_{g,n}$.
Charts will be homeomorphic to $\RR_+^{6g-6+2n}$, and with gluing rules encoded by graphs.

\subsection{Strebel graphs}
\label{secstrebelquadraticdif}

Let $(\genus,n)$ such that $2\genus-2+n>0$.
Rather than considering $\mathcal M_{\genus,n}$, let us consider the product $\mathcal M_{\genus,n}\times \RR_+^n$, that we prefer to view as a trivial bundle over $\mathcal M_{\genus,n}$ whose fiber is $\RR_+^n$:
\beq
\tilde{\mathcal M}_{\genus,n} = \mathcal M_{\genus,n} \times \mathbb R_+^n \to \mathcal M_{\genus,n} .
\eeq
Let $\spcurve=(\curve,p_1,\dots,p_n,L_1,\dots,L_n)\in \tilde{\mathcal M}_{\genus,n}$, where $\curve$ is a smooth Riemann surface of genus $\genus$, and $p_1,\dots,p_n$ are $n$ labeled distinct marked points on $\curve$, and $L_1,\dots,L_n$ are positive real numbers.

Let us consider the set $\Omega_{\spcurve}$ of \textbf{quadratic differentials}\index{quadratic differential} $\omega$ (see def. II-\ref{defkthorderform}), having double poles at the marked points, and no other poles, and behaving (in some chart of $\curve$ with coordinate $\phi$) near the marked point $p_i$ as:
\beq
\omega(p) \mathop{\sim}_{p\to p_i} \frac{-L_i^2}{(\phi(p)-\phi(p_i))^2} \ \left(1+O(\phi(p)-\phi(p_i))\right) d\phi(p)^2 \ .
\eeq
$\Omega_{\spcurve}$ is an affine space, whose underlying linear space is the vector space of quadratic differentials with at most simple poles at the $p_i$s.

\medskip
$\bullet$ Example when $(\genus,n)=(0,3)$, and $\curve=\mathbb CP^1$,
we must have
\beq
\omega(z) = \frac{-L_\infty^2 z^2+(L_\infty^2+L_0^2-L_1^2) z - L_0^2}{z^2(z-1)^2} \ dz^2 
\eeq
and $\Omega_{\spcurve}$ consists of a unique quadratic differential, $\dim \Omega_{\spcurve}=0$.

$\bullet$ Example when $(\genus,n)=(0,4)$, and $\spcurve=(\mathbb CP^1,0,1,\infty,p,L_0,L_1,L_\infty,L_p)$,
we have
\beq
\omega(z) = \frac{-dz^2}{z(z-1)(z-p)} \ \left( L_\infty^2 z + \frac{L_0^2}{z} + \frac{L_1^2}{z-1} + \frac{L_p^2}{z-p} + c
\right)
\eeq
where $c\in \mathbb C$ can be any constant, in other words $\dim \Omega_{\spcurve}=1$.

$\bullet$ Example when $\genus=0$ and $n\geq 4$, and $\spcurve=(\mathbb CP^1,p_1,\dots,p_{n};L_\infty,L_1,\dots,L_{n})$,
\beq\label{eq:OmSg0}
\omega(z) = \frac{-dz}{\prod_{i=1}^n (z-p_i)} \left(  \sum_{i=1}^n \frac{L_i^2\prod_{j\neq i}(p_i-p_j)}{z-p_i}  + \sum_{j=0}^{n-4} c_j z^j \right)
\eeq
where $c_0,c_1,\dots,c_{n-4}$ are arbitrary complex numbers, in other words
$\Omega_{\spcurve} \sim \mathbb C^{n-3}$.

$\bullet$ Example when $(\genus,n)=(1,1)$, and $\spcurve=(T_\tau,0,L_0)$,
we must have
\beq
\omega(z) = \left( -L_0^2 \ \wp(z;\tau) + c\right) \ dz^2
\eeq
where $\wp$ is the Weierstrass function and $c\in \mathbb C$, so that $\dim \Omega_{\spcurve}=1$.

\medskip

\bt
\beq
\dim \Omega_{\spcurve}=d_{\genus,n}=3\genus-3+n.
\eeq
\et

\proof
$\bullet$ If $\genus=0$, on the Riemann sphere, $\Omega_{\spcurve}$ is the set of forms given in \eqref{eq:OmSg0}. Indeed $\omega/dz^2$ must be a rational function with $n$ double poles of given leading behavior, and that behaves as $O(1/z^4)$ at $\infty$.
Therefore $f(z)=\frac{\omega(z)}{dz^2} \prod_{i=1}^n (z-p_i)^2$ must be a polynomial, of degree at most $2n-4$, and such that $f(p_i)=-L_i^2$. The space of such polynomials has dimension $n-3$.

\smallskip

$\bullet$ Now assume $\genus=1$, and $\curve=\mathbb C/\mathbb Z+\tau\mathbb Z$. Then the following 1-form
\beq
\tilde \omega(z) = \frac{\omega(z)}{dz}+ \sum_{i=1}^n L_i^2 \wp(z-p_i) dz 
\eeq
must be a meromorphic 1-form with at most simple poles at the $p_i$s.
Let $r_i=\Res_{p_i} \tilde\omega$. 
Then the 1-form  (where $\sigma$ is a primitive of $\wp$, i.e. $d\sigma=\wp$)
\beq
\tilde \omega(z) - \sum_{i=1}^n r_i \sigma(z-p_i)dz
\eeq
must be holomorphic, i.e. proportional to $dz$, therefore there must exist $r_1,\dots,r_n$, such that $\sum_i r_i=0$, and $C$ such that
\beq
\omega(z) = \left( C -\sum_{i=1}^n L_i^2 \wp(z-p_i) -r_i \sigma(z-p_i)\right) dz^2.
\eeq
Vice versa, every such quadratic form is in $\Omega_{\spcurve}$, therefore
\beq
\dim \Omega_{\spcurve} = n.
\eeq

\smallskip

$\bullet$ Now assume $\genus\geq 1$.
First, $\Omega_\spcurve$ is not empty, indeed the quadratic form
\beq
\omega(z)= - \sum_{i=1}^n L_i^2 \ B(z,p_i) \ \frac{\omega_1(z)}{\omega_1(p_i)}
\quad \in \Omega_{\spcurve}.
\eeq
Any element of $\Omega_\spcurve$ is of the form
\beq
\omega+\omega'
\eeq
where $\omega'$ is a quadratic form having at most simple poles at the $p_i$s, i.e. in the linear underlying space of the affine space $\Omega_{\spcurve}$.

Let us choose once for all, a generic (having no zero at the $p_i$s) holomorphic 1-form $\nu\in \mathcal O^1(\curve)$ (typically choose $\nu=\omega_1$), and let $o$ a zero of $\nu$, and let
\beq
r_i = \Res_{p_i} \frac{\omega'}{\nu}.
\eeq

The holomorphic quadratic form
\beq
\tilde \omega(z) = \omega'(z) - \sum_{i=1}^n r_i \ \omega'_{p_i,o}(z)\ \nu(z)
\eeq
has no poles (where $\omega_{p,q}(z)$ is the 3rd kind form introduced in cor.II-\ref{corexistthridkindforms} or def. III-\ref{defthirdkindfromB}).
Since the ratio of quadratic forms is a meromorphic function, and since meromorphic functions have the same number of poles and zeros, this implies that $\tilde\omega$ must have the same number of zeros as any quadratic form, in particular as $\nu^2$, and therefore $\tilde\omega$ must have $4\genus-4$ zeros.

Among these zeros, choose $\genus-1$ of them, and choose the unique (up to scalar multiplication) holomorphic 1-form $\mu$ that vanishes at those $\genus-1$ points.
Therefore $\frac{\tilde\omega}{\mu}$ is a meromorphic 1-form with $3\genus-3$ zeros and $\genus-1$ poles (the other $\genus-1$ zeros of $\mu$) that we name $q_1,\dots,q_{\genus-1}$.
Let $s_i = \Res_{q_i} \frac{\tilde \omega}{\mu}$, then
\beq
\frac{\tilde \omega}{\mu} - \sum_{i=1}^{\genus-1} s_i \ \omega_{q_i,o}
\eeq
must be a holomorphic 1-form, therefore a linear combination of $\omega_1,\dots,\omega_{\genus}$.
In the end we may write
\beq
\omega' = \nu\sum_{i=1}^n r_i \ \omega_{p_i,o} + \mu \left( \sum_{i=1}^{\genus-1} s_i \ \omega_{q_i,o} + \sum_{i=1}^\genus \alpha_i \ \omega_i \right).
\eeq
The decomposition is not unique, as we could have chosen any $\genus-1$ zeros of $\tilde\omega$, but this is a discrete ambiguity.
Vice versa, for any choice of $r_i$, $s_i$ (subject to $\sum_i s_i=0$), $\mu$, $\alpha_i$, we get an element of $\Omega_{\spcurve}$.
The only redundency is that we can multiply $\mu$ by a scalar and divide $s_i$ and $\alpha_i$s by the same scalar.
This shows that
\beq\label{dimOmSdecomp}
\dim \Omega_{\spcurve} = (\overbrace{n}^{r_i\in \CC^n}+\overbrace{\genus-2}^{s_i\in \CC^{\genus-2}}+\overbrace{\genus}^{\mu\in \mathcal O^1}+
\overbrace{\genus}^{\sum_i \alpha_i\omega_i\in \mathcal O^1}) -\overbrace{1}^{\text{scalar redundency}} = 3\genus-3+n.
\eeq

\eproof

\bd
Given $\omega\in \Omega_{\spcurve}$, a \textbf{horizontal trajectory}\index{horizontal trajectory} is a maximal connected set $\gamma\subset (\curve-\{p_1,\dots,p_n\})^{\text{universal cover}}$, on which the map
\beq
 p \mapsto \Im \left( \int^p \sqrt{\omega(z)} \right)
\eeq
is constant. 
This is independent of a choice of initial point of integration, and of a choice of a sign of the square root.
\ed

Horizontal trajectories have the following properties:

\begin{itemize}

\item Locally, in a neighborhood of a point where $\omega$ has neither pole nor zero,
 horizontal trajectories are $C^\infty$ Jordan arcs.

\item In a neighborhood of a pole $p_i$, we have
\beq
\sqrt{\omega(z)} \sim \ii L_i \ \frac{dz}{z-p_i} = \ii L_i \ d\log{(z-p_i)}
\eeq
so that horizontal trajectories are circles $|z-p_i|\sim $constant.
$$
\includegraphics[scale=0.55]{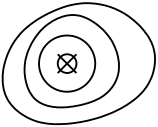}
$$

\item Horizontal trajectories can not cross except at zeros of $\omega$.

\item 
In a neighborhood of a zero $a$ of $\omega$, of order $k_a$, we have
\beq
\omega(z) \sim c_a \ (z-a)^{k_a} \ dz^2
\eeq
and thus the horizontal trajectories going through $a$ are locally the rays
\beq
\sim a+ e^{\ii (-\operatorname{Arg} c_a + \pi j ) \frac{2}{k_a+2} } \ \mathbb R_+
\qquad , \quad j=1,\dots,k_a+2
\eeq
they  form a star with $k_a+2$ branches.
These are called "critical trajectories".
Generically, zeros are simple, $k_a=1$, so that critical trajectories have generically trivalent vertices.
$$
\includegraphics[scale=0.55]{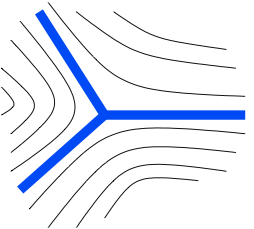}
$$

\item a critical trajectory starting from a vertex (zero of $\omega$) can either meet another (or the same) vertex, it is then called a finite trajectory, or not, it is then called an infinite trajectory.

A finite critical trajectory starting at $a$, is a Jordan arc $\gamma: [0,l]\to \curve$ such that $\frac{1}{2\pi}\int_a^{\gamma(t)} |\sqrt\omega| = t$.

An infinite critical trajectory starting at $a$, is a Jordan arc $\gamma: [0,\infty[\to \curve$ such that $\frac{1}{2\pi}\int_a^{\gamma(t)} |\sqrt\omega| = t$.

\item infinite trajectories have an adhrerence, which is also a horizontal trajectory.
Indeed, let $\gamma:\RR_+\to\curve$ an infinite trajectory, and let $\bar\gamma=\{p\in \curve \ | \ \forall \ \epsilon>0, \ \exists q\in \gamma \ | d_{|\sqrt{\omega}|}(q,p)<\epsilon\}$, where the distance is defined by the metric $|\sqrt\omega|$.
This adherence is non empty, because if we take an infinite sequence $t_1,t_2,\dots$ in $\RR_+$ tending to $+\infty$, then the sequence $\gamma(t_n)$ must have  a limit on $\curve$ (because $\curve$ is compact), so $\hat\gamma$ is not empty. Moreover, if $\gamma(t_n)$ converges to a point $p\in\hat\gamma$, then $\omega(\gamma(t_n))$ converges to $\omega(p)$, and thus $\hat\gamma$ has a tangent vector, it must be a compact  $C^1$ curve.
Moreover, it must be a horizontal trajectory.
We call it a critical horizontal trajectory.
It is compact and has finite length, but doesn't need to go through a vertex.

There are at most 3 critical trajectories per vertex, thus the number of compact critical trajectories is finite.

\item The set of compact critical trajectories forms a graph $\Gamma$ embedded in $\curve$. The graph is not necessarily connected. Its vertices are zeros of $\omega$.

\item The connected components of $\curve-\Gamma$ are called faces.
For each $i=1,\dots,n$, there is a unique face containing $p_i$, and it is topologically a disc.

\item Faces that do not contain any $p_i$, have the topology of cylinders.

\proof
Choose a finite number $\geq 1$ of marked points on each edge of the graph.
Consider the vertical trajectories emanating from them, and continue them until they reach another edge or a point $p_i$.
Those trajectories can have finite or infinite length. 
If a trajectory has infinite length, this means that $|\Im \int \sqrt\omega|$ becomes larger than the distance (measured with the metric $|\sqrt\omega|$) between any 2 edges, and thus an infinite vertical trajectory must necessarily enter one of the discs around one $p_i$.
The vertical trajectories inside the faces not containing the $p_i$s have finite length.
Consider the graph $\td\Gamma$ of all these finite vertical trajectories.

Consider the connected components of $\curve\setminus (\Gamma\cup \td\Gamma)$.
Every component $f$ not containing a $p_i$ has at least one marked point on its boundary, call it $q_f$.
The map $g_f: f \to \CC$,  $p\mapsto \int_{q_f}^p \sqrt\omega$ is bounded, i.e. $\Re g_f$ and $\Im g_f$ have a minimum and maximum in $f$, defining a rectangle $R_f$ in $\CC$.
The map $g_f$ is analytic in $f$, and thus the image of $f$ is an open set of $R_f$, whose boundary can be made only of horizontal and vertical lines, i.e. must be a rectangle in $\CC$.

This construction provides an atlas of $\curve$, whose charts are either discs centered around the $p_i$s and a finite number of rectangles. The transition functions are translations $x\mapsto \pm x+c$, up to the choice of sign $\pm$ for the square root.

Consider a compact critical trajectory, and the set of all rectangles bordering its left (reps. right) side. Since $\curve$ is orientable, the choice of sign of the square root can be chosen in such a way that all transition maps from a rectangle to its neighbour, are $c\mapsto x+c$ with $c\in \RR$.
Moreover the other horizontal boundary of the face, must be a horizontal trajectory, i.e. all rectangles must have the same height.
The gluing of all rectangles bordering the trajectory is then the gluing of a finite number of rectangles of the same height along their parallel sides, it is a cylinder.

\eproof

$$
\includegraphics[scale=0.45]{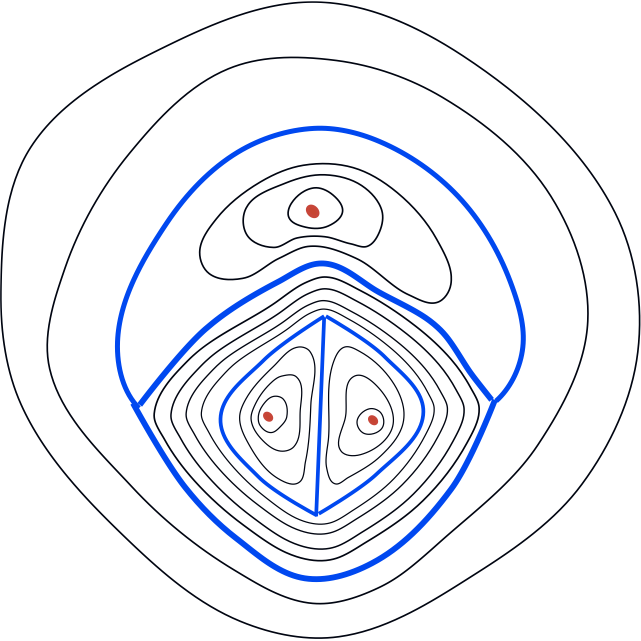}
$$

\end{itemize}

\bt[Strebel]
There exists a unique element $\omega\in \Omega_{\spcurve}$ such that the graph is cellular, i.e. all faces are discs (no cylinder).
$\omega$ is called the \textbf{Strebel differential}\index{Strebel differential}, and the graph of its critical horizontal trajectories is called the \textbf{Strebel graph}\index{Strebel graph}.

The Strebel graph map:
\bea
\tilde{\mathcal M}_{\genus,n} & \to & \oplus_{\Gamma\in \mathcal G_{\genus,n}} \mathbb R_+^{\{\text{edges}(\Gamma)\}} \cr
\spcurve &\mapsto & \left(\Gamma,\{ \ell_e\}_{e\in\text{edges}(\Gamma)}\right)
\qquad , \quad \ell_e = \left| \int_e \sqrt{\omega} \right|
\eea
is an isomorphism of orbifolds (it sends $\operatorname{Aut}\spcurve \to \operatorname{Aut}\Gamma$).
Here $\mathcal G_{\genus,n}$ is the set of trivalent cellular graphs of genus $\genus$, with $n$ faces.
$\ell_e$ is the length of edge $e$, measured with the metric $|\sqrt\omega|$.

\et
$$
\includegraphics[scale=0.45]{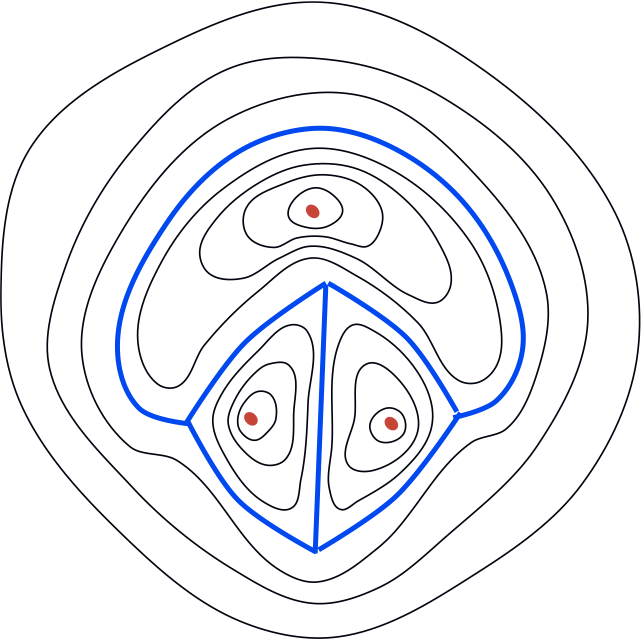}
$$

\proof
Given $\omega\in \Omega_{\spcurve} $, for each $p_i$ and a given chart, we define the unique primitive of $\sqrt\omega$ such that
\beq
dg_{p_i}(z)  = \sqrt{\omega(z)}
\qquad , \qquad
g_{p_i}(z) \sim_{z\to p_i} \ii L_i \log{(\phi(z)-\phi(p_i))} + O((\phi(z)-\phi(p_i)))
\eeq
(in other words we have fixed the integration constant so that the is no term of order 0).
Now, given some positive real numers $r_1,\dots,r_n$, we define:
\beq
\mathcal A(\omega) 
= \int_{\curve\setminus \cup_i \{\Im g_{p_i}(z)<\log{r_i}\} } |\sqrt\omega|^2
+ 2\pi \sum_i L_i \log{r_i} .
\eeq
One easily checks (Stokes theorem) that $\mathcal A(\omega)$ is actually independent of $r_i$, provided that $r_i$ is sufficiently small so that the discs $\Im g_{p_i}(z)<\log{r_i}$ do not intersect $\Gamma$.

We have
\beq
\mathcal A(\omega) \geq 2\pi\sum_i L_i \log{R_i(\omega)}
\eeq
where $R_i(\omega)$ is the largest radius such that $\Im g_{p_i}(z)<\log{R_i}$ does not intersect $\Gamma$, so that
\beq
\mathcal A(\omega) - 2\pi\sum_i L_i \log{R_i(\omega)}
= \int_{\cup\ \text{cylinders}} |\sqrt\omega|^2.
\eeq
Moreover $\mathcal A$ is a convex functional of $\omega$, with second derivative
\beq
\mathcal A''(\delta \omega, \tilde\delta\omega) = \int_{\curve} |\sqrt\omega|^2 \ \Im \frac{\delta\omega}{\omega} \ \Im \frac{\tilde \delta\omega}{\omega} \ 
\eeq
which is a positive definite quadratic form.
Therefore $\mathcal A$ possesses a minimum, at which the gradient of $\mathcal A$ vanishes. The gradient is
\beq
\mathcal A'(\delta \omega) = \int_{\curve} |\sqrt\omega|^2 \ \Im \frac{\delta\omega}{\omega} .
\eeq
\beq
\delta \log{R_i(\omega)} = \frac12 \Im \int_{p_i}^{a} \frac{\delta \omega}{\sqrt\omega}
\eeq
Therefore the minimum is the Strebel differential.

\eproof

\subsubsection{Example}

$\td{\mathcal M}_{0,3}=\mathcal M_{0,3}\times \RR_+^3$
is a sum of 4 graphs times $\RR_+^3$.
$$
\includegraphics[scale=0.35]{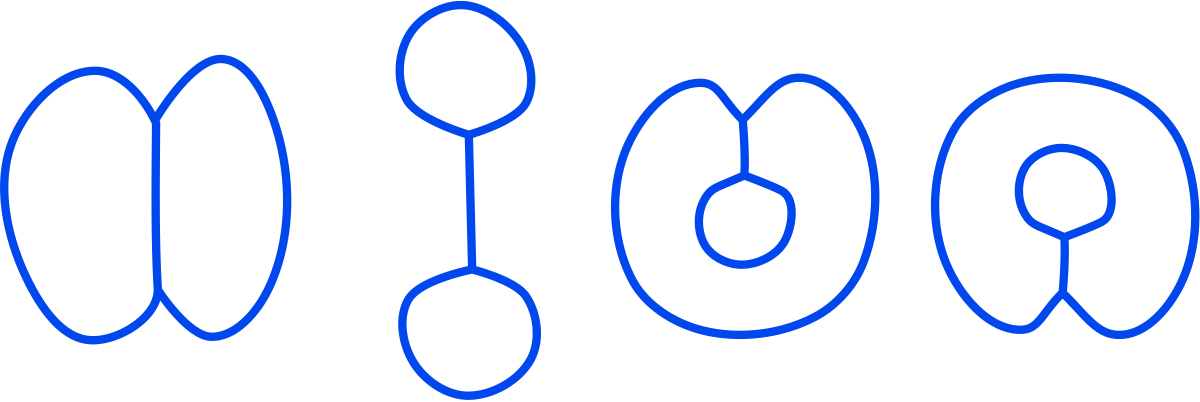}
$$
In each graph there are 3 lengths corresponding to the 3 edges.

In the 2nd, 3rd, 4rth graph we have a triangular inequality respectively $L_\infty\geq L_0+L_1$, $L_1\geq L_0+L_\infty$, $L_0\geq L_1+L_\infty$, whereas in the 1st graph no triangular inequality is satisfied.
This cuts $\RR_+^3$ into 4 disjoint regions, each labelled by a graph
$$
\includegraphics[scale=0.4]{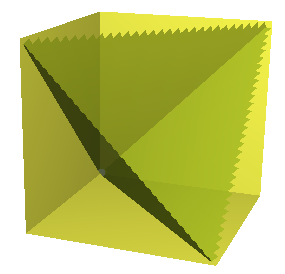}
$$

In the 1st graph we have $L_0=\ell_1+\ell_2$, $L_1=\ell_2+\ell_3$ and $L_\infty=\ell_3+\ell_1$, whereas in the second graph we have
$L_0=\ell_1$, $L_1=\ell_2$ and $L_\infty=\ell_1+\ell_2+2\ell_3$.

\subsection{Topology of the moduli space}

An atlas of $\td{\mathcal M}_{\genus,n}$ is thus made of charts labelled by 3-valent graphs of genus $\genus$ with $n$ faces.

The boundary between the domains correspond to one or several edge lengths vanishing. Shrinking an edge amounts to merge 2 trivalent vertices into a 4-valent vertex. The gluing of charts amounts to glue together all graphs whose shrinking edges give the same higher valence graph.

This atlas makes $\td{\mathcal M}_{\genus,n}$ a smooth real manifold of dimension $6\genus-6+3n$ (number of edges), equipped with the topology inherited from $\RR_+^{6\genus-6+3n}$. It is connected.

\medskip

\subsubsection{Bounday and compactification}

Not all vanishing edge lengths correspond to graphs embedded on a smooth surface, some of them can be embedded only in nodal surfaces, and these correspond to the boundary of $\td{\mathcal M}_{\genus,n}$.

Adding the "nodal graphs" makes $\overline{{\mathcal M}_{\genus,n}}$  a Deligne--Mumford compact space, but not a manifold because there are pieces of different dimensions.
It is connected.

\subsubsection{Complex structure}

Instead of the real lengths $\ell_e$, we can parametrize the Strebel differential as a point in $\Omega_\spcurve$, as in \eqref{dimOmSdecomp}:
\beq
\Omega_\spcurve \sim \CC^{n+\genus-2} \times (\mathcal O^1(\curve)\times \mathcal O^1(\curve))/\CC \sim \CC^{3\genus-3+n}.
\eeq
The Strebel differential has complex coordinates in that space, and these complex coordinates can be used as coordinates of $\mathcal M_{\genus,n}$.

A basis of $\mathcal O^1(\curve)$ can be defined locally in a chart where a symplectic basis of cycles can be held fixed. 
Changing charts changes the symplectic basis, and the transition functions to glue coordinates are obtained from the transition functions of the bundle with fiber $\mathcal O^1(\curve)\to \mathcal M_{\genus,n}$, and they are analytic.

This provides a complex structure to $\mathcal M_{\genus,n}$.

\section{Uniformization theorem}

\textbf{Question: Poincar\'e metric}

{\it{
Let $(\genus,n)$ be such that $2-2\genus-n<0$.
Let $(\curve,p_1,\dots,p_n)\in \mathcal M_{\genus,n}$ a compact Riemann surface of genus $\genus$ with $n$ marked points.
Let $\alpha_1,\dots,\alpha_n$ be $n$ real numbers.

Does there exists a Riemannian metric of constant curvature $-1$, that vanishes at order $2\alpha_i$ at marked point $p_i$ ? 
Is is unique ?
}}

%

\medskip

In a chart with coordinate $z$, the \textbf{Poincar\'e metric}\index{Poincar\'e metric} (if it exists) can be written
\beq
e^{-\phi(z,\bar z)} \ |dz|
\qquad , \qquad
e^{-\phi(z,\bar z)} \mathop{\sim}_{z\to p_i} C_i \ |z-p_i|^{2\alpha_i} \ (1+o(1))
\eeq
where $\phi$ is a real valued function which we write as a function of $z$ and $\bar z$ instead of $\Re z$ and $\Im z$ in charts  $U\subset \RR^2$ identified with $\CC$.
Under a holomorphic change of chart and coordinates, i.e. under a holomorphic transition function $z\to \td z=\psi(z)$, $\phi(z,\bar z)$ changes as 
\beq\label{eqchangephicoord}
\td\phi(\psi(z),\bar{\psi(z)}) = \phi(z,\bar  z) + \log\left|\psi'(z)\right|= \phi(z,\bar  z) + \frac12 \log \psi'(z) +\frac12\log \overline{\psi'(z)} .
\eeq

The curvature is
\beq
-1=R(z,\bar z)=e^{2\phi(z,\bar z)} \ \Delta\phi(z,\bar z) .
\eeq
Finding a metric with constant curvature $=-1 $ thus amounts to solving \textbf{Liouville's equation}\index{Liouville's equation}
\beq\label{Liouvilleeq}
\Delta\phi(z,\bar z) = 4 \partial \bar\partial \phi(z,\bar z) = -e^{-2\phi(z,\bar z)}.
\eeq

\subsubsection{Stress energy tensor and projective connection}

From \eqref{Liouvilleeq} we have
\bea
\bar\partial(\partial^2\phi+(\partial\phi)^2)
&=& \bar\partial\partial^2\phi+2 \partial\phi \bar\partial\partial \phi \cr
&=& \partial(\bar\partial\partial\phi) -\frac12 e^{-2\phi} \partial\phi \cr
&=& \frac{-1}{4} \left( \partial  e^{-2\phi}+2 e^{-2\phi} \partial \phi \right) \cr
&=& 0,
\eea
which we rewrite as
\beq
\bar\partial T(z)=0
\qquad \text{where} \quad
T(z)= \partial^2\phi(z,\bar z) + (\partial \phi(z,\bar z))^2 .
\eeq
\beq
T(z) \mathop{\sim}_{z\to p_i} \frac{-\Delta_i}{(z-p_i)^2}\ (1+o(1))
\qquad , \quad
\Delta_i=\alpha_i(1-\alpha_i).
\eeq
$T(z)$ is called the \textbf{stress energy tensor}\index{stress energy tensor}, and we see that it must be analytic outside of the marked points. We may drop the $\bar z$ dependence because $\bar\partial T=0$.

Under a change of chart $z\to \td z=\psi(z)$, $T(z)$ changes (using \eqref{eqchangephicoord}) as
\beq\label{eqchangeT}
\td T(\psi(z)) = \frac{1}{\psi'(z)^2} \ \left( T(z) +\frac12  \{\psi,z\} \right)
\eeq
where $\{\psi,z\}$ is called the \textbf{Schwartzian derivative}\index{Schwartzian derivative} of $\psi$:
\beq
\{\psi,z\} = \frac{\psi'''}{\psi'} -\frac32 \left(\frac{\psi''}{\psi}\right)^2
\eeq
A quadratic differential form
\beq
2T(z)dz^2
\eeq
with transitions given by \eqref{eqchangeT}
is called a \textbf{projective connexion}\index{projective connexion}.

If $B(z_1,z_2)$ is the Bergman kernel, the fundamental second kind differential on $\curve$ normalized on a chosen symplectic basis $\acycle_i,\bcycle_i$ of cycles, and $f$ a meromorphic function of $\curve$, then
\beq
S_f(z) =  -6 \ df(z)^2 \lim_{z'\to z} \left( \frac{B(z,z')}{df(z) df(z')} - \frac{1}{(f(z)-f(z'))^2} \right)
\eeq
is a projective connection. It has poles at the zeros of $df$.

It follows that, for any choice of a given projective connection $S$ independent of $p_i$s and $\alpha_i$s (for instance $S_f$ as above), then
\beq
\omega(z) =T(z)dz^2 - \frac12 S(z)
\eeq
is a meromorphic quadratic differential on $\curve$.

It has poles at the poles of $S$, and it has double poles at the $p_i$s:
\beq
\omega(z) \mathop{\sim}_{z\to p_i} \frac{-\Delta_i\ dz^2}{(z-p_i)^2}\ (1+o(1))
\qquad , \quad
\Delta_i=\alpha_i(1-\alpha_i).
\eeq
It belongs to an affine space, whose underlying linear space is the space of quadratic differentials introduced in section \ref{secstrebelquadraticdif}, i.e.
\beq
\omega \in \omega_S+\Omega'_{(\curve,p_1,\dots,p_n;\Delta_1,\dots,\Delta_n)}.
\eeq
This is a space of real dimension $6\genus-6+3n$.

\subsubsection{Oper}

If $T(z)$ would be known, we would recover $f(z,\bar z)=e^{\phi(z,\bar z)} $ by solving the Schr\"odinger equation in each chart:
\beq
\partial^2 f(z,\bar z) =-T(z) \ f(z,\bar z).
\eeq
The operator:
\beq
dz^2 (\partial^2 + T(z))
\eeq
is in fact independent of a choice of chart and coordinate (we leave to the reader to verify that it transforms well under chart transitions), it is called an \textbf{oper}\index{oper}.

Notice that $f(z,\bar z)=e^{\phi(z,\bar z)}$ is not a function, according to \eqref{eqchangephicoord} it transforms as a $(\frac{-1}2,\frac{-1}2)$ spinor form, so the oper does not act in the space of functions but in the space of spinors.
It sends a $\frac{-1}{2}$ spinor to a $\frac{3}{2}$ spinor.

\subsubsection{Monodromies}

This ODE in $z$ (at fixed $\bar z$, since the oper is independent of $\bar z$), is a second order ODE, it has 2 linearly independent solutions, call a choice of basis $f_1(z),f_2(z)$.
Doing the same thing for the $\bar z$ dependence, since $
\bar\partial^2 f(z,\bar z) = -\bar T(\bar z) f(z,\bar z)
$,
we see that there must exist 4 complex constants $c_{i,j}$ such that
\beq
f(z,\bar z) = c_{1,1} f_1(z) \bar f_1(\bar z) + c_{1,2} f_1(z) \bar f_2(\bar z) + c_{2,1} f_2(z) \bar f_1(\bar z) + c_{2,2} f_2(z) \bar f_2(\bar z)
\eeq
They form a $2\times 2$ matrix
\beq
C=\begin{pmatrix}
c_{1,1} & c_{1,2} \cr c_{2,1} & c_{2,2}
\end{pmatrix}.
\eeq
The choice of constants must be such that $f(z,\bar z)$ is a real monovalued function.
$f$ real implies that the matrix $C$ must be hermitian
\beq
C^\dagger= C.
\eeq
Up to a change of basis we may choose $C$ to be diagonal and real, and in fact we can choose $C=\text{Id}$.

Solutions of ODE usually have monodromies while going around a closed cycle $\gamma$: the vector space of solutions remains unchanged, but solutions can be replaced by linear combinations, so that the monodromy around a closed contour $\gamma$ is encoded by a matrix:
\beq
\begin{pmatrix}f_1(z+\gamma) \cr f_2(z+\gamma)\end{pmatrix}
= M(\gamma) \ \begin{pmatrix}f_1(z) \cr f_2(z)\end{pmatrix}.
\eeq
The $2\times 2$ monodromy matrix $M(\gamma)$ is independent of $z$ and  actually depends only on the homotopy class of $\gamma$.
We have $M(-\gamma)=M(\gamma)^{-1}$ and $M(\gamma_1+\gamma_2)=M(\gamma_2)M(\gamma_1)$, so that  monodromies provide a representation of the fundamental group $\pi_1(\curve-\{p_1,\dots,p_N\})$ into $Sl_2(\CC)$:
\bea
\pi_1(\curve-\{p_1,\dots,p_N\}) & \to & Sl_2(\CC) \cr
\gamma & \mapsto & M(\gamma).
\eea
$M(\gamma)\in Sl_2(\CC)$ rather than $Gl_2(\CC)$, i.e. 
$\det M(\gamma)=1$, thanks to the fact that the Wronskian $f'_1(z)f_2(z)-f_1(z)f'_2(z)$ is constant independent of $z$, and in particular remains constant after going around a cycle.

Requiring that $f(z,\bar z)$ is monovalued, i.e. has no monodromy,
implies that $\forall \ \gamma$:
\beq\label{eqMdaggerMunif}
M(\gamma)^\dagger C M(\gamma)=C.
\eeq

The rank of $\pi_1(\curve-\{p_1,\dots,p_n\})$ is $2\genus-2+n$, and therefore \eqref{eqMdaggerMunif} impose  $3\times(2\genus-2+n)=6\genus-6+3n$ real constraints on the choice of $\omega \in \omega_S+\Omega'_{(\curve,p_1,\dots,p_n;\Delta_1,\Delta_n)}$, which is precisely of that dimension.
We admit that this fixes a unique choice of $\omega$, and thus this determines uniquely the stress energy tensor $T(z)$ and then the function $\phi(z,\bar z)$.

\subsubsection{Mapping class group}

Therefore, for every $\alpha_1,\dots,\alpha_n\in \RR^n$ and every $(\curve,p_1,\dots,p_n)\in \mathcal M_{\genus,n}$, there is a unique Riemannian metric on $\curve$ of constant curvature $-1$, which has zeros (or poles) of order $\alpha_i$ at $p_i$.

This implies that a universal cover of $\curve-\{p_1,\dots,p_n\}$ is the hyperbolic plane, i.e. the upper complex plane $\CC_+$.
We recover $\curve$ by quotienting the universal cover, by the fundamental group $\pi_1(\curve-\{p_1,\dots,p_n\})$.

If we homotopically move a neighborhood $U\in \curve-\{p_1,\dots,p_n\} $ around a closed cycle $\gamma$, it should come back to itself in $\curve$, and to an isometric copy in the universal cover.
In other words, to each closed contour $\gamma$ is associated an isometry in $\CC_+$, and the fundamental group has a representation into the group of isometries of the hyperbolic plane.

The \textbf{Fuchsian group}\index{Fuchsian group} $\mathfrak K$ is the discrete subgroup of the hyperbolic isometries (called $PSL(2,\RR)$) of $\CC_+$, generated by $\pi_1(\curve-\{p_1,\dots,p_n\})$.
We have
\beq
\curve-\{p_1,\dots,p_n\} \sim \CC_+/\mathfrak K.
\eeq

We shall admit that it is possible to find a fundamental domain of $\CC_+$, bounded by geodesics, i.e. a polygon in the hyperbolic plane.
The quotient by $\mathfrak K$ then amounts to glue together some sides of the polygon to recover the surface $\curve-\{p_1,\dots,p_n\}$.
The points $p_1,\dots,p_n$ sit at the boundary of $\CC_+$ (i.e. on $\RR\cup\{\infty\}$), and are corners of the polygon, of angles $2\pi\alpha_i$, and all other angles are $\pi/2$.

It is possible to prove that the Fuchsian group is always a torsion free (no finite order element) discrete subgroup of the group $PSL(2,\RR)$ of hyperbolic isometries. And vice--versa, every such group is the Fuchsian group of a Riemann surface.

\subsubsection{Uniformization theorem}

This leads to
\bt[Uniformization theorem]
Every compact Riemann surface of genus $\genus=0$ is isomorphic to the Riemann sphere, every compact Riemann surface of genus $\genus=1$ is isomorphic to the standard torus $T_\tau$ (its Jacobian), and if $2\genus-2+n>0$:

For every $\alpha_1,\dots,\alpha_n\in \RR^n$ and every $(\curve,p_1,\dots,p_n)\in \mathcal M_{\genus,n}$, there is a unique Riemannian metric on $\curve$ of constant curvature $-1$, which has zeros (or poles) of order $2\alpha_i$ at $p_i$.

This allows to identify $\curve$ with a polygon  in the hyperbolic plane $\CC_+$, whose sides are glued pairwise, i.e. to $\CC_+/\mathfrak K$ where $\mathfrak K$ is a Fuchsian group, a discrete subgroup of isometries of $\CC_+$
\beq
\curve-\{p_1,\dots,p_n\} \sim \CC_+/\mathfrak K.
\eeq

\et

\br
The actual uniformization theorem is slightly stronger than the one we have written here, in particular it also considers surfaces with boundaries, and it characterizes Fuchsian groups in deeper details.
\er

\br
An interesting fact is that the uniformization theorem strongly relies on the Liouville equation and the stress energy tensor, in a way very similar to classical conformal field theory.
\er

\br
The Stress energy tensor, or more precisely the projective connexion, or more precisely the projective connexion shifted by a fixed projective connection, is found as a unique element of the space of quadratic differentials $\Omega_{(\curve,p_1,\dots,p_n;\Delta_1,\Delta_n)}$, like the Strebel differential.
It is similar but slightly different, indeed the Strebel differential was found by requiring that closed cycle-integrals $\oint_\gamma \sqrt\omega$ had to be real, whereas the stress energy tensor is found by requiring that the monodromy $M(\gamma)$ had to be unitary.
In a "heavy limit", where all $\alpha_i$ would be "large", the solutions of Schr\"odinger equation could be approximated by the WKB approximation, and the monodromies in that approximation would have eigenvalues of the form
\beq
\text{eigenvalues of }  M(\gamma) \ \sim e^{\pm \ii \oint_\gamma \sqrt{\omega}}
\eeq
and saying that the matrix be unitary implies that the integrals in the exponential are real.
In other words, in the heavy limit, the stress energy tensor tends to the Strebel quadratic differential.
\er

\br
To each choice of $(\curve,p_1,\dots,p_n,\alpha_1,\dots,\alpha_n)\in \mathcal M_{\genus,n}\times \RR^n$ corresponds a $SU(2)$ representation (by the monodromies) of the fundamental group:
\beq
\mathcal M_{\genus,n}\times \RR^n \to \text{Betti}
\eeq
where the \textbf{Betti space}\index{Betti space} is the set of representations of the fundamental group into $SU(2)$ (with monodromies of given eigenvalues $e^{\pm 2\pi\ii \alpha_i}$ on the small cycles $\mathcal C_{p_i}$):
\beq
\text{Betti} = \operatorname{Hom}(\pi_1(\curve-\{p_1,\dots,p_n\}),SU(2)) \ / \ (\operatorname{sp}(M(\mathcal C_{p_i}))=(e^{2\pi\ii \alpha_i},e^{-2\pi\ii \alpha_i}) ).
\eeq

\er

\br
An infinitesimal change of point in the moduli space $\mathcal M_{\genus,n}$, i.e. an infinitesimal change of complex structure, i.e. a cotangent vector to $\mathcal M_{\genus,n}$, corresponds to an infinitesimal change in the uniformization.
This can be seen as an infinitesimal change also in the Betti space.

In other words, this shows that the cotangent space to the moduli space, as well as the cotangent space to the Betti space, and the cotangent space to the space of opers, or also to the space of flat $SU(2)$ connections, are all isomorphic to the space of quadratic differentials with double poles at $p_i$. Their common real dimension is
\beq
2 d_{\genus,n} +n .
\eeq
where the addition of $n$ actually corresponds to the trivial factor by $\RR^n$.

\er

\section{Teichm\"uller space}

\bd[Teichm\"uller space]\index{Teichm\"uller space}

Let $S_\genus$ a smooth orientable surface of genus $\genus$.
The Teichm\"uller space $T(S_\genus)$ is the set of all complex structures on $S_\genus$, modulo  diffeomorphism isotopic to identity. 
An element of $T(S_\genus)$, i.e. a surface with a class of complex structures, is called a \textbf{marked}\index{marking} surface.

Due to the uniformization theorem, for $\genus\geq 2$, $T(S_\genus)$ is also the set of complete hyperbolic (curvature $R=-1$) Riemannian metrics on $S_\genus$, modulo  diffeomorphism isotopic to identity (there is a similar statement for $\genus=1$, with parabolic metric $R=0$, and for $\genus=0$ with elliptic metric $R=1$). 

The \textbf{mapping class group}\index{mapping class group}  $\Gamma(S_\genus)$ is the quotient of the group of all diffeomorphisms of $S_\genus$, by the subgroup of diffeomorphisms isotopic to identity.
The moduli space is the quotient
\beq
\mathcal M_{\genus,0} = T(S_\genus)/\Gamma(S_\genus).
\eeq
$T(S_\genus)$ is a universal cover of the moduli space $\mathcal M_{\genus,0}$. 
\ed

There are many ways of putting a topology on $T(S_\genus)$.

\subsection{Fenchel--Nielsen coordinates}
\index{Fenchel-Nielsen coordinates}

Consider
\bd
$\mathcal M_{g,n}(L_1,\dots,L_n)$ be the moduli space of hyperbolic metrics on a connected surface of genus $\genus$, with $n$ labelled boundaries, such that the boundaries are geodesic of respective lengths $L_1,\dots,L_n\in \RR_+^n $.
\ed

We shall admit that it is always possible to find $3\genus-3+n$ non intersecting closed geodesic curves, that cut $\curve$ into $2\genus-2+n$ disjoint pairs of pants.

A pant decomposition is not unique.
$$
\includegraphics[scale=0.6]{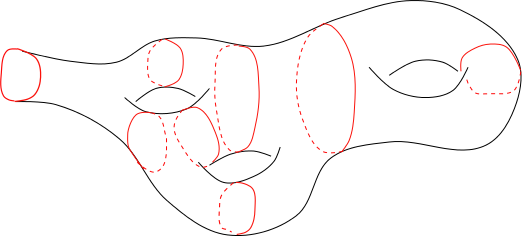}
$$

\bl[Pair of pants]
The moduli space $\mathcal M_{0,3}(L_1,L_2,L_3)$ contains a single element.
In other words, there is a unique (up to isometries) pair of pants, with 3 given boundary lengths $L_1,L_2,L_3$ (this can be extended if some boundary length is 0, to given cusp angle rather than given length).
It is built by gluing the unique hyperbolic right-angles hexagon with 3 edge lengths $L_1/2,L_2/2,L_3/2$ (the other intermediate 3 edge lengths are then uniquely determined as functions of $L_1/2,L_2/2,L_3/2$), and its mirror image, along the 3 other edges (see figure).
\el

$$
\includegraphics[scale=0.3]{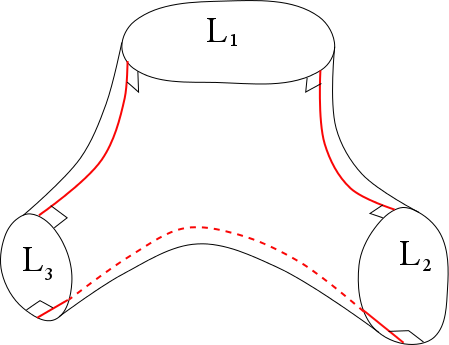}
$$

$$
\includegraphics[scale=0.2]{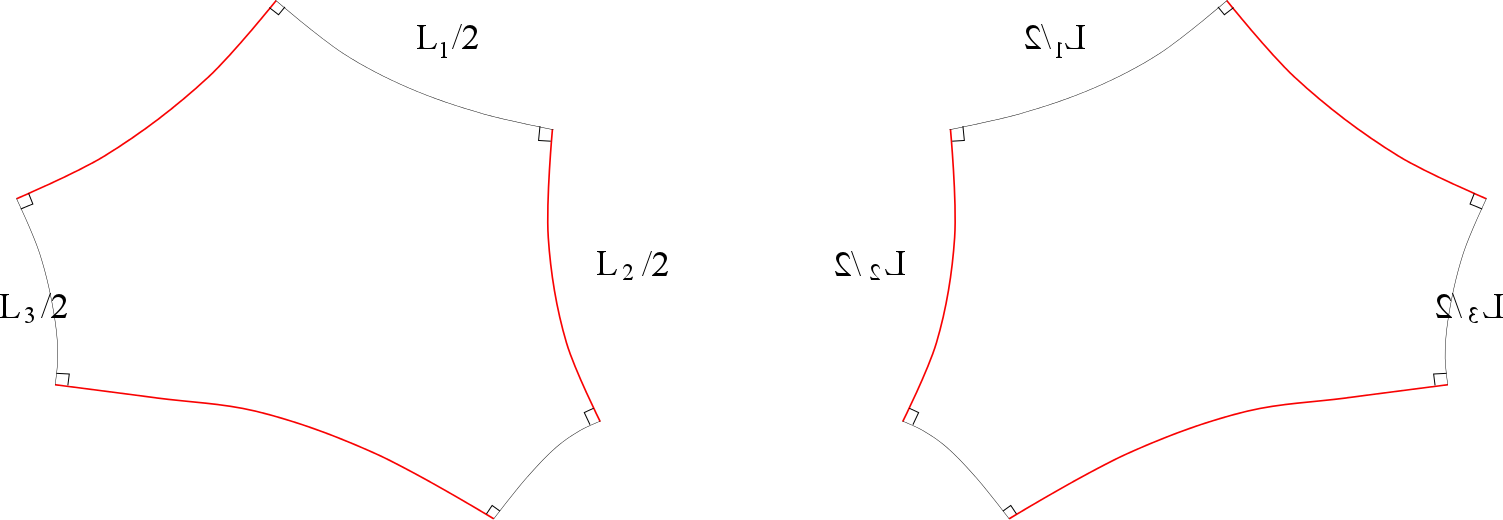}
$$
Notice that each pair of pants carries marked points on its boundary (the points at which geodesics orthogonal to 2 boundaries meet the boundary).

Hyperbolic surfaces can be built by gluing pairs of pants along their geodesic boundaries, provided that the glued boundaries have the same lengths, but then the boundaries can be glued rotated by an arbitrary twisting angle (angle between marked points).
$$
\includegraphics[scale=0.3]{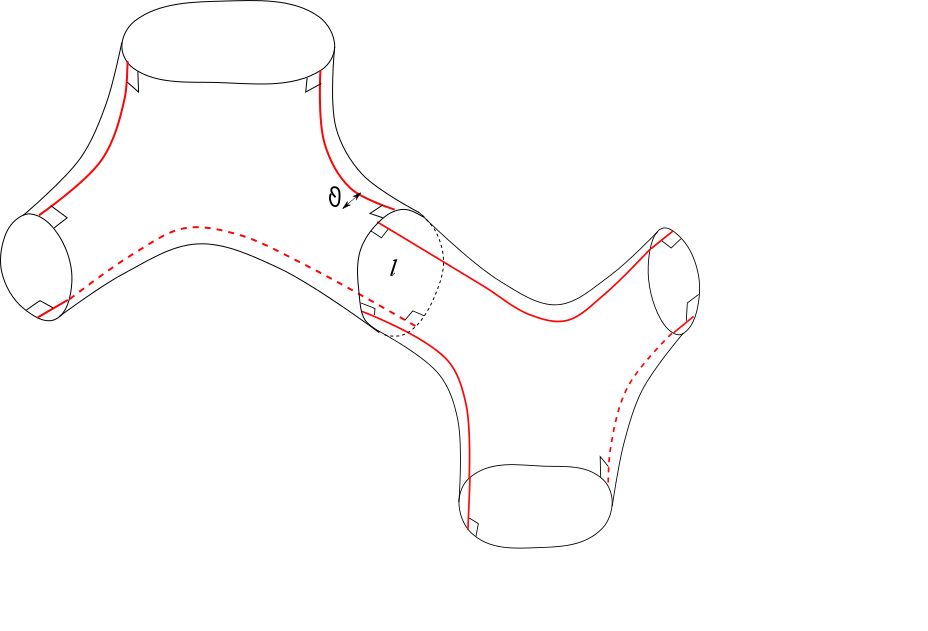}
$$
Every hyperbolic surfaces can be obtained in that way (not uniquely, because of the many ways of cutting the same surface into pairs of pants, but this is a discrete ambiguity). 
This leads to introduce

\bd[Fenchel-Nielsen coordinates]
\index{Fenchel-Nielsen coordinates}
Every hyperbolic surface $\curve\in \mathcal M_{\genus,n}(L_1,\dots,L_n) $, can be built by gluing $2\genus-2+n$ pairs of pants along $3\genus-3+n$ non--intersecting geodesic closed curves.
The $3\genus-3+n$ pairs $(\ell_i,\theta_i)$ of geodesic lengths and twisting angles at the cutting geodesics, are the Fenchel--Nielsen coordinates of $\curve$.

They are local coordinates in $\mathcal M_{\genus,n}(L_1,\dots,L_n) $ (but not global because of the non--uniqueness of the pant decomposition).
\ed
Therefore locally $\mathcal M_{\genus,n}(L_1,\dots,L_n)\sim \RR^{6\genus-6+2n} $, which defines a topology and metric on $\mathcal M_{\genus,n}(L_1,\dots,L_n) $.

It was proved by Weil and Petersson, that the transition maps from a pant decomposition to another, are symplectic transformations in $\RR^{6\genus-6+2n}$ (equipped with the canonical symplectic form), and this allows to define:

\bd[Weil-Petersson form]
The following 2-form on $\mathcal M_{\genus,n}(L_1,\dots,L_n) $:
\beq
\omega = \sum_{i=1}^{3\genus-3+n} d\ell_i \wedge d\theta_i
\eeq
is independent of the pair of pant decomposition, it is a globally defined 2-form on $\mathcal M_{\genus,n}(L_1,\dots,L_n) $.
It is called the \textbf{Weil-Petersson form}\index{Weil-Petersson form}.

Notice that $\omega^{3\genus-3+n}$ is a top--dimensional form on $\mathcal M_{\genus,n}(L_1,\dots,L_n)$, and we define
the \textbf{Weil-Petersson volume}\index{Weil-Petersson volume} of $\mathcal M_{\genus,n}(L_1,\dots,L_n) $ as
\beq
\mathcal V_{\genus,n}(L_1,\dots,L_n) = \frac{1}{(3\genus-3+n)!} \int_{\mathcal M_{\genus,n}(L_1,\dots,L_n) } \omega^{3\genus-3+n}.
\eeq

\ed
It can be proved that the volume is finite, and is a polynomial in the $L_i^2$, moreover, the coefficients of the polynomial, are powers of $\pi^2$ times rational numbers, i.e.
\beq
\mathcal V_{\genus,n}(L_1,\dots,L_n) \in \mathbb Q[L_1^2,\dots,L_n^2,\pi^2]
\eeq
is a homogeneous polynomial of $L_1^2,L_2^2,\dots,L_n^2,\pi^2$ with rational coefficients, of total degree $3\genus-3+n$.
For example
\bea
\mathcal V_{0,3}(L_1,L_2,L_3) = 1
\quad , \quad
\mathcal V_{1,1}(L_1) = \frac{1}{48}\left(4\pi^2+ L_1^2 \right)  \ ,
\cr
\mathcal V_{0,4}(L_1,L_2,L_3,L_4) = 2\pi^2+\frac12 \sum_{i=1}^4 L_i^2.
\eea

\subsubsection{Mirzakhani's recursion}

Maryam Mirzakhani won the Fields medal in 2014 for having found a recursion relation that computes all volumes (recursion on $2\genus-2+n$).

Let us introduce the Laplace transforms of the volumes:
\beq
W_{g,n}(z_1,\dots,z_n) = \int_0^\infty\dots \int_0^\infty \mathcal V_{g,n}(L_1,\dots,L_n) \prod_{i=1}^n e^{-z_i L_i} L_i dL_i,
\eeq
for example
\bea
W_{0,3}(z_1,z_2,z_3) = \frac{1}{z_1^2 z_2^2 z_3^2}
\quad , \quad
W_{1,1}(z_1) = \frac{1}{24z_1^2}\left(2\pi^2+ \frac{3}{z_1^2} \right) \ ,
\cr
W_{0,4}(z_1,z_2,z_3,z_4) = \frac{1}{\prod_{i=1}^4 z_i^2} \left( 2\pi^2+ \sum_{i=1}^4 \frac{3}{z_i^2}\right).
\eea
Observe that these are polynomials of $1/z_i^2$.

Mirzakhani's theorem, restated in Laplace transform is the following recursion
\bt[Mirzakhani's recursion, Laplace transformed]

\bea
W_{g,n+1}(z_1,\dots,z_n,z_{n+1})
&=& \Res_{z\to 0} \frac{dz}{z_{n+1}^2-z^2} \ \frac{\pi}{\sin{2\pi z}}
\ \Big[ W_{g,n-1}(z,-z,z_1,\dots,z_n) \cr
&& + \sum'_{\stackrel{g_1+g_2=g}{I_1\sqcup I_2=\{z_1,\dots,z_n\}}}
W_{g_1,1+|I_1|}(z,I_1)W_{g_2,1+|I_2|}(-z,I_2) \Big] \cr
\eea
where $\sum'$ means that we exclude the terms $(g_1,I_1)=(0,\emptyset)$ and $(g_2,I_2)=(0,\emptyset)$, and where we defined (not the Laplace transform of an hyperbolic volume):
\beq
W_{0,2}(z_1,z_2) = \frac{1}{(z_1-z_2)^2}.
\eeq
\et

This theorem efficiently computes all volumes recursively.
In particular it easily proves that the Laplace transforms are indeed polynomials of $1/z_i^2$, and therefore that the volumes are polynomials of $L_i^2$.

%

\chapter{Eigenvector bundles and solutions of Lax equations}

A good introduction can be found in \cite{BBT}.
Many known integrable systems, can be put in "Lax form", i.e. the Hamilton equations of motions, can be generated by a single matrix equation, called \textbf{Lax equation}\index{Lax equation}
\beq\label{eqLaxintro}
\frac{\partial}{\partial t} L(x,t) = [M(x,t),L(x,t)]
\eeq
where $L(x,t)$ and $M(x,t)$ depend rationally on an auxiliary parameter $x$, that generates the equations, for instance the Taylor expansion in powers of $x$ generates a sequence of matrix equations for the Taylor coefficients.

We shall see now that such equations can be solved by algebraic geometry methods, their solutions can be expressed in terms of $\Theta$--functions.

Equation \eqref{eqLaxintro} implies that the eigenvalues of $L(x,t)$ do not depend on $t$, they are conserved, indeed
\bea
\frac{\partial}{\partial t} \log\det(y-L(x,t)) 
&=& \frac{\partial}{\partial t} \Tr \log(y-L(x,t)) \cr
&=& -  \Tr \ [M(x,t),L(x,t)]\ \  (y-L(x,t))^{-1}  \cr
&=& -  \Tr \ M(x,t)  \ [L(x,t)],  (y-L(x,t))^{-1}]  \cr
&=& 0.
\eea
The \textbf{conserved quantities}\index{conserved quantities} are the Taylor coefficients in the $x$ expansion, of the eigenvalues, or of symmetric polynomials of the eigenvalues, in particular coefficients of the characteristic polynomial:
\beq
\det(y-L(x,t)) = \sum_{k,l} x^k y^l \ P_{k,l}(t)
\qquad \implies \ \frac{\partial}{\partial t} P_{k,l}=0.
\eeq

The time dependence is thus only in the eigenvectors of $L(x,t)$.
As we shall see, the fact that $L(x,t)$ is a rational fraction of $x$, implies that the eigenvalues are algebraic functions of $x$, and the eigenvectors are also algebraic functions of $x$.
Algebraic functions, can be thought of as meromorphic functions on an algebraic curve, i.e. on a Riemann surface.
Meromorphic functions are determined by their behavior at their poles, and thus characterized by a small number of parameters, they can also be decomposed on the basis of $\Theta$-functions.
This will allow to entirely characterize the eigenvectors, and actually find an explicit formula for eigenvectors using $\Theta$-functions.
This is called \textbf{Baker-Akhiezer}\index{Baker-Akhiezer functions} functions.

\section{Eigenvalues and eigenvectors}

Let us for the moment work at fixed time $t$.
The question we want to solve is the following:
let $L(x)$ an $n\times n$ matrix, rational function of $x$:
\beq
L(x)\in M_n(\mathbb C(x)).
\eeq
The eigenvalues and eigenvectors are functions of $x$, what can these functions be ?

\medskip

\subsection{The spectral curve}

Let $P(x,y)=\det(y-L(x))$ be the characteristic polynomial, and $\tilde\curve=\{(x,y) \ | \ P(x,y)=0\}\subset \mathbb CP^1\times \mathbb CP^1$. Let us call $\curve$ its desingularisation, i.e. a compact Riemann surface.
$\curve$ has a projection to $\tilde\curve$, and an immersion into $\mathbb CP^1\times \mathbb CP^1$, and two projections $x$ and $y$ to $\mathbb CP^1$:
\beq
\begin{array}{lll}
 \curve &\to \tilde\curve & \hookrightarrow \mathbb CP^1\times \mathbb CP^1 \cr
 &x\searrow & \searrow \downarrow \cr
&& \mathbb CP^1
\end{array}
\eeq

The eigenvalues of $L(x)$ are thus points $(x,y)\in\tilde\curve$, and should be thought of as points $z\in\curve$.

Locally, in some neighborhood, we may label the preimages of $x$:
\beq
z_1(x),\dots,z_n(x),
\eeq
and thus locally, we may label the eigenvalues $Y_1(x),\dots,Y_n(x)$, with $Y_i(x)=y(z_i(x))$  and define the diagonal matrix
\beq
Y(x) = \operatorname{diag}(Y_1(x),\dots,Y_n(x)).
\eeq

The eigenvalues are algebraic functions of $x\in\CC P^1 $, and thus they are meromorphic functions on $\curve$.

\smallskip

\br
The 1-form $y(z)dx(z)$ is a meromorphic 1-form on $\curve$, it is called the \textbf{Liouville form}\index{Liouville form}.
In fact the 1-form $ydx$ is defined in the whole $\CC P^1 \times \CC P^1$, it is called the \textbf{tautological form}\index{tautological form}, its differential is the 2-form $dy\wedge dx$ the canonical symplectic 2-form in $\CC P^1\times \CC P^1$.
The Liouville form is thus the restriction of the tautological form to the locus of the immersion of the spectral curve.
The immersion of the spectral curve is a Lagrangian with respect to the symplectic form $dy\wedge dx$ of $\CC P^1 \times \CC P^1$.
\er

\subsection{Eigenvectors and principal bundle}

Let $Y_j(x)$ be an eigenvalue of $L(x)$, and $V_j(x)=\{V_{i,j}(x)\}_{i=1,\dots,n}$ be a non--vanishing eigenvector for that eigenvalue.
With $j=1,\dots,n$ we define a complete set of eigenvectors, and define a matrix $V(x)=\{V_{i,j}(x)\} \in GL_n$, with
\beq
\det V(x)\neq 0.
\eeq
We then have
\beq
L(x) =  V(x) Y(x) V(x)^{-1}.
\eeq
However, eigenvectors are not uniquely defined, we may rescale them arbitrarily, and in particular rescale them by a non-vanishing $x$--dependent factor.
This is equivalent to say that we may right--multiply $V(x)$ by an arbitrary $x$--dependent invertible diagonal matrix.

We say that the eigenvector matrix $V(x)$ is a section of a bundle over $\mathbb CP^1$, whose fiber over each point $x$ is the group $GL_n$.
Moreover, when we represent $V(x)$ as a matrix, we assume a choice of basis, and we could change our choice of basis, i.e. conjugate $L(x)$ by an arbitrary matrix, $L(x)\to U L(x) U^{-1}$, equivalent to $V(x)\to U V(x)$.
In other words we are interested in $GL_n$ only modulo left-multiplication, this is called modulo gauge transformation. Somehow we may fix the identity matrix in $Gl_n$ to our will, this is called an affine group.
A bundle whose fiber is an affine Lie group, is called a \textbf{principal bundle}\index{principal bundle}.

\br[Other Lie groups]

So far we have not assumed that $L(x,t)$ had any particular symmetry, we could also require some symmetries conserved under time evolution. This would imply that eigenvectors matrices would belong to a subgroup of $Gl_n$. 
We can obtain any Lie group in this way.
It is thus possible to consider any principal bundle.

The spectral curve also gets extra symmetries, not all coefficients of the characteristic polynomial are independent. The set of independent coefficients is called the \textbf{Hitchin base}\index{Hitchin base}.
The good notion to describe a spectral curve with those extra symmetries, is the notion of \textbf{cameral curve}\index{cameral curve}, beyond the scope of these lectures.

\er

\subsection{Monodromies}

The labelling of eigenvalues can only be local, in a small neighborhood, and when we move $x$ around a closed cycle $\gamma$ (which may surround branchpoints), the eigenvalues get permuted by a permutation $\sigma_\gamma$ (we shall identify the permutation group $\mathfrak S_n$ with its representation as matrices in $Gl_n$), and the eigenvectors get right multiplied:
\bea\label{eqmonodromiesev}
Y(x+\gamma)  &=& \sigma_\gamma^{-1} Y(x) \sigma_\gamma, \cr
V(x+\gamma)  &=& V(x) \sigma_\gamma.
\eea
In other words, the eigenvector bundle has monodromies, and these monodromies are permutations, they are precisely the deck transformations of the spectral curve.

\br
In case the group is a Lie subgroup $G$ of $Gl_n$, the monodromies form a subgroup of $\mathfrak S_n$, in fact they are in the \textbf{Weyl group}\index{Weyl group} of $G$.
\er

\subsection{Algebraic eigenvectors}

Let us first show that it is possible to choose $V(x)$ as an algebraic function of $x$.

More precisely, let $y$ an eigenvalue, i.e. $(x,y)=(x(z),y(z))$ a point of $\tilde\curve$ for $z\in\curve$, and $V(z)=(V_1(z),\dots,V_n(z))$ a corresponding eigenvector.
Since $V(z)\neq 0$, there must exist at least one $i$ such that $V_{i}(z)\neq 0$, and let us assume here that, up to relabelling, $i=n$.
In a neighborhood, we may choose the normalization $V_{n}(z)=1$.

The equation $L(x(z))V(z)=y(z)V(z)$ can  then be written as an $(n-1)\times (n-1)$ linear system
\beq
\forall \ i=1,\dots,n-1 \ ,
\quad
\sum_{k=1}^{n-1} L_{i,k}(x(z)) V_{k}(z)  - y(z) V_{i}(z) = - L_{i,n}(x(z))V_{n}(z) = -L_{i,n}(x(z)).
\eeq
This linear system is solved by Kramers formula
\beq
V_{i}(z) = (-1)^{n-i} \ \frac{M_{i,n}(L(x(z))-y(z)\text{Id})}{M_{n,n}(L(x(z))-y(z)\text{Id})}
\eeq
where $M_{u,v}(A)$ is the minor of the matrix $A$ obtained by removing the $u^{\rm th}$ line and $v^{\rm th}$ column and taking the determinant.
This expression is a rational function of $x(z)$ and $y(z)$, it is thus a meromorphic function of $z$.
Therefore we see that there exists some meromorphic functions $\tilde \psi_i$ such that
\bea
V_{i}(z)
&=& \tilde\psi_i(z) \quad \in \mathfrak M^1(\curve).
\eea

{\it{Remark:}
this formula automatically has the monodromies \eqref{eqmonodromiesev} because monodromies are the deck transformations of $\curve$.}

\smallskip

Instead of having to look for an everywhere invertible $n\times n$ matrix of algebraic functions $V_{i,j}(x)$ for $x\in\mathbb C$, we have to look for a everywhere non--vanishing vector of meromorphic functions $\tilde\psi_i(z)$ on $\curve$.

This is called the "\textbf{Abelianization procedure}\index{abelianization}": we have transformed the problem of finding an algebraic section of a principal bundle of a non-Abelian group over the Riemann sphere $\mathbb CP^1$, into the problem of finding a meromorphic section of a projective vector bundle whose fiber is a projective vector space ($\mathbb CP^n$), over a compact Riemann surface $\curve$ covering $\mathbb CP^1$.
Instead of matrices (thus in a non-Abelian space), we have vectors of functions, the problem has somehow become Abelian.
The price to pay is to have replaced the Riemann sphere by a higher genus  Riemann surface $\curve$ that is a covering of $\CC P^1$.

\smallskip

The next idea is that meromorphic functions are entirely determined by their behavior at their poles and zeros, and their periods.
The Riemann--Roch theorem says what is the dimension (the number of parameters to choose) to characterize all such functions.
The main difficulty is the invertibility of the matrix, or the non-vanishing of the vector.
The full solution was explicitely found by Russian mathematicians (see \cite{Kri77,Dub81, BBT}), and we shall now present the final solution, starting from the end.
We shall "reconstruct" the integrable system, the Lax matrix $L(x,t)$ from the spectral curve.

\subsection{Geometric reconstruction method}
\index{reconstruction method}

Let us now start from a spectral curve $\tilde\curve=\{(x,y) \ | \ P(x,y)=0\}\subset \mathbb CP^1\times \mathbb CP^1$, with $\curve$ its desingularisation.
$\curve$ has a projection to $\tilde\curve$, and an immersion into $\mathbb CP^1\times \mathbb CP^1$, and a projection to $\mathbb CP^1$ by $x$:
\beq
\begin{array}{lll}
 \curve &\to \tilde\curve & \hookrightarrow \mathbb CP^1\times \mathbb CP^1 \cr
 &x\searrow & \searrow \downarrow \cr
&& \mathbb CP^1
\end{array}
\eeq
Let us assume that $\curve$ has a genus $\genus>0$, and choose a symplectic basis of cycles, $\acycle_i,\bcycle_j$, and define the Abel map $z\mapsto \mathbf u(z)$.
Let $c$ a non--singular odd half characteristic (and thus a zero of $\Theta$), and $E$ the corresponding prime form.
Let $\Omega$ be a meromorphic 1-form on $\curve$, of the second kind,  having no residues at its poles.
Let $\zeta(\Omega)\in\mathbb C^\genus$ be the vector with coordinates
\beq
\zeta_i(\Omega) = \oint_{\bcycle_i} \Omega - \sum_j \tau_{i,j} \oint_{\acycle_j} \Omega,
\eeq
which for short we denote
\beq
\zeta(\Omega)=\oint_{\bcycle-\tau\acycle} \Omega.
\eeq

Let us define:
\bd
We define the \textbf{Szeg\"o kernel}\index{Szeg\"o kernel}, for $z$ and $z'$ two distinct points of $\curve$
\beq
\psi(\Omega;z',z) = \frac{e^{\int_{z'}^z \Omega}}{E(z,z')} \ \frac{\Theta(\mathbf u(z)-\mathbf u(z') + \zeta(\Omega)+c)}{\Theta( \zeta(\Omega)+c)}.
\eeq
It is a $\frac12\otimes \frac12$ bi-spinor form on $\curve\times \curve$.
\ed

\bd
For $x$ and $x'$ two distinct points of $\CC P^1$, let us define the $n\times n$ matrix in a neighborhood where is defined an ordering of preimages of $x$ and $x'$
\beq
\Psi(\Omega;x',x)_{i,j} = \psi(\Omega;z_i(x'),z_j(x)).
\eeq
It is a matrix--valued $\frac12\otimes \frac12$ bi-spinor form on $\CC P^1\times \CC P^1 $.
\ed

\bp
It satisfies
\beq
\Psi(\Omega;x_1,x) \Psi(\Omega;x,x_2) = \frac{(x_1-x_2)\ dx}{(x-x_1)(x-x_2)} \ \Psi(\Omega;x_1,x_2)
\eeq
In particular taking the limit $x_1\to x_2=x'$ this gives
\beq
\Psi(\Omega;x',x) \Psi(\Omega;x,x') = \frac{dx \ dx'}{(x-x')^2} \ \text{Id}
\eeq
which shows that for $x\neq x'$, $\Psi(\Omega;x',x)$ is invertible.

\ep

\bp
The matrix
\beq
L(\Omega;x',x) = \Psi(\Omega;x_1,x) Y(x) \Psi(\Omega;x_1,x)^{-1}
\eeq
is a $n\times n$ matrix rational in $x$.
It is algebraic in $x'$, and it depends on $\Omega$.

Remark that changing the choice of $x'$ amounts to a conjugation by an $x$--independent matrix:
\beq
L(\Omega;x'',x) = \Psi(\Omega;x'',x') L(\Omega;x',x) \Psi(\Omega;x'',x')^{-1},
\eeq
in other words $x'$ plays the role of a choice of gauge, i.e. a choice of basis for $GL_n$.

\ep

\bp
Let us choose a basis $\{\Omega_i\}$ (or an independent family) of meromorphic 1-forms of the second kind, and let
\beq
\Omega_t = \sum_i t_i \Omega_i
\eeq
where $t=\{t_i\}$ is called the "time" or more precisely the "times".

Let us define
\beq
L(x';x,t) = \Psi(\Omega_t;x_1,x) Y(x) \Psi(\Omega_t;x_1,x)^{-1}
\eeq
and
\beq
M_i(x';x,t) = \left(\frac{\partial}{\partial_{t_i}}\Psi(\Omega_t;x',x) \right) \  \Psi(\Omega_t;x',x)^{-1}.
\eeq
$M_i(x';x,t)$ is a rational function of $x$ (its poles are the $x$--projections of points where $\Omega_i$ has poles, and of at most the same degrees), and we have the Lax equations
\beq
\frac{\partial}{\partial_{t_i}} L(x';x,t)  = [M_i(x';x,t),L(x';x,t)].
\eeq

\ep

In fact every (finite dimension $n$) solution of Lax equation, can be obtained in this way.
What we see, is that the time dependence, is encoded in the choice of $\Omega_t\in \mathfrak M^1(\curve)$, i.e. times are some linear coordinates in the affine space of meromorphic 1-forms.
This means that, under this parametrization the motion, in the space $\mathfrak M^1(\curve)$ is linear at constant velocity.

The $\genus$--dimensional vector $\zeta(\Omega_t)$ is called the \textbf{angle variables}\index{angle variable}.
It follows a linear motion at constant velocity in $\CC^\genus$.
The velocity is:
\beq
\nu_i=\frac{\partial}{\partial t_i} \zeta(\Omega_t) = \oint_{\bcycle-\tau\acycle} \Omega_i .
\eeq

The \textbf{action variables}\index{action variable} parametrize the spectral curve, and it is usual to choose the $\genus$ dimensional vector of $\acycle$--cycles periods of the Liouville 1-form $ydx$:
\beq
\epsilon_i = \frac{1}{2\ii\pi} \oint_{\acycle_i} ydx.
\eeq
This $\genus$ dimensional vector parametrizes the spectral curve, i.e. the polynomial $P(x,y)=0$, or more precisely it parametrizes all the coefficients of $P$ that are interior of the convex envelope of the Newton's polygon.
The coefficients that are at the boundary of the convex envelope, are called \textbf{Casimirs}\index{Casimir} of our integrable system. 
We have
\beq
ydx = 2\pi\ii \sum_{i=1}^\genus \epsilon_i \ \omega_i + \sum_{(k,l)\in \partial \mathcal N(P)} c_{k,l}\ \omega_{(k-1,l-1)}.
\eeq

\subsection{Genus 0 case}

The previous section assumed that $P(x,y)$ was generic, with the genus of $\curve$ equal to the number of points inside the Newton's polygon, i.e. no cycle pinched to a nodal point.

When some cycles are pinched into nodal points, $\Theta$ functions of genus $\genus$ degenerate and become polynomial combinations of $\Theta$ functions of lower genus.

The extreme case is when  all non--contractible cycles of $\curve$ have been pinched into nodal points of $\tilde\curve$, the genus of $\curve$ is then 0.

Let $(p_{i,+},p_{i,-})_{i=1,\dots,N}$ be the $N$ nodal points ($N=\#\stackrel{\circ}{\mathcal N}$ is the genus of the unpinched curve) i.e. all the pairs of distinct points of $\curve$ that have the same projection to $\tilde\curve$:
\beq
x(p_{i,+})=x(p_{i,-})
\quad \text{and} \quad
y(p_{i,+})=y(p_{i,-}).
\eeq

The theta functions of a pinched curve degenerate into determinants of rational functions, and the Szeg\"o kernel degenerates into
\beq
\psi(\Omega;z',z) =  \frac{\det_{0\leq i,j\leq N} \frac{e^{\int_{p_{j,-}}^{p_{i,+}} \Omega}}{p_{i,+}-p_{i,-}}}
{\det_{1\leq i,j\leq N} \frac{e^{\int_{p_{j,-}}^{p_{i,+}} \Omega}}{p_{i,+}-p_{i,-}}}
\eeq
where we defined $p_{0,+}=z$ and $p_{0,-}=z'$.

\subsection{Tau function, Sato and Hirota relation}

Let us come back to the non--degenerate case where there is no pinched cycle.

\bd[Tau function]
\beq
\Tau(\Omega) = e^{\frac12 \sum_{i,j} Q_{i,j} t_i t_j  } \ \Theta(\zeta(\Omega)+c)
\eeq
where, introducing the generalized cycle $\Omega_i^*\in \mathfrak M_1(\curve)$ that generates $\Omega_i = \oint_{\Omega^*_i} B$ in theorem III-\ref{BtoM1} (using  the meromorphic function $f=x$ in theorem III-\ref{BtoM1}), we define the quadratic form as the integral (i.e. the Poincar\'e pairing):
\beq
Q_{i,j} = \oint_{\Omega^*_i}\Omega_j = <\Omega^*_i,\Omega_j>.
\eeq
We thus have
\beq
\sum_{i,j} Q_{i,j} t_i t_j = \sum_i t_i \oint_{\Omega^*_i}\Omega= \oint_{\sum_i t_i \Omega^*_i}\Omega = \oint_{\Omega^*}\Omega =  \oint_{\Omega^*} \oint_{\Omega^*} B.
\eeq
\ed

Notice that
\beq
\zeta(\Omega) = \oint_{\bcycle-\tau\acycle} \Omega = \oint_{\bcycle-\tau\acycle}  \oint_{\Omega^*} B = 2\pi\ii (\bcycle-\tau\acycle) \cap \Omega^*.
\eeq
so that
\beq
\Tau(\Omega) = e^{\frac12 \oint_{\Omega^*} \oint_{\Omega^*} B  } \ \Theta(c+2\pi\ii (\bcycle-\tau\acycle) \cap \Omega^*)
\eeq

\br
We see that in fact, using the form--cycle duality, it seems easier and more natural to define the Tau function in the space of cycles $\Omega^*$ rather than the space of 1-forms $\Omega$.
What is hidden here, is that the map $\hat B:\Omega^* \mapsto \Omega=\oint_{\Omega^*} B$ is not invertible, it has a kernel (a huge kernel).
The map $\Omega\mapsto \Omega^*$ is ill-defined, it can be defined only by choosing  representents of equivalence classes modulo $\Ker \hat B$, i.e. as in 
 theorem III-\ref{BtoM1} make an explicit choice of basis of $\mathfrak M_1(\curve)$.
 We could change this basis by shifting with elements of $\Ker \hat B$. 
 Doing so would change the quadratic form, and would change the Tau function by multplication by a phase.
The choice of basis is in fact a choice of a Lagrangian \textbf{polarization}\index{polarization} in $\mathfrak M_1(\curve)$, and thus the Tau function is not unique, it depends on a choice of Lagrangian polarization.
Under a --time independent-- change of Lagrangian polarization, $\Tau$ gets multiplied by $e^S$ where $S$ is the generating function of the Lagrangian change of polarization.

\er

\bt[Sato]
\index{Sato}
The Baker Akhiezer function is a ratio of the Tau function shifted by a 3rd kind form
\beq
\psi(\Omega;z',z) = \frac{\Tau(\Omega+\omega_{z',z})}{\Tau(\Omega)}.
\eeq
\et
\proof
It is obvious by explicit computation.
\eproof

Notice that the ratio of $\Tau$-functions is independent of a "choice of polarization".

\bt[Fay identities and Pl\"ucker relations]
\index{Fay}
\index{Pl\"ucker}

\beq
\frac{\Tau(\Omega+\omega_{z_1,z_2}+\omega_{z_3,z_4})}{\Tau(\Omega)} = \frac{\Tau(\Omega+\omega_{z_1,z_2})}{\Tau(\Omega)} \ \frac{\Tau(\Omega+\omega_{z_3,z_4})}{\Tau(\Omega)} - \frac{\Tau(\Omega+\omega_{z_1,z_4})}{\Tau(\Omega)} \ \frac{\Tau(\Omega+\omega_{z_3,z_2})}{\Tau(\Omega)}.
\eeq
More generally
\beq
\frac{\Tau(\Omega+\sum_{i=1}^n \omega_{z_i,z'_i})}{\Tau(\Omega)} =
\det_{1\leq i,j\leq n} \left( \frac{\Tau(\Omega+\omega_{z_i,z'_j})}{\Tau(\Omega)} \right) .
\eeq

\et

\proof
These are \textbf{Fay identities} for Theta functions \cite{Fay}.
This can be proved by showing that the ratio of the right and left side, is a well defined meromorphic function on $\curve$ (in particular there is no phase when some $z_i$ goes around a cycle), and has no pole, therefore it must be a constant.
The constant is seen to be 1 in a limit $z_i\to z_j$.
\eproof

\bd[Hirota derivative]
For a  function $f$ on $\mathfrak M^1(\curve)$, and for any $z\in\curve$, choosing a chart and local coordinate $\phi$ in a neighborhood of $z$, we defined (in theorem III-\ref{BtoM1} using $\phi$) the 1-form $\omega_{\phi,z,1}$ that has a double pole at $z$.
We define
\beq
\Delta_z f(\Omega) = d\phi(z) \  \lim_{\epsilon\to 0} \frac{1}{\epsilon} \ \left( f(\Omega+\epsilon \ \omega_{\phi,z,1}) - f(\Omega) \right)
\eeq
It is a meromorphic 1-form of $z$ on $\curve$, it is independent of the choice of chart and coordinate.
\ed

\bp
Let $p\in\curve$ in a chart $U$, and a coordinate $\phi$ in $U$.
For $z$ in a neighborhood of $p$ we define the KP times as the negative part coefficients of the Taylor--Laurent expansion
\beq
\Omega \sim \sum_{k=0}^{-\order_p ydx} t_{p,k} \ \frac{d\phi(z)}{(\phi(z)-\phi(p))^{k+1}} + \text{analytic at }p.
\eeq
Then the Hirota derivative can be locally written as the following series of times derivatives
\beq\label{eqHirotausual}
\Delta_z \sim d\phi(z) \sum_{k=1}^\infty k \ (\phi(z)-\phi(p))^{k-1} \ \frac{\partial}{\partial t_{p,k}} 
\eeq
\ep

\eqref{eqHirotausual} is the usual way of writing the Hirota operator for KP hierachies, but as we see here, this is just an asymptotic Taylor expansion in a neighborhood of $p$, whereas the Hirota operator is globally defined on $\curve$.
In other words, for $z$ in a neighborhood of $p$
\beq
\Delta_z
- d\phi(z) \sum_{k=1}^n k \ (\phi(z)-\phi(p))^{k-1} \ \frac{\partial}{\partial t_{p,k}} 
= O((\phi(z)-\phi(p))^{n}) d\phi(z).
\eeq

\proof
In a local coordinate $\phi$, if $|\phi(z)-\phi(p)|<|\phi(q)-\phi(p)|$ we have
\bea
\omega_{\phi,z,1}(q)
& \sim & \sum_{k=1}^\infty k \ \frac{(\phi(z)-\phi(p))^{k-1}}{(\phi(q)-\phi(p))^{k+1}} \ d\phi(q) \cr 
& \sim & \sum_{k=1}^\infty k \ (\phi(z)-\phi(p))^{k-1} \ \omega_{p,k}(q) .
\eea
from which we see that the Hirota derivative acts as \eqref{eqHirotausual}.

\eproof

\bt[Hirota equations]
\index{Hirota}
\beq
\Delta_z\frac{\Tau(\Omega+\omega_{z_1,z_2})}{\Tau(\Omega)} = -\  \frac{\Tau(\Omega+\omega_{z_1,z})}{\Tau(\Omega)} \ \frac{\Tau(\Omega+\omega_{z,z_2})}{\Tau(\Omega)} ,
\eeq
this can also be written
\beq
\Delta_z \psi(\Omega;z_1,z_2)
 = - \psi(\Omega;z_1,z) \ \psi(\Omega;z_1,z) .
\eeq

\et

\proof
This is the limit $z_3\to z_4=z$ of the Fay identities.
\eproof

\bp[Sato formula as a shift of times]
Let $p\in\curve$ in a chart $U$, and a coordinate $\phi$ in $U$.
For $z$  in a neighborhood of $p$ the Sato formula can be written
as the Taylor expansion
\beq\label{eqSatousual}
\Tau(\Omega+\omega_{z,z'})
\sim \Tau(\Omega+\omega_{p,z'}+\sum_{k=1}^\infty (\phi(z)-\phi(p))^{k} \ \omega_{p,k})
\eeq
in other words, writing $\Omega=\sum_{k} t_{p,k} \omega_{p,k} + \text{analytic at}\ p$, the Sato shift is equivalent to
\beq
t_{p,k} \to t_{p,k} + (\phi(z)-\phi(p))^k.
\eeq
Similarly, if $z'$ is in a neighborhood of $p$ the Sato formula can be written
as the Taylor expansion
\beq\label{eqSatousual1}
\Tau(\Omega+\omega_{z,z'})
\sim \Tau(\Omega-\omega_{p,z}-\sum_{k=1}^\infty (\phi(z')-\phi(p))^{k} \ \omega_{p,k})
\eeq
in other words, the Sato shift is equivalent to
\beq
t_{p,k} \to t_{p,k} - (\phi(z')-\phi(p))^k.
\eeq
And if both $z$ and $z'$ are in a neighborhood of $p$ the Sato formula can be written as the Taylor expansion
\beq\label{eqSatousual2}
\Tau(\Omega+\omega_{z,z'})
\sim \Tau(\Omega+\sum_{k=1}^\infty (\phi(z)-\phi(p))^{k} \ \omega_{p,k}-\sum_{k=1}^\infty (\phi(z')-\phi(p))^{k} \ \omega_{p,k})
\eeq
in other words,  the Sato shift is equivalent to
\beq
t_{p,k} \to t_{p,k} + (\phi(z)-\phi(p))^k- (\phi(z')-\phi(p))^k.
\eeq
\ep

\proof
In a local coordinate $\phi$, if $|\phi(z)-\phi(p)|<|\phi(q)-\phi(p)|$ we have
\bea
\omega_{z,z'}(q)
& \sim & \sum_{k=0}^\infty \ \frac{(\phi(z)-\phi(p))^{k}}{(\phi(q)-\phi(p))^{k+1}} \ d\phi(q) \cr 
& \sim & \omega_{p,z'}(q) + \sum_{k=1}^\infty  \ (\phi(z)-\phi(p))^{k} \ \omega_{p,k}(q) .
\eea
\eproof

\subsubsection{To go further}

Readers interested in learning more about this way of presenting the algebraic reconstruction method, and in particular in the space of cycles $\mathfrak M_1(\curve)$ can see \cite{EynISlectures}.

\end{document}